\documentclass{aa}
\usepackage{dcolumn}
\usepackage{natbib,amssymb}
\usepackage[pdftex]{graphicx,color}
\newcolumntype{.}{D{.}{.}{-1}}
\newcolumntype{;}{D{;}{.}{7}}

\begin{document}

\authorrunning{Zamaninasab et al.}
\titlerunning{ NIR flares of Sgr A*}
\title{\huge Near infrared flares of Sagittarius~A*} \subtitle{Importance of near infrared polarimetry}
\author{M. Zamaninasab\inst{1,2}
          \and
           ~A.~Eckart\inst{1,2}
          \and
           ~G.~Witzel\inst{1}
          \and
           ~M.~Dovciak\inst{3}
          \and
           ~V.~Karas\inst{3}
          \and
           ~R.~Sch\"odel\inst{4}
          \and
           ~R.~Gie{\ss}\"{u}bel\inst{1,2}
          \and
           ~M.~Bremer\inst{1}
          \and
           ~M.~Garc\'{\i}a-Mar\'{\i}n\inst{1}
          \and
           ~D.~Kunneriath\inst{1,2}
          \and
           ~K.~Mu\v{z}i\'{c}\inst{1}
          \and
           ~S.~Nishiyama\inst{5}
          \and
           ~N.~Sabha\inst{1}
          \and
           ~C.~Straubmeier\inst{1}
          \and
           ~A.~Zensus\inst{2,1}   }
\offprints{M. Zamaninasab (zamani@ph1.uni-koeln.de)}

\institute{I.Physikalisches Institut, Universit\"at zu K\"oln,
             Z\"ulpicher Str.77, 50937 K\"oln, Germany
         \and
             Max-Planck-Institut f\"ur Radioastronomie,
             Auf dem H\"ugel 69, 53121 Bonn, Germany
         \and
             Astronomical Institute, Academy of Sciences,
             Bo\v{c}n\'{i} II, CZ-14131 Prague, Czech Republic
         \and
             Instituto de Astrof\'{\i}sica de Andaluc\'{\i}a (IAA)-CSIC,
             Camino Bajo de Hu\'{e}tor 50, 18008 Granada, Spain
         \and
             Department of Astronomy, Kyoto University, Kyoto 606-8502, Japan
}

\abstract{ We report on the results of new simulations of
near-infrared (NIR) observations of the Sagittarius~A* (Sgr~A*)
counterpart associated with the super-massive black hole at the
Galactic Center.}{Our goal is to investigate and understand the
physical processes behind the variability associated with the NIR
 flaring emission from Sgr~A*. }
{ The observations have been carried out using the  NACO
adaptive optics (AO) instrument at the European Southern
Observatory's Very Large Telescope and CIAO NIR camera on the Subaru
telescope (13 June 2004, 30 July 2005, 1 June 2006, 15 May 2007, 17
May 2007  and 28 May 2008). We used a model of synchrotron emission
from relativistic electrons in the inner parts of an accretion disk.
The relativistic simulations have been carried out using the
Karas-Yaqoob (KY) ray-tracing code.} {We probe the
existence of a correlation between the modulations of the observed
flux density light curves and changes in polarimetric data.
Furthermore, we confirm that the same correlation is also predicted
by the hot spot model. Correlations between intensity and
polarimetric parameters of the observed light curves as well as a
comparison of predicted and observed light curve features through a
pattern recognition algorithm result in the detection of a signature
of orbiting matter under the influence of strong gravity. This
pattern is detected statistically significant against randomly
polarized red noise.
Expected results from future observations of VLT interferometry like
GRAVITY experiment are also discussed.}{ The observed correlations
between flux modulations and changes in linear polarization degree
and angle can be a sign that the NIR flares have properties that are
not expected from purely random red-noise. We find that the
geometric shape of the emission region plays a major role in the
predictions of the model. From fully relativistic simulations of a
spiral shape emitting region, we conclude that the observed swings
of the polarization angle during NIR flares support the idea of compact
orbiting spots instead of extended patterns. The effects of
gravitational shearing, fast synchrotron cooling of the components
and confusion from a variable accretion disk have been taken into
account. Simulated centroids of NIR images lead us to the conclusion
that a clear observation of the position wander of the center of NIR
images with future infrared interferometers will prove the existence
of orbiting hot spots in the vicinity of our Galactic super-massive
black hole.}


\maketitle

\section{Introduction}

\begin{figure*}[!htb]
\begin{minipage}{\textwidth}
\centering{\includegraphics[width=\textwidth]{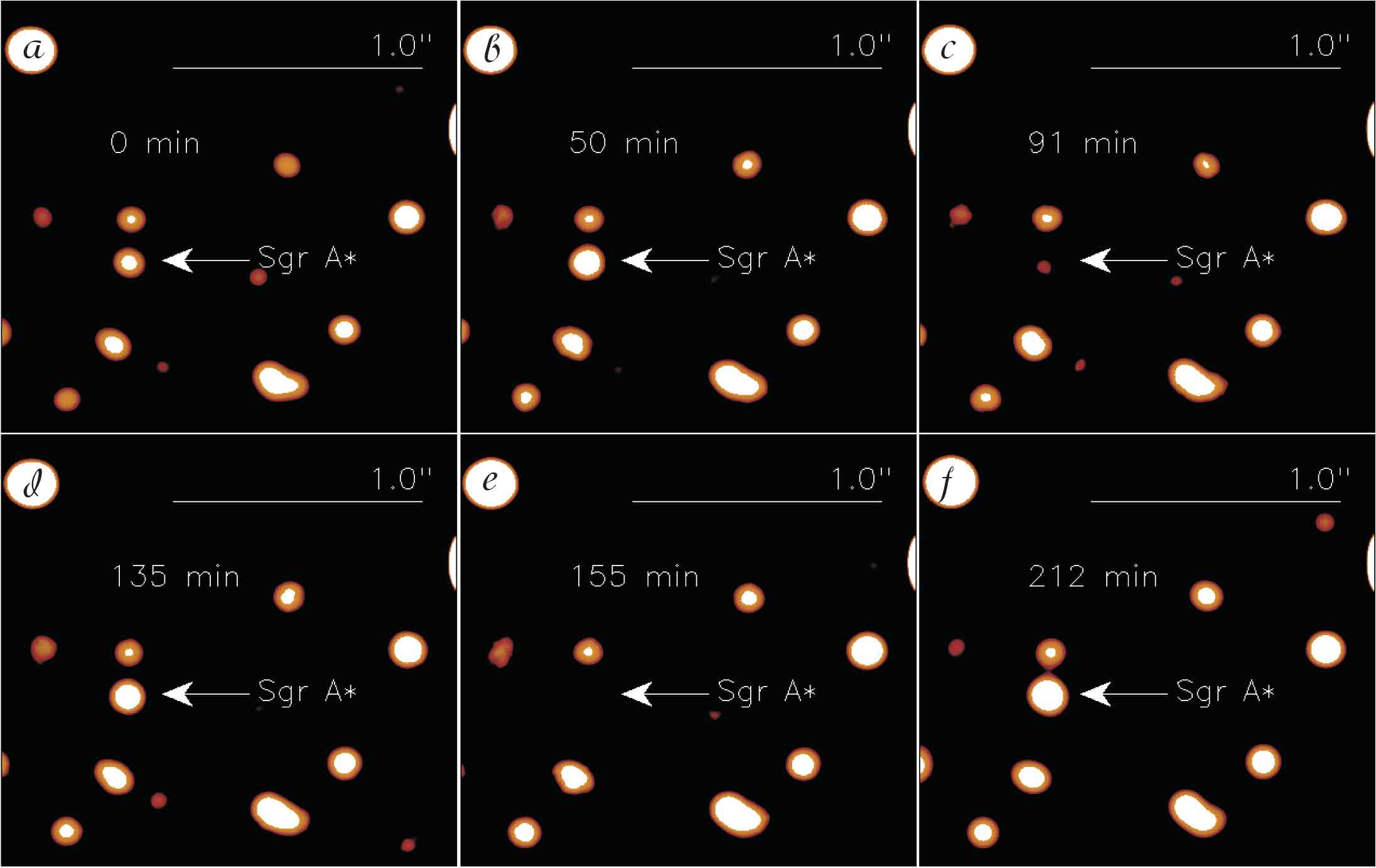}}
\end{minipage}
\caption{Sgr A* as it was observed in  NIR $L^{'}$-band
(3.8 $\mu$m) on 3 June 2008 between 05:29:00 - 09:42:00 ~(UT time).
({\it a-f}) show the observed images of Sgr A* after 0, 50, 91, 135,
155 and 212 minutes off the start of observation. The images show
that when Sgr A* is in its flaring state the flux changes up to
100\% in time intervals  of the order of only tens of minutes.}
\label{label}
\end{figure*}

The nearest super-massive black hole candidate ($\sim~4\times10^6
M_\odot$) lies at the center of our galaxy, as inferred from motions
of stars near the Galactic Center (Eckart \& Genzel 1996, 1997;
Eckart et al. 2002; Sch\"{o}del et al. 2002; Eisenhauer et al. 2003;
Ghez et al. 2000, 2005, 2008; Gillessen et al. 2009). With a
luminosity of $10^{-9}-10^{-10} L_{Edd}$, where $L_{Edd}$ is its
limiting Eddington luminosity, Sagittarius A*, the radio source
associated with this SMBH, is one of the most extreme sub-Eddington
sources accessible to observations. However, X-ray and near-infrared
(NIR) flares are routinely detected with high spatial and spectral
resolution observations (Baganoff et al. 2001; Porquet et al. 2003,
2008; Genzel et al. 2003; Eckart et al. 2004, 2006a-c, 2008a-c;
Meyer et al. 2006a,b, 2007; Yusef-Zadeh et al. 2006a,b, 2007, 2008).
These short bursts of increased radiation last normally for about
100 minutes and occur four to five times a day (see Fig. 1 for a typical
behavior of Sgr~A* during a flaring phase in NIR bandwidth).

Recent NIR and X-ray observations have revealed the non-thermal
nature of  high frequency radiation from Sgr~A* (Eckart et al.
2006a-c, 2008a-c; Gillessen et al. 2006; Hornstein et al. 2007).
Sgr~A* is probably visible in the NIR regime only during its flaring
state. The short time scale variabilities seen during several
observed NIR and X-ray flares argue for an emitting region not
bigger than about ten Schwarzschild radii
($r_s=\frac{2GM}{c^2}=2r_g=1.2\times10^{12}\big(\frac{M}{4\times10^6
M_{\odot} }\big)$ cm) of the associated super-massive black hole
(Baganoff et al. 2001; Genzel et al. 2003). We have scaled the
relevant physical distances according to the gravitational radius
($r_g$) throughout this paper. The NIR flares are highly polarized
and normally have X-ray counterparts, which strongly suggests a
synchrotron-self-Compton (SSC) or inverse Compton emission as the
responsible radiation mechanism (Eckart et al. 2004, 2006a,b; Yuan
et al. 2004; Liu et al. 2006). Several observations have already
confirmed the existence of a time lag between the simultaneous
NIR/X-ray flares and the flares in the lower frequencies. This is
interpreted as a sign for cooling down via adiabatic expansion
(Eckart et al. 2006a, 2008b,c; Yusef-Zadeh et al. 2006a,b, 2007,
2008; Marrone et al. 2008, Zamaninasab et al. 2008a).

\begin{figure*}[!t]
\begin{minipage}{1.\textwidth}
\centering{\includegraphics[width=\textwidth]{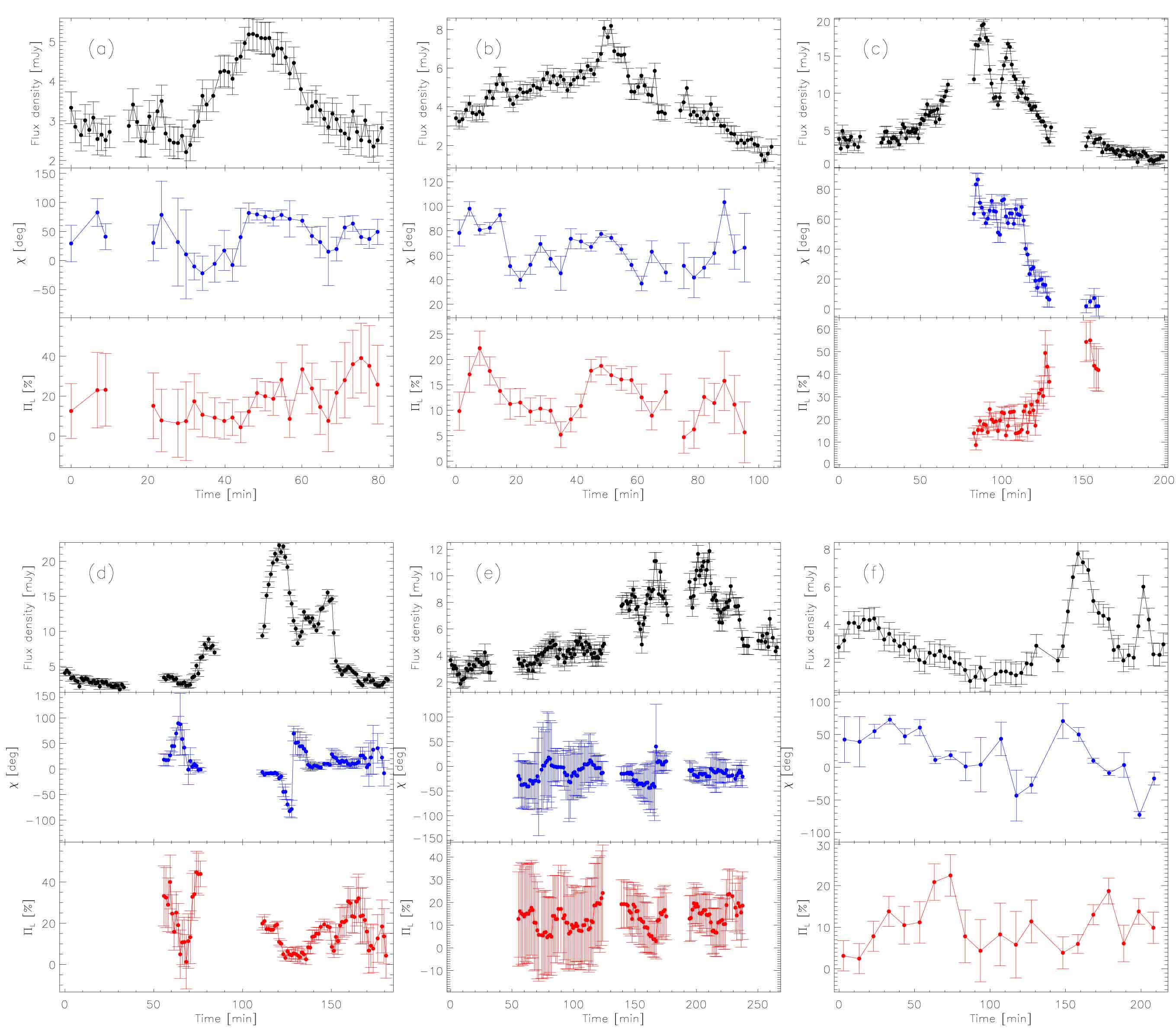}}
\end{minipage}
\caption{Our sample of light curves of Sgr~A* flares
observed in NIR $K_s$ band (2.2$\mu$m) polarimetry mode. The events
were observed on  2004 June 13 (a), 2005 July 30 (b), 2006 June 1
(c), 2007 May 15 (d), 2007 May 17 (e) and  2008 May 28 (f). In each
panel, the top shows the de-reddened flux density measured in mJy
(black), the middle shows the polarization angle (blue) and the
bottom shows the degree of linear polarization (red). The gaps in
the light curves are due to the sky background measurements.}
\label{events}
\end{figure*}

\begin{table*}[btp]
\begin{center}
\begin{tabular}{lcccccccr}\hline \hline
Date (Telescope) & Spectral domain & UT start & UT stop & Max. flux  & Min. flux& Average flux & Average polarization  \\
 &  & time & time &  & & sampling rate & sampling rate  \\
       \hline \hline \\
2004 June 13 (NACO)& 2.2$\mu$m & 07:20:02 & 09:15:08  & 5.19 mJy & 2.21 mJy & 1.2 min &  3 min \\
      \\
       &        &          &       &       \\
2005 July 30 (NACO)& 2.2$\mu$m & 02:07:36 & 07:03:39 &  8.19 mJy & 1.23 mJy & 1.2 min & 3 min \\
             &        &          &       &      \\
       &        &          &       &       \\
2006 June 1 (NACO)& 2.2$\mu$m & 04:26:03 & 10:44:27 & 19.33 mJy & 0.72 mJy & 1.5 min & 2 min \\
             &        &          &       &      \\
             \\
2007 May 15 (NACO)& 2.2$\mu$m & 09:08:14 & 09:42:12 & 22.27 mJy & 1.70 mJy & 1.5 min & 2 min \\
             &        &          &       &      \\
             \\
2007 May 17 (NACO)& 2.2$\mu$m & 04:42:14 & 09:34:40 & 11.87 mJy & 1.86 mJy & 1.5 min & 2 min \\
             &        &          &       &      \\
             \\
2008 May 28 (CIAO)& 2.15$\mu$m & 09:22:51 & 13:00:37 & 7.70 mJy & 0.97 mJy & 3.3 min & 10 min \\
             &        &          &       &      \\
\hline \hline
\end{tabular}
\end{center}
\caption{Observations log.} \label{table1}
\end{table*}

The other feature related to these NIR/X-ray flares are the
claimed quasi-periodic oscillations (QPOs) with a period of $20\pm5$
minutes, which have been reported  in several of these events
(Genzel et al. 2003; Belanger et al. 2006; Eckart et al. 2006b,c;
2008a; Meyer et al. 2006a,b, Hamaus et al. (2009)). Short periods of
increased radiation (the so called "NIR flares", normally around 100
minutes) seem to be accompanied by QPOs. All the studies mentioned
above probed this 20$\pm$5 minutes quasi-periodicity, by performing a sliding window analysis with
window lengths of the order of the flaring time. Recently, Do et al.
(2009) argued that they did not find any significant periodicity at
any time scale while probing their sample of observations for a
periodic signal. Their method is based on the Lomb-Scargle
periodogram analysis of a sample of six light curves and comparing
them with several thousands of artificial light curves with the same
underlying red-noise. One must note that the suggested
QPOs are transient phenomena, lasting for only very few cycles
(50-100 minutes). This kind of behavior, along with the inevitable
uncertainty  in the red noise power law index determination, makes a
clear and unambiguous detection ($>5\sigma$) of a periodic signal very
difficult. Whenever a flare of Sgr~A* was observed with polarimetry,
it was found that it is accompanied by significant polarization which
varies on similarly short timescales as the light curve itself. By carrying
out an analysis only on the total flux, some pieces of
the observed information are ignored. It is already known that
polarimetric data have been shown to be able to reveal substructure
in flares, even when the light curve appears largely featureless
(e.g. see Fig. 4 in Eckart et al. 2006b). The other main advantage
of polarimetric observations is that, in addition to the flux
density light curve, one can analyze the changes in the observed
degree of polarization and the changes in polarization angle
during flaring time as well.

The claimed quasi-periodicity has been interpreted as being related to the orbital time
scale of the matter in the inner parts of the accretion disk. According
to well-known observed high frequency quasi-periodic oscillations
(HFQPOs) in X-ray light curves of stellar mass black holes and
binaries (Nowak \& Lehr 1998), this interpretation is of special
interest since it shows a way to better understand  the behavior of
accretion disks for a wide range of black hole masses. The recent
unambiguous discovery of a one hour (quasi-)periodicity in the X-ray
emission light curve of the active galaxy RE J1034+396 provides further
support to this idea and extends the similarity between
stellar-mass and super-massive black holes to a new territory
(Gierli\'{n}ski et al. 2008).

Although the origin of the observed QPOs in the sources associated
with black holes is still a matter of debate, several
magnetohydrodynamic (MHD) simulations confirmed that it could be
related to instabilities in the inner parts of the accretion disks,
very close to the marginally stable orbit of the black hole
($r_{mso}$), and also possibly connected with the so-called "stress
edge" (Hawley 1991; Chan et al. 2009b). If the flux modulations are
related to a single azimuthal compact over-dense region (hereafter:
"hot spot"), orbiting with the same speed as the underlying
accretion disk, one can constrain the spin of the black hole by
connecting the observed time scales of QPOs to the orbital time
scale of matter around the black hole: $T = 2.07
(r^{\frac{3}{2}}+a)(\frac{M}{4\times10^6M_\odot})$~min (Bardeen et
al. 1972, where $ -1\leq a\leq1$ is the black hole dimensionless
spin parameter and $r$ is the distance of the spot from the black
hole). The characteristic behavior of general relativistic flux
modulations produced via the orbiting hot spots have been discussed
in several papers (see e.g. Cunningham \& Bardeen 1973; Abramowicz
et al. 1991; Karas \& Bao 1992; Hollywood et al. 1995; Dov\v{c}iak
2004, 2007). In this paper, we have used a spotted accretion disk
 scenario to model the observed patterns of our sample of NIR
light curves.

Several authors have proposed different models in order to explain
the flaring activity of Sgr~A*. These models cover a wide range of
hypotheses like disk-star interactions (Nayakshin et al. 2004),
stochastic acceleration of electrons in the inner region of the disk
(Liu et al. 2006), sudden changes in the accretion rate of the black
hole (Liu et al. 2002), heating of electrons close to the core of a
jet (Markoff et al. 2001; Yuan et al. 2002), trapped oscillatory
modes in the inner regions of the accretion disk in the form of
spiral patterns or Rossby waves (Tagger \& Melia 2006; Falanga et
al. 2007; Karas et al. 2008), non-axisymmetric density perturbations
which emerge as the disk evolves in time (Chan et al. 2009b),
non-Keplerian orbiting spots falling inward inside the plunging
region created via magnetic reconnections (Falanga et al. 2008), and
also  comet like objects trapped and tidally disrupted by the black
hole (\v{C}ade\v{z} et al. 2006; Kosti\'{c} et al. 2009).

Observational data render some of these models unlikely. The
star-disk interaction model is unable to produce the repeated flux
modulations and also the high degree of polarization since it mainly
deals with thermal emission. Tidal disruption of  comet-like objects
are also unable to reproduce the observed rate of  flares per  day,
since the estimated capture rate of such objects for the Sgr~A*
environment is at least one order of magnitude lower (\v{C}ade\v{z}
et al. 2006). Nevertheless, several viable models exist and make
different predictions that can be distinguished observationally. For
example, one important characteristic prediction of hot spot models
is about the wobbling of the center of the images (Broderick \& Loeb
2006a,b; Paumard et al. 2006; Zamaninasab et al. 2008b; Hamaus et
al. 2009). Significant  effort has been already devoted to measure
this possible position wander of Sgr~A* in the mm, sub-mm and NIR
regimes (Eisenhauer 2005b, 2008, Gillesen 2006, Reid et al. 2008,
Doelleman et al. 2008). In this paper we discuss how NIR polarimetry
and the next generations of VLT interferometry (VLTI) and Very Long
Baseline Interferometry (VLBI) experiments can provide data to
support or reject certain models for the accretion flow/outflow
related to Sgr~A*. Obtaining accurate data on the accretion flow of
Sgr~A* can lead us to a better understanding of the physics of
strong gravitational regimes, formation of black holes and their
possible relation to the galaxy formation process in a cosmological
context.

 In Sect. 2 we present a complete sample of NIR light curves
observed in  the polarimetry mode. A brief description about the
details of the observation and data reduction methods is provided. We
discuss  the quasi-periodicity detection methods and present the
results of a correlation analysis between the flux and polarimetric
parameters. A general description of our model setup and results of
simulations are discussed in Sect. 3. We show how NIR polarimetry can be
used as a way to constrain physical parameters of the emitting
region (like its geometrical shape) in Sect. 4 and Sect. 5. In Sect. 6 we mainly
discuss the predictions that the future NIR interferometer (GRAVITY)
is expected to reveal and how different assumptions in the model
parameters can modify the results. In Sect. 7 we summarize the main
results of the paper and draw our conclusions.

\section{NIR polarimetry}
\subsection{Observations and data reduction}
All observations we refer to in this paper have been carried
out in the NIR $K_s$ band with the NIR camera CONICA and the
adaptive optics (AO) module NAOS (NACO) at the ESO VLT unit
telescope 4 (YEPUN) on Paranal, Chile\footnote{Based on observations
at the Very Large Telescope (VLT) of the European Southern
Observatory (ESO) on Paranal in Chile; Programs:075.B-0093 and
271.B-5019(A).} and CIAO NIR camera on the Subaru telescope
\footnote{Based on data collected at Subaru Telescope, which is
operated by the National Astronomical Observatory of Japan.}. These
facilities are suited for both time resolved observations of total
intensity and  polarimetric degree and angle  with a sampling
of about two to three minutes. The NAOS/CONICA NIR camera installed on
UT4, VLT allows for a simultaneous measurement of two orthogonal
directions of electric field vector via a Wollaston prism. The
combination with a rotary half-wave plate allows the rapid alternation
 between measurements of different angles of the electric
vector. This is crucial for  determining the linear polarization
characteristics of a time-varying source. The CIAO camera uses a
rotating half-wave plane combined with a fixed wire grid polarizer
for measuring the linear polarization.

Since the
first NIR  polarimetric observation of Sgr A* in 2004 (Eckart et al.
2006a), several polarized flares have been observed (Meyer et al.
2006 a,b, 2007;  Eckart et al. 2008a).  In the VLT observations, the
infrared wavefront sensor of NAOS was used to lock the AO loop on
the NIR bright (K-band magnitude $\sim6.5$) supergiant IRS~7,
located about $5.6''$ north of Sgr~A*. Atmospheric conditions (and
consequently the AO correction) were stable enough during the
observations for doing high angular resolution photometry and
polarimetry (with a typical coherence time of 2 milliseconds and larger).
The exposures have been dithered. The reductions of the 1 June 2006
and 15 May 2007 data presented here have been repeated for this
publication to confirm the significance of the discussed features
(see also Eckart et al. 2008a). The 17 May
 2007 data have not been published before. All exposures were sky
subtracted, flat-fielded, and corrected for dead or bad pixels. As
the most important improvement of the new reduction the dithered
exposures have been aligned with sub-pixel accuracy by a
cross-correlation method (Devillard 1999). PSFs were extracted from
these images with StarFinder (Diolaiti et al. 2000). The images were
deconvolved with the Lucy-Richardson (LR) algorithm. Beam
restoration was carried out with a Gaussian beam of FWHM
corresponding to the respective wavelength. The final resolution at
$2.2\mu$m is about 60 milli-arcseconds (mas). Flux densities of the
sources were measured by aperture photometry. Because of the high
accuracy of the image alignment it was possible to separate Sgr~A*
from the nearby stars S17 and S13 by choosing a circular aperture of
about 52 mas radius (see Fig. 15 of Eckart et al. 2006a  and
discussion therein), resulting in a better correction for the flux
contribution of these stars. The data was corrected for extinction
using $A_K = 2.8$ (Eisenhauer et al. 2005a, Sch\"{o}del et al.
2007). The relative flux density calibration was carried out using
known K-band flux densities and positions of 14 sources in the IRS16
cluster by
 R. Sch\"{o}del (private communications). This results in a K-band flux of
the high velocity star S2 of $22\pm1$ mJy, which compares well with
the magnitudes and fluxes for S2 quoted by Ghez et al. (2005b) and
Genzel et al. (2003). The measurement uncertainties for Sgr~A* were
obtained from the reference star S2. For more details about CIAO
observations see Nishiyama et al. (2009).

\begin{figure*}[t]
\begin{minipage}{\textwidth}
\centering{\includegraphics[width=\textwidth]{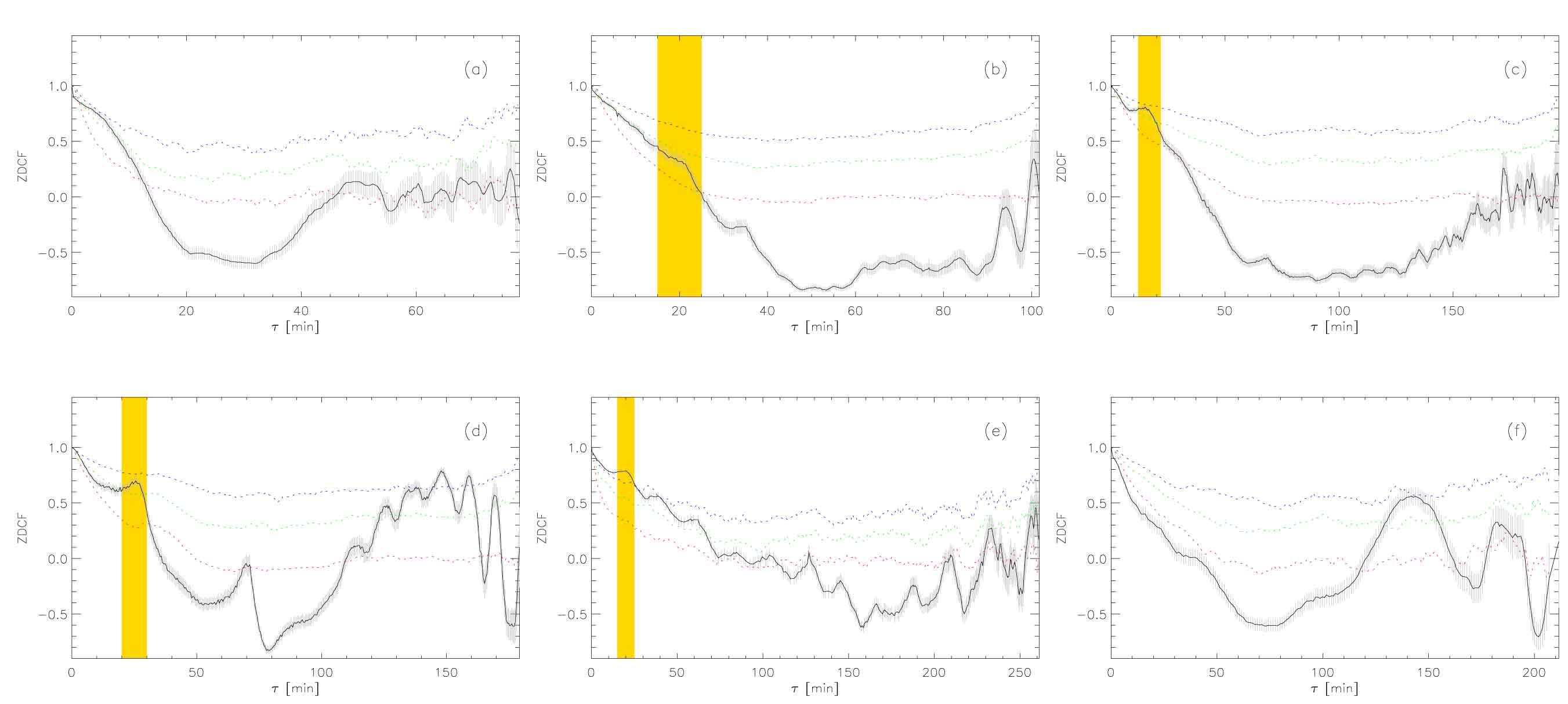}}
\end{minipage}
\caption{ZDCF of the flux light curves of 13 June 2004 (a),
30 July 2005 (b), 1 June 2006 (c), 15 May 2007 (d),  17 May 2007 (e)
and 28 May 2008 (f). The vertical colored boxes indicate the
position of the closest peaks to the zero time lags. Dotted lines
show the median (red), $68.3\%$ (green) and $99.8\%$ (blue)
thresholds derived from $10^4$ red noise simulated light curves.}
\label{zdaf}
\end{figure*}

\subsection{Data analysis}
\subsubsection{Periodicity}
Figure \ref{events} shows  flare events observed in the NIR K-band
(2.2$\mu$m) on 13 June 2004, 30 July 2005, 1 June 2006, 15 May 2007,
17 May 2007 and 28 May 2008. This sample includes all flare events
observed in NIR polarimetry during recent years according to the
  knowledge of the authors. The flux densities rise and come back to their
quiescent level in time intervals of roughly 100 minutes. The
measured values of total flux, degree of polarization, and angle of
polarization vary significantly on time scales of $\sim$10 minutes.
These abrupt changes can be more clearly detected when the events
are in their brightest state. Even if the flares are different in
some aspects (e.g. the ratio of the changes of the flux, the maximum
brightness achieved, or the degree of linear polarization), there
could exist some features that are repeated
 in our sample. Here we perform a quantitative analysis in
order to probe such features. We focus first on detecting periodic
signatures in flux densities and then a possible correlation between
changes in flux and polarimetric light curves.

The autocorrelation function and Lomb-Scargle periodograms
can be used to detect signatures of time periodic structures. For
the autocorrelation analysis we used the z-transformed discrete
correlation function (ZDCF) algorithm (Alexander 1997), which is
particularly useful for analyzing sparse, unevenly sampled light
curves. For Lomb-Scargle periodograms we followed Press and Rybicki
(1989).

Figure \ref{zdaf} shows the cross-correlation of the flux
density light curves of our sample with themselves by using ZDCF
method. The ZDCF of the 30 July 2005, 1 June 2006, 15 and 17 May 2007
flares show peaks around $20\pm5$ min time-lag, and specially for 15
and 17 May 2007 the peaks look significant. These peaks can be signs
of a possible periodicity.

\begin{figure}[!t]
\centering{\includegraphics[width=0.45\textwidth]{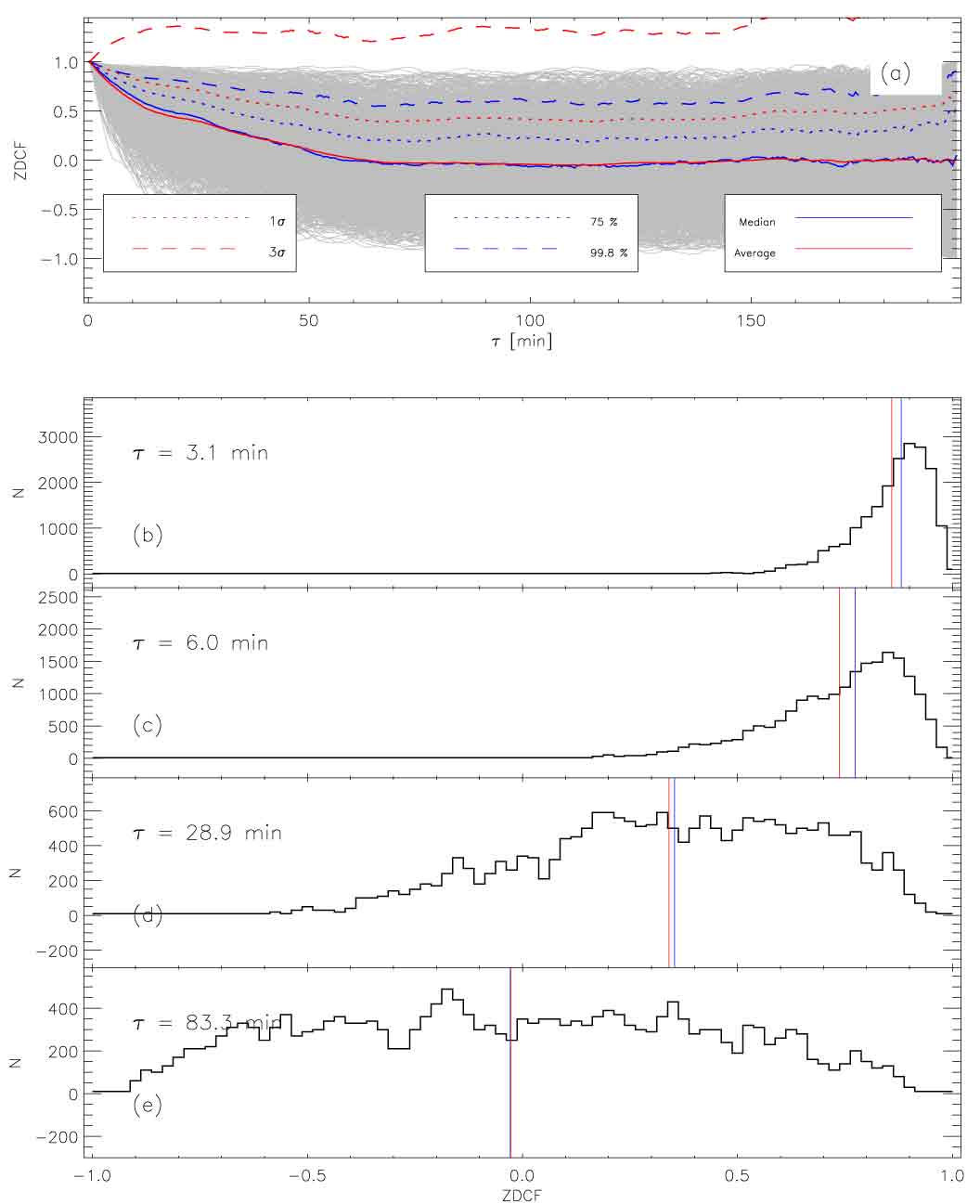}}
\caption{ZDCF of $10^4$ simulated red noise light curves
overplotted in one image (a). The resulted average and standard
deviation as well as the median and percentile values are presented.
The distribution of the ZDCF values for different time lags ($\tau$)
are also depicted (b-e), which clearly show non-Gaussian
distributions, specially for small values of $\tau$. }
\label{non-gauss}
\end{figure}

As one can see, only 17 May 2007's ZDCF shows a peak above
the $99.8\%$ significance threshold. The false alarm values have
been derived by repeating the same ZDCF analysis on the $10^4$
random red-noise light curves. The criteria for this comparison and
the way the light curves are produced are described below. Here we
must note that since the ZDCF values are normalized and distributed
in a [-1:1] interval, the distribution of the values for each time lag
($\tau$) is not Gaussian (see Fig. \ref{non-gauss}). As a result we
have used the median and percentile nomenculture instead of the
normal standard deviation formalism for deriving the false alarm
levels.

We have highlighted the time windows in which ZDCFs show a peak
around $20\pm5$ min, coinciding with the claimed quasi-periodicity.
Even though the 1 June 2006 and 15 May 2007's ZDCFs do not show peaks
above $99.8\%$ significance, they  however still reach  $99\%$ and
$95.4\%$ levels, respectively. Also, by a rough estimate, the
significance of the ZDCF peaks are correlated to the brightness of
the flares, and for the faint events, as is observed for example in
2004, there is no detectable peak.

\begin{figure*}[!htb]
\begin{minipage}{\textwidth}
\centering{\includegraphics[width=\textwidth]{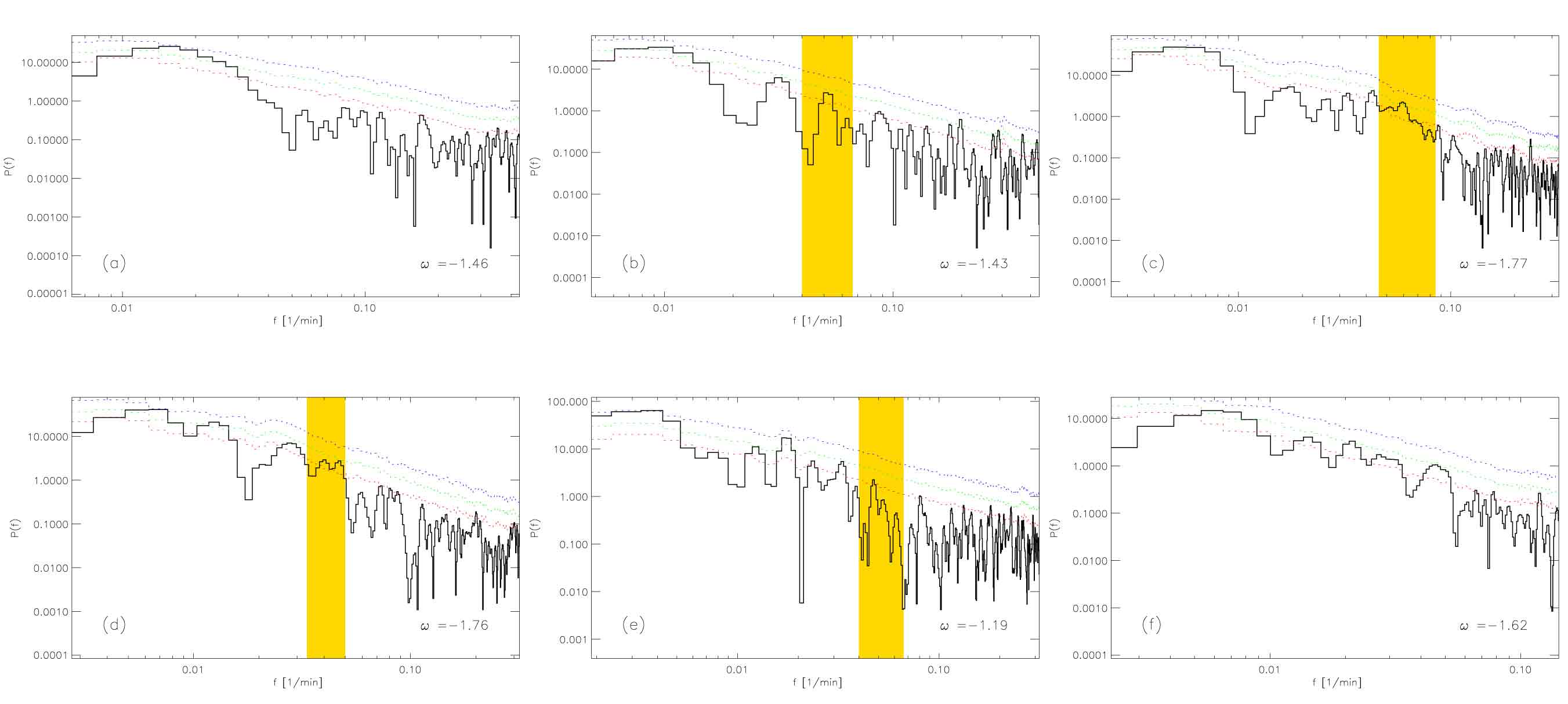}}
\end{minipage}
\caption{Lomb-Scargle periodograms of the flux light curves
for 13 June 2004 (a), 30 July 2005 (b), 1 June 2006 (c), 15 May 2007
(d), 17 May 2007 (e) and 28 May 2008 (f). The dashed lines show the
median (red) and , $68.3\%$ (green) and $99.8\%$ (blue) thresholds
derived from the $10^4$ red noise simulated light curves. The
highlighted boxes show the band of frequencies that have
corresponding ZDCF peaks.} \label{ls}
\end{figure*}

Figure \ref{ls} shows the Lomb-Scargle periodogram for each
flux light curve. The usual factor of four times over-sampling was
applied in order to increase the sensitivity. All  periodograms show the
power law behavior, $\mathcal{P}\propto f^{\omega}$,  showing
greater power at lower frequencies. Here $\mathcal{P}$ is power, $f$
is frequency and $\omega$ is power law index. All light curves
follow a red noise under-lying power spectral density (PSD) with
$-1.19\le\omega\le-1.77$. To test the significance of the peaks in
the periodogram we have repeated the Lomb-Scargle analysis for
$10^4$ simulated red noise light curves.  For simulating the red
noise we have followed the algorithm by Trimmer and K\"{o}nig (1995)
(see Figs. \ref{alpha0}, \ref{alpha1} and \ref{alpha2} for
examples). The red noise light curves were produced for each event
separately following the procedure below: (1) Determining the slope
of the observed PSD for each event using a linear fit to the
periodogram in log-log space and also by fitting a first-order
autoregressive function to the PSD (Schulz and Mudelsee 2002) and
averaging the results of both methods (the resulted $\omega$ is
presented in the lower right corner of each plot in Fig. \ref{ls}).
(2) $10^4$ red noise light curves produced with the related $\omega$
following the method of Timmer  and K\"{o}nig (1995). The light
curves have been produced by selecting a middle part of a light
curve with a length at least ten times longer than the observed light
curve following the argument by Uttley et al. (2002). The selected
segment of the light curve is sampled to the same time bin of the
corresponding observation. This will correct for any artificial
effects caused by time lags which exist on our light curves and also
the uneven sampling of our observations. (3) The value of the flux
has multiplied by one factor as the mean of the simulated flux has the
same value as the observational one for each night.

\begin{figure}[!t]
\centering{\includegraphics[width=0.47\textwidth]{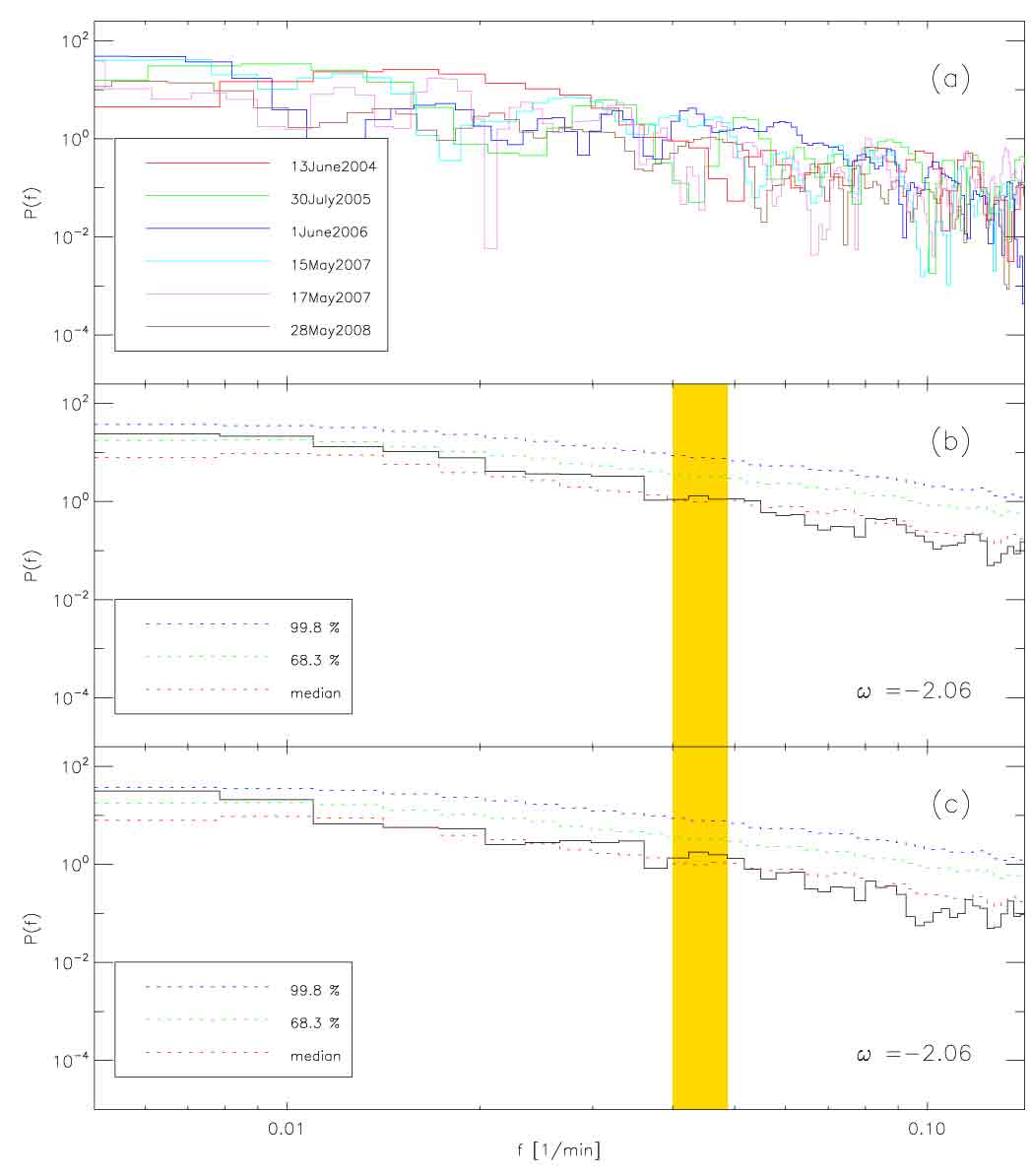}}
\caption{Lomb-Scargle periodograms of the flux light curves
of our sample overplotted in the same plot (a). Averaged result of
all periodograms  (b) and excluding the 13 June 2004 flare
(c). Highlighted box shows the expected $20\pm5$ minutes periodicity
band.} \label{ls_ave}
\end{figure}

\begin{figure}[!t]
\centering{\includegraphics[width=0.47\textwidth]{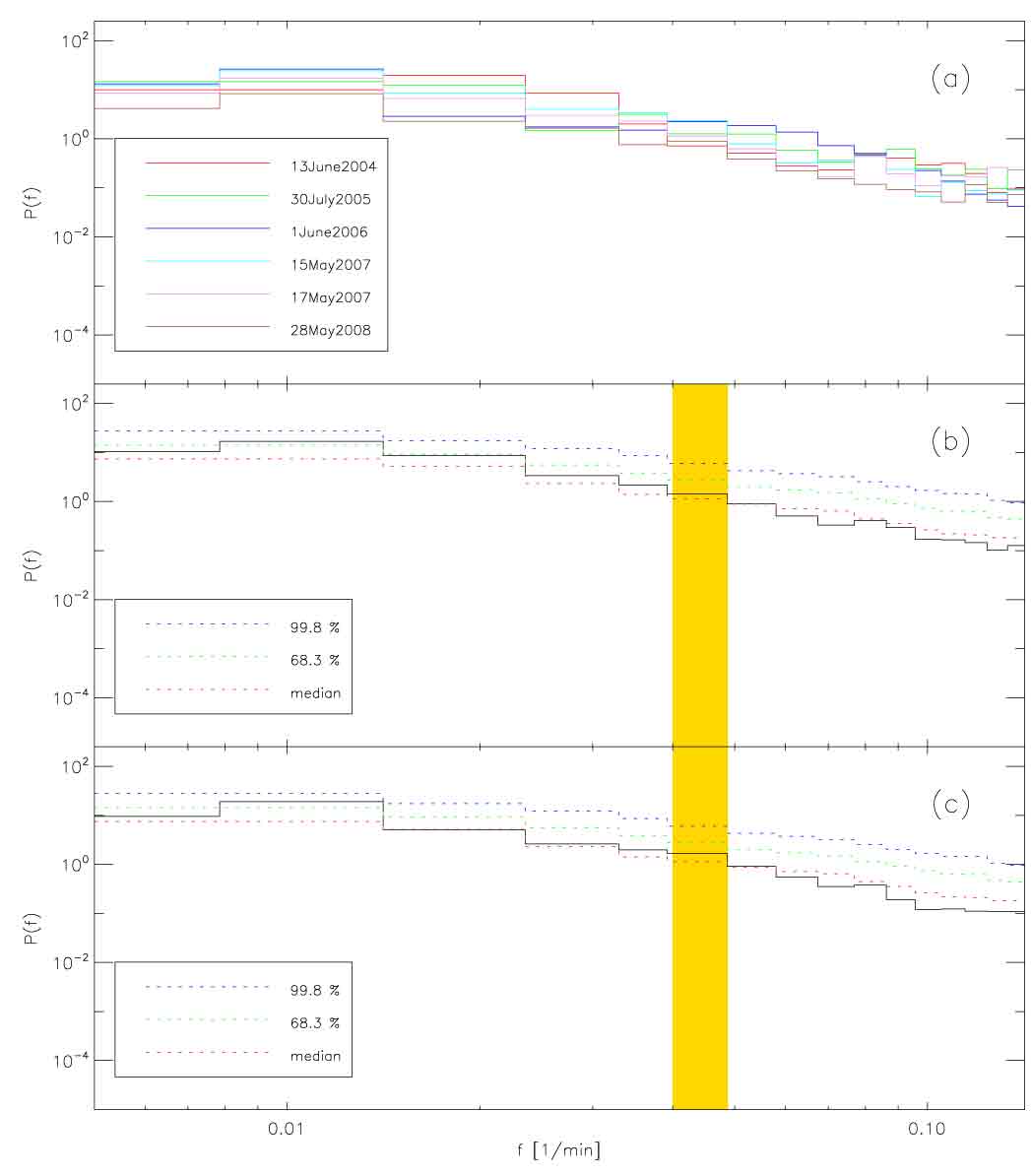}}
\caption{Same as Fig. \ref{ls_ave} but averaged for a band
of frequencies with the same size as the highlighted region.}
\label{ls_ave_bin}
\end{figure}

\begin{figure*}[!t]
\begin{minipage}{\textwidth}
\centering{\includegraphics[width=\textwidth]{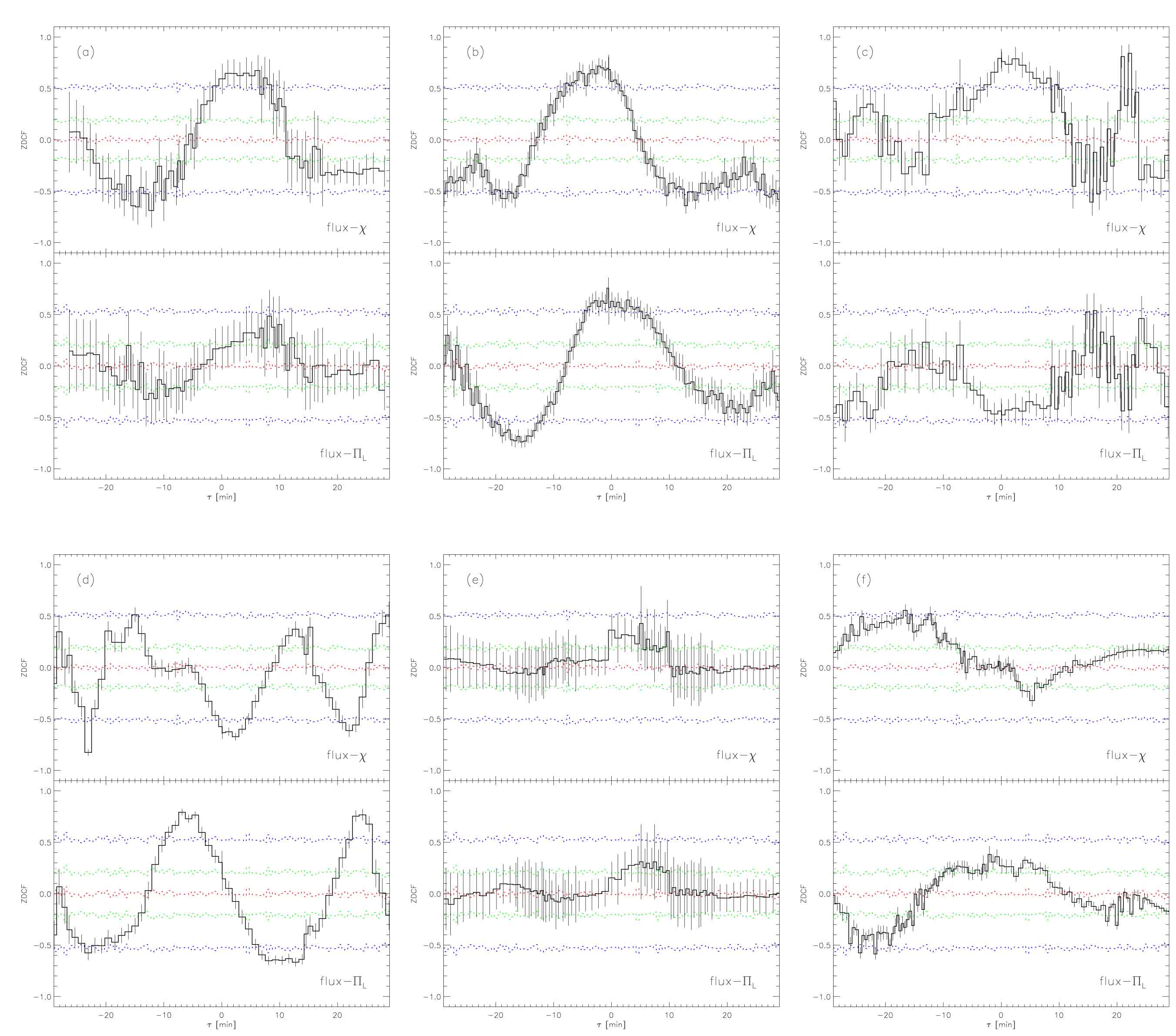}}
\end{minipage}
\caption{Cross-correlation between the flux and polarization
angle (degree) light curves of Fig. \ref{events} [13 June 2004 (a),
30 July 2005 (b), 1 June 2006 (c), 15 May 2007 (d), 17 May 2007 (e)
and 28 May 2008 (f)]. In each panel, top (bottom) shows the
correlation between the flux and the polarization angle  (degree of
linear polarization). Dashed lines indicate the position of the
median (red), $68.3 \%$ (green) and $99.8\%$ (blue) significance
levels derived from $10^4$ simulated red noise light curves. }
\label{zdcf}
\end{figure*}

None of the periodograms of Fig. \ref{ls} show a peak that
exceeds the $99.8\%$ threshold except the ones correlated to the
lengths of the flare events ($\sim$100 minutes). We specially
highlighted the windows in which corresponding ZDCFs peak. In order
to see if there exists a persistent frequency peak  in the PSDs, we
averaged all  observed periodograms (Fig. \ref{ls_ave}
(b)). The average periodogram doesn't show a significant peak in
comparison with the false alarm level  even if we exclude the 13
June 2004 light curve (Fig. \ref{ls_ave} (c)). As we mentioned
before the short time scale of the flares and limited number of
cycles make a significant detection of any periodicity very
difficult. In addition, the possible quasi-periodic structure can vary
 during the time of the flare if it is connected to the falling
clumps of matter into the black hole (Falanga et al. 2008). In this
case the Lomb-Scargle algorithm finds different frequencies and
allocates them separate values of power. This effect can result in a
periodogram in which the values of power are higher in a band of
frequencies instead of a specific value. As one can see in
Fig. \ref{ls}, 1 June 2006 and 15 May 2007 can be the candidates for
such an effect since the value of the PSD function remains high in the
highlighted window. To test for such an effect we averaged the
periodograms with a new bin size and repeated the same procedure for
the random red noise PSDs (Fig. \ref{ls_ave_bin}). The resulting
averaged periodogram does not show any significant peak again (Fig.
\ref{ls_ave_bin} (b)), even if we exclude the 13 June 2004 event
(Fig. \ref{ls_ave_bin} (c)). Since this procedure is very sensitive
to the size of the chosen window (which can vary from event to
event), a more detailed analysis is needed to study the possible
evolution of any periodic signal.

\subsubsection{Importance of polarimetry}
We have also performed  cross correlation analysis between
variations of flux and degree (angle) of polarization. A search for
any short time-lag correlation has been carried out by scanning the
light curves using a sliding window method (see Fig.
\ref{zdcf_sketch} panel (a)). The size of the scanning window
($\epsilon$) is fixed on 40 minutes since we are interested in
magnifying any short lag correlation related to the possible
$20\pm5$ minutes quasi-periodicity. Figure \ref{zdcf} shows the
cross-correlations of the two sets of mentioned light curves for all
events in our sample. The time steps of the scans were fixed to be five
minutes. The presented results are the average of all scans weighted
by the flare signal to noise ratio (i.e. each part of the flare that
is brighter, has more weight). Since the ZDCF algorithm needs at
least 11 points per bin for reliable results (specially for the
error estimation; see Alexander 1997)  a linear interpolation of the
polarimetric data points has been performed. We must mention that
due to the confusion from nearby stars and the diffuse background
emission the accuracy of measuring polarimetric parameters is
related to the brightness of Sgr~A*. This means that the most
reliable polarimetric data are measured when strong flares happen,
specially when the source is in its brightest state.

\begin{figure}[t]
\centering{\includegraphics[width=0.47\textwidth]{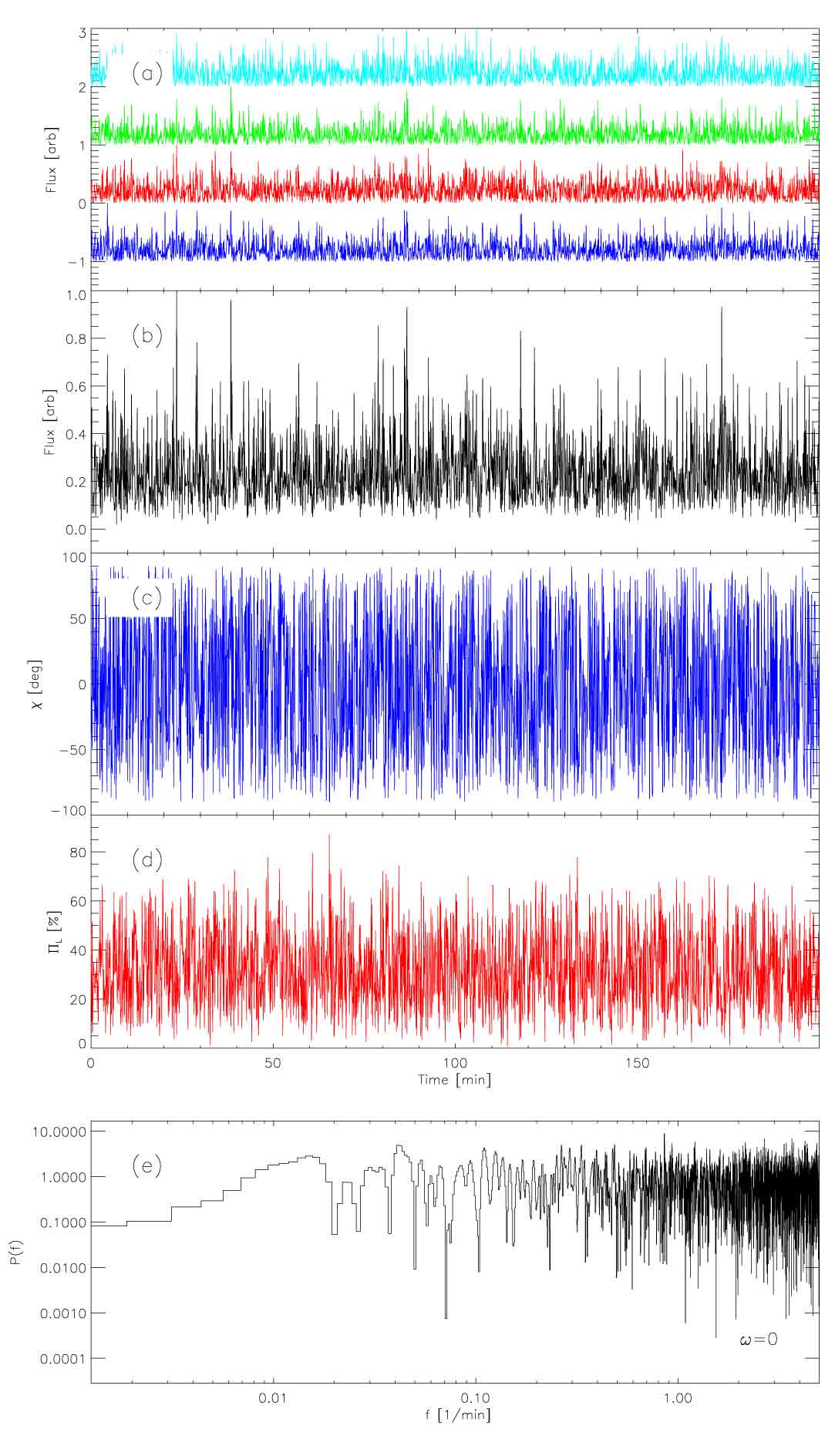}}
\caption{(a): Simulated light curves of four different
polarimetric channels all showing white noise behavior ($\omega=0$).
(b): Total flux. (c): Angle of polarization.
(d): Degree of polarization.  (e):
Lomb-Scargle periodograms of the total flux.} \label{alpha0}
\end{figure}

\begin{figure}[t]
\centering{\includegraphics[width=0.47\textwidth]{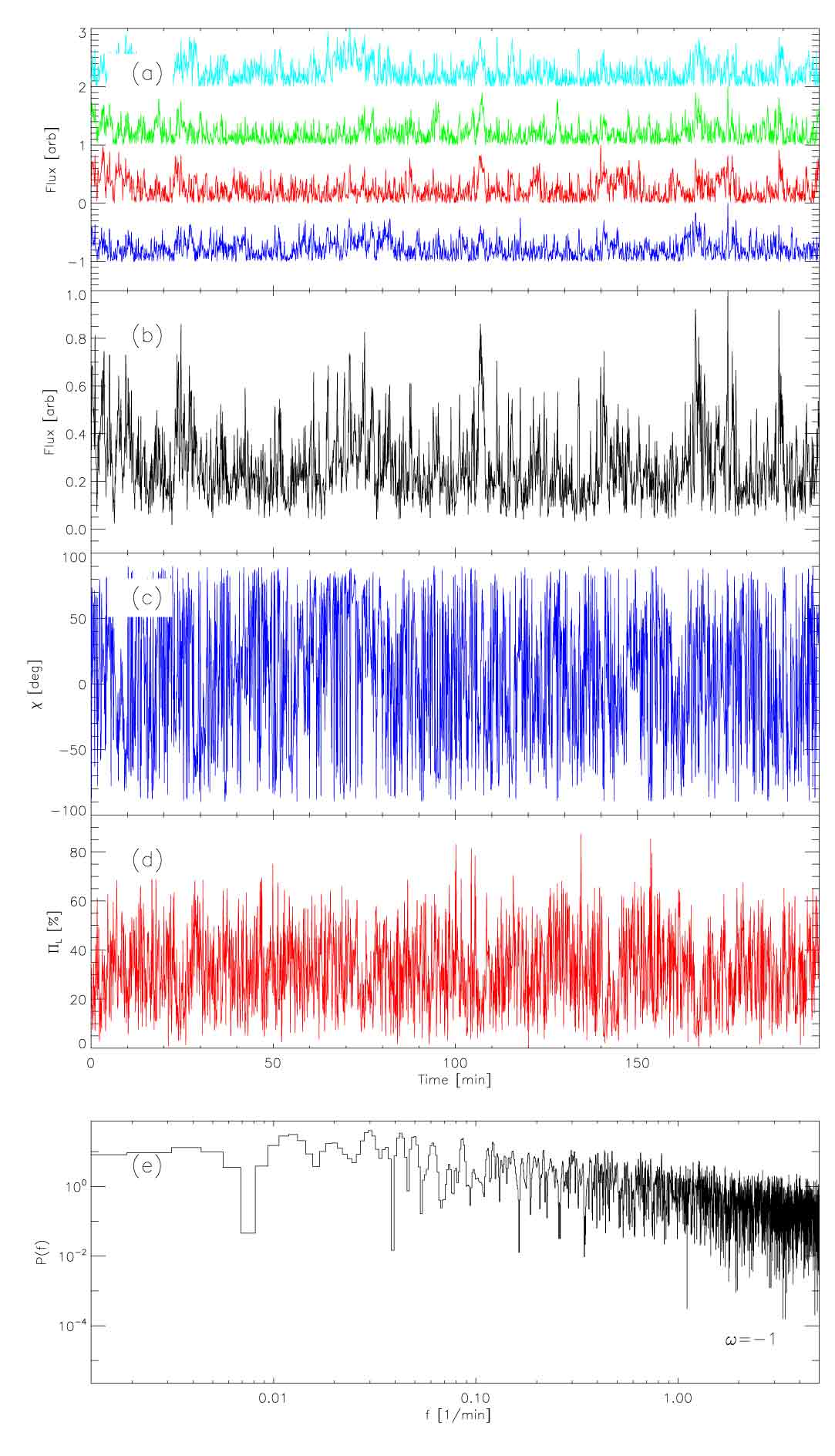}}
\caption{Same as Fig. \ref{alpha0} for $\omega=-1$ (flicker noise).}
\label{alpha1}
\end{figure}

\begin{figure}[t]
\centering{\includegraphics[width=0.47\textwidth]{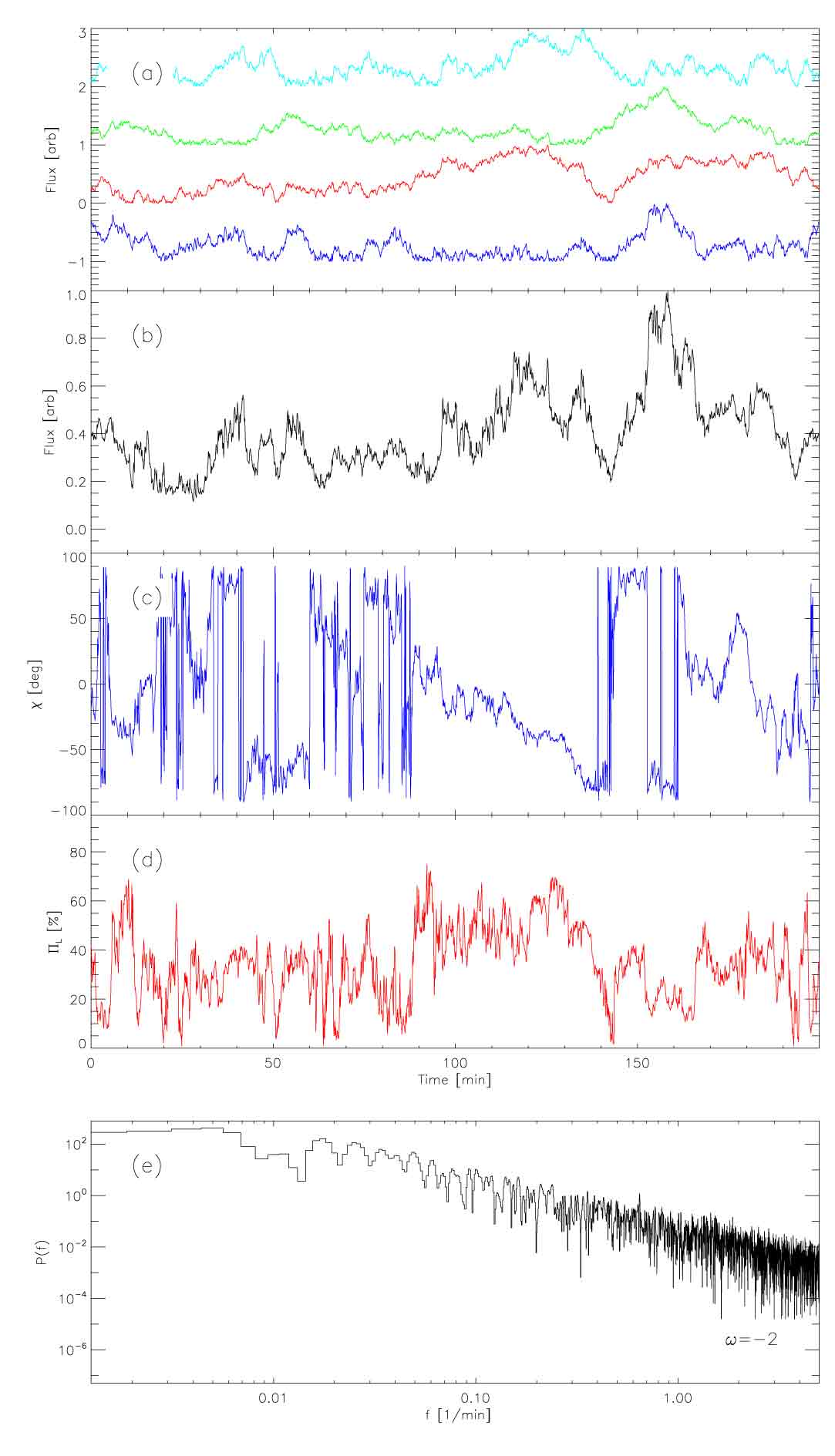}}
\caption{Same as Fig. \ref{alpha0} for $\omega=-2$ (red noise).}
\label{alpha2}
\end{figure}

\begin{figure}[t]
\centering{\includegraphics[width=0.47\textwidth]{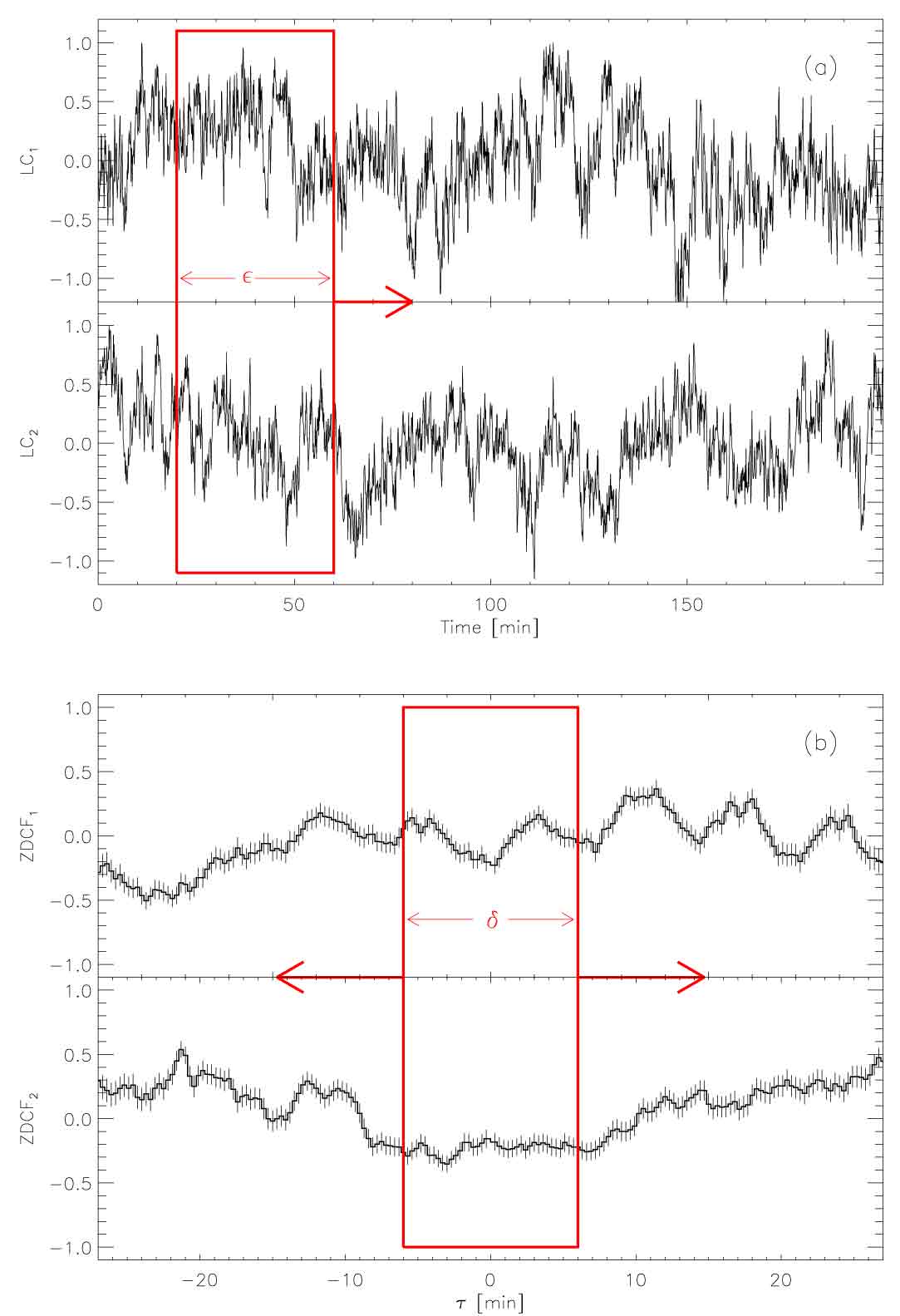}}
\caption{(a) Sketch showing how the cross-correlations of
Fig. \ref{zdcf} have been derived. A moving window of the size
$\epsilon$ scans the flux and polarization angle (degree) light curves.
The final cross-correlation is the average of all windows. (b)
Sketch showing how the probability function in Fig. \ref{prob} has
been derived. The algorithm finds the number of events in which two
ZDCFs show at least one point over $3\sigma$ threshold of red noise
correlation in a window the size of $\delta$. By repeating the same
procedure for $10^4$ cross-correlation of red noise light curves the
probability  that a significant correlation is expected to happen in
that window is derived.} \label{zdcf_sketch}
\end{figure}

In order to test whether  a random red noise model can
produce the same correlation patterns we need to simulate red noise
light curves including polarimetric data. For this purpose, we have
simulated random $E$ vectors for four perpendicular directions:
\begin{eqnarray}
\overrightarrow{E_{\theta}}=E_{\theta} e^{-i\phi t}\hat{\theta}
\end{eqnarray}
where  $\theta=0^{\rm o}$, $45^{\rm o}$, $90^{\rm o}$, $135^{\rm
o}$, $\hat{\theta}$ is the corresponding normal vector, $\phi$ is
the phase and $E_{\theta}$ is a random value following the algorithm
by Trimmer and K\"{o}nig (1995). Following the detection of the
signal including the cross-talk from non-orthogonal neighboring
channels, the polarized flux in each channel is then produced by
using the Mueller matrix formalism. The resultant four different
channels ($F_0, F_{45}, F_{90}$ and $F_{135}$) all show the same
power law index (Figs. \ref{alpha0}, \ref{alpha1} and \ref{alpha2}
(a)). Using
\begin{eqnarray}
F &=& F_0+F_{90}\\
Q &=& F_0-F_{90}\\
U &=& F_{45}-F_{135}\\
\chi &=& \frac{1}{2}\arctan{(\frac{U}{Q})}\\
\Pi_L &=& \frac{\sqrt{Q^2+U^2}}{F}
\end{eqnarray}
(where $F$  is the total flux, $Q$ and $U$ are the Stokes
parameters, $\chi$ is the polarization angle and $\Pi_L$ is the
degree of linear polarization) the flux and polarization light
curves for each set can be derived. Figures
\ref{alpha0}, \ref{alpha1} and \ref{alpha2} show examples of such light
curves with three different
PSD slopes: $\omega=0$ (white noise), $\omega=-1$ (flicker noise) and
$\omega=-2$ (red noise).

As one can see in Fig. \ref{alpha2} for $\omega=-2$ some random
correlations between the changes in the total flux and the polarimetric
parameters can occur. In order to see whether the observed
correlations in our sample are just the same random coincidences or
 if they show signs of a more subtle process, we have simulated $10^4$
artificial light curves for each value of $\omega$ as derived from
the average PSD of observations. Then for each set of  light curves
the correlation function between flux and polarization angle
(degree) has been calculated. Dashed lines in Fig.
\ref{zdcf} show the median, $68.3\%$ and $99.8\%$ false alarm
values. One can see that  some of the observed correlations are
above $99.8\%$ significance level. Even though  most of the
mentioned cross-correlation peaks happen around the same value
(around zero time-lag) not all of them are exactly in the same
$\tau$. To examine  how probable  it is that a strong deviation from
the average in the red noise simulation repeatedly happens in a
specific window we have calculated the probability that the
mentioned cross correlations show significant peaks (above $99.8\%$)
in a window of a size $\delta$ (see Fig. \ref{zdcf_sketch} (b)).

\begin{figure}[t]
\centering{\includegraphics[width=0.47\textwidth]{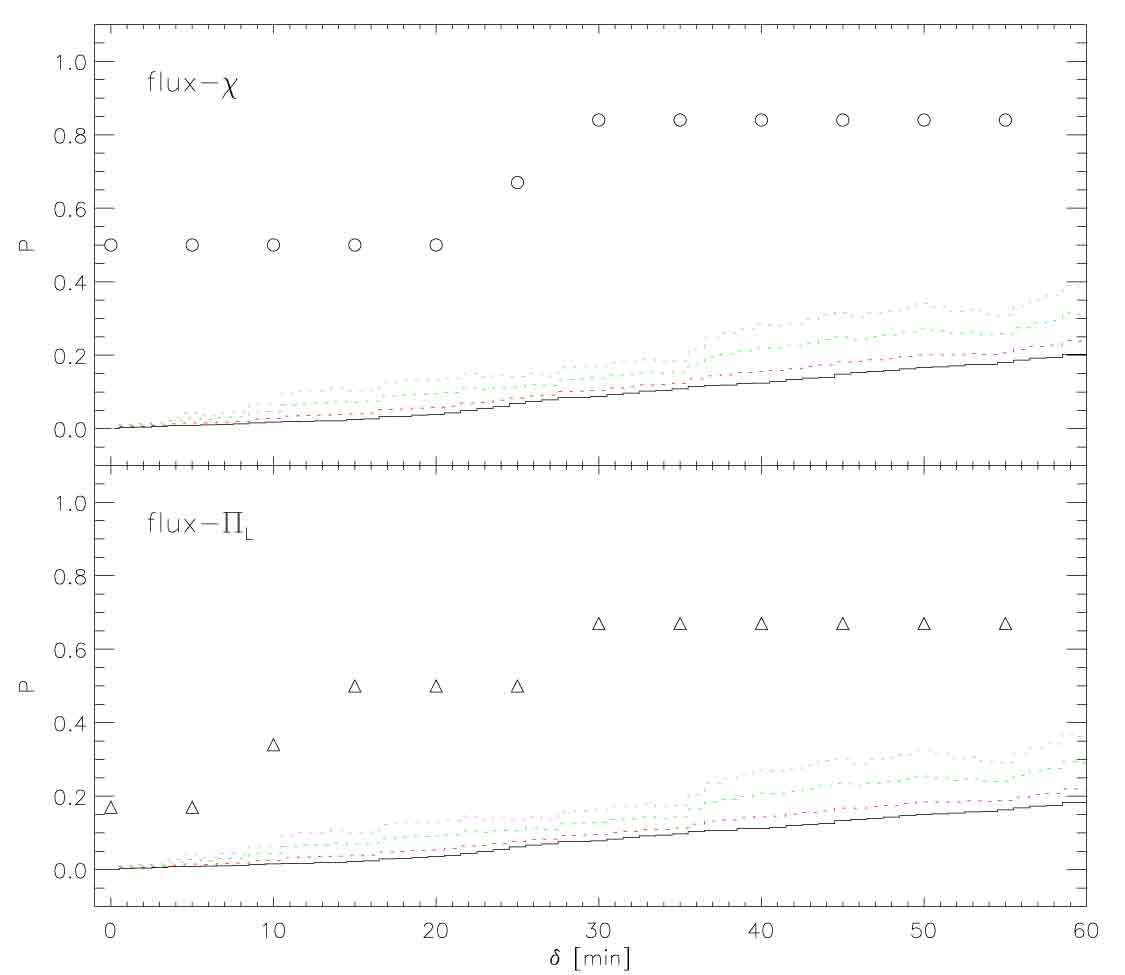}}
\caption{Top: Probability that two sets of $10^4$ simulated
red noise light curves show significant correlation between total
flux and polarization angle in a window of the size of $\delta$ minutes
(solid line). Bottom: Same as top panel for the correlation between
flux and polarization degree. Dotted lines show $1\sigma$ (red),
$3\sigma$ (green) and $5\sigma$ (violet) levels of confidence for
the probability function, calculated by repeating the whole analysis
100 times. The circles and triangles indicate the probabilities
derived from our sample of NIR light curves (see Fig. \ref{events}
and \ref{zdcf}).} \label{prob}
\end{figure}

Figure \ref{prob} shows this probability as derived for the simulated
 light curves (solid line) and observations (circles and triangles).
This analysis shows that it is very unlikely to observe a correlation
between total flux and polarimetric parameters approximately at the
same time lags (small $\delta$), while our observations show that a
strong correlations exist in the light curves of Sgr~A* and that they also 
repeat themselves for approximately the same time lag. We must note here
that the value derived for the observed probability in Fig.
\ref{prob} (circles and triangles) are derived from a sample of only
six sets of light curves. In order to make a more reliable statistical
analysis, more NIR observations of Sgr~A* in polarimetric mode needs to
be done in the future. Furthermore, the method described here can  be used in
principle for polarimetric observations of other sources
showing the same variability;  which may  help in understanding the
general underlying physical process causing this kind of behavior.

Without GR effects being taken into account, the physical models which have been
already proposed to simulate the observed red-noise light curves of
AGNs (Lyubarskii 1997; Armitage \& Reynolds 2003, Vaughan et al.
2003) would have difficulties in reproducing this type of correlation
between the behavior of polarimetric parameters and the total flux.
As a result, the observed correlation between changes in flux and
the polarimetric data  suggests a way to distinguish between the
possible physical processes responsible for the overall red-noise
behavior. A semi-analytical study by Pech\'{a}\v{c}ek et al. (2008)
showed that a signal generated by an ensemble of spots randomly
created on the accretion disk surface can produce red noise signals
with PSD slopes of the order of -2. In their simulations the spot
generation is governed by Poisson or Hawkes processes. In
combination with our observations of Sgr~A* the spotted disk
scenario is a possible explanation for this commonly observed red
noise behavior, while some exceptionally luminous events can show
their signature in polarized light. This could point out to the
transient occurrence of QPOs which may appear repeatedly during the
bright flares, which seems to be a rather natural possibility. This
will be discussed in the next sections in more detail.

\section{Modeling}
In this section, we first describe in detail our emission model,
which is mainly  based on  synchrotron emission from  accelerated electrons
in the inner parts  of a relativistic accretion disk.  We  also
describe the ray-tracing method  used and the predictions of the model.

\subsection{Fluctuations of the inner parts of an accretion disk: A possible description for the observed signal?}
\subsubsection{Emission model}
Since the discovery of X-ray and NIR flares from Sgr~A*, several
flaring regions theories tried to describe the physics behind them,
varying from abrupt changes in the accretion rate of a Keplerian
disk (Melia et al. 2001) to the interaction of the accretion disk with
nearby stars (Nayakshin et al. 2004).  Although none of these
scenarios can be ruled out, there are observational evidences that
give more support to some of them.

\begin{figure}[!b]
\includegraphics[width=0.45\textwidth]{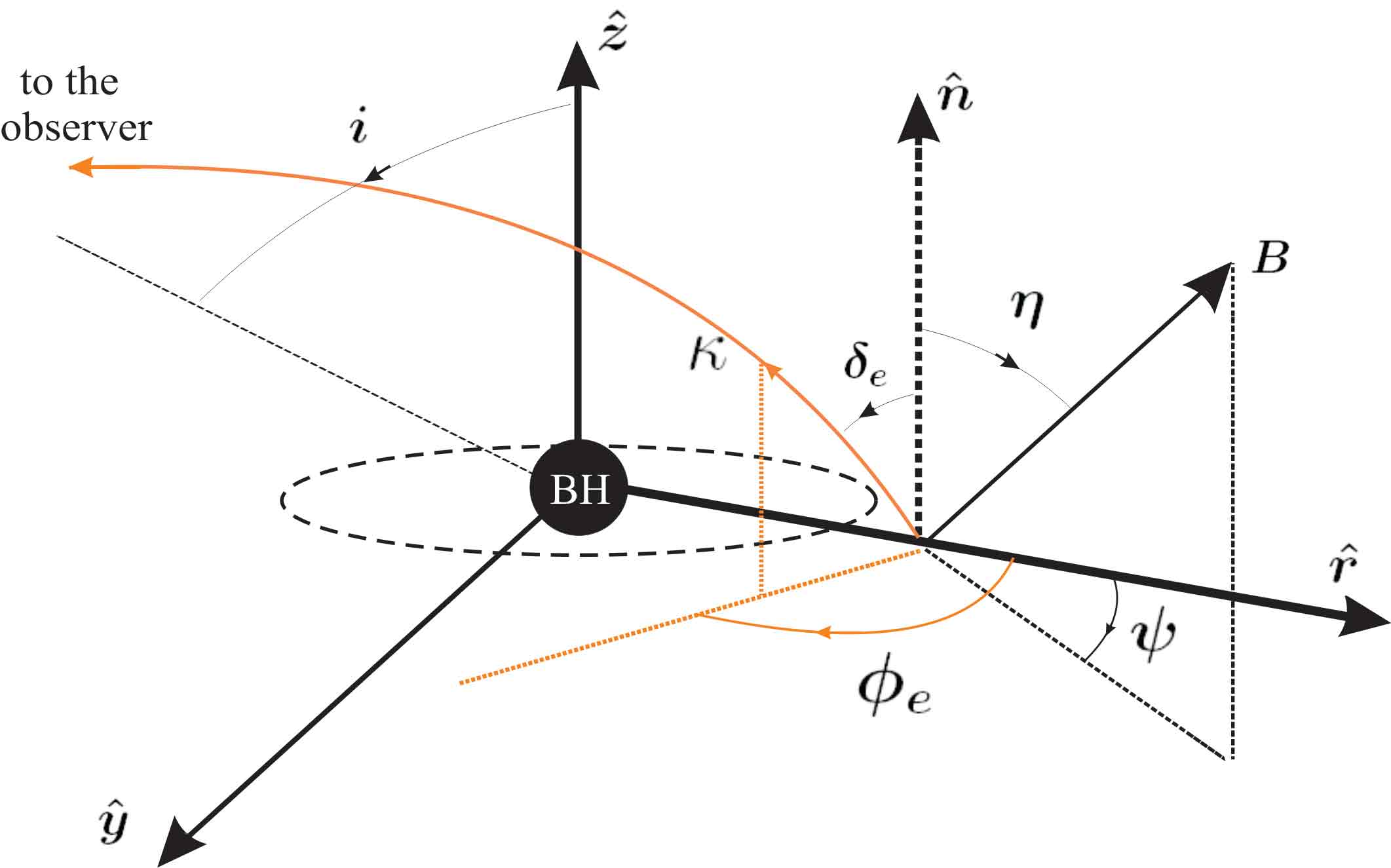}
\caption{The geometry we considered in our emission model. The
accretion disk around the black hole lies on the $\hat{y}-\hat{r}$
plane and $\hat{n}$ is the unit normal vector of the disk. The direction of the
magnetic field lines ($B$) corresponding to the disk frame is
defined by two angles $\eta$ and $\psi$. ${\kappa}$ represents the
momentum of the emitted photon and its direction determined by a set
of angles [$\delta_e$,$\phi_e$]. The distant observer is looking into
the system along a line of sight inclined by a certain angle $i$.}
\label{magnetic}
\end{figure}

For example, as we mentioned before, the frequently observed rate of
NIR flares (four to five flares/day) makes it hard for disk-star interaction
or tidal capture scenarios to describe the events. Of special
interest to us are the observed quasi-periodic  flux modulation
during the NIR and X-ray flares. The recent unambiguous discovery of
(quasi-)periodicity in an active galaxy (RE J1034+396) reported by
Gierli\'{n}ski et al. (2008) brings more support to the idea that
the similarity in the behavior of black holes extends from stellar-mass
black holes to super-massive ones.  The most interesting scenario
could be a relation to the orbital time scale of the accretion disk,
with a possible connection to the plunging region which feeds the
black hole through a channel of inflow or a possible clumpy
infalling flow. As we describe  here and in the next section, our
interpretation of these variable signals (which relates flux
modulations mainly to  lensing and boosting effects) can open a new
window to study  physics in very strong gravitational regimes, very
close to the event horizon of  black holes.

\begin{figure*}[!t]
\centering{\includegraphics[width=\textwidth]{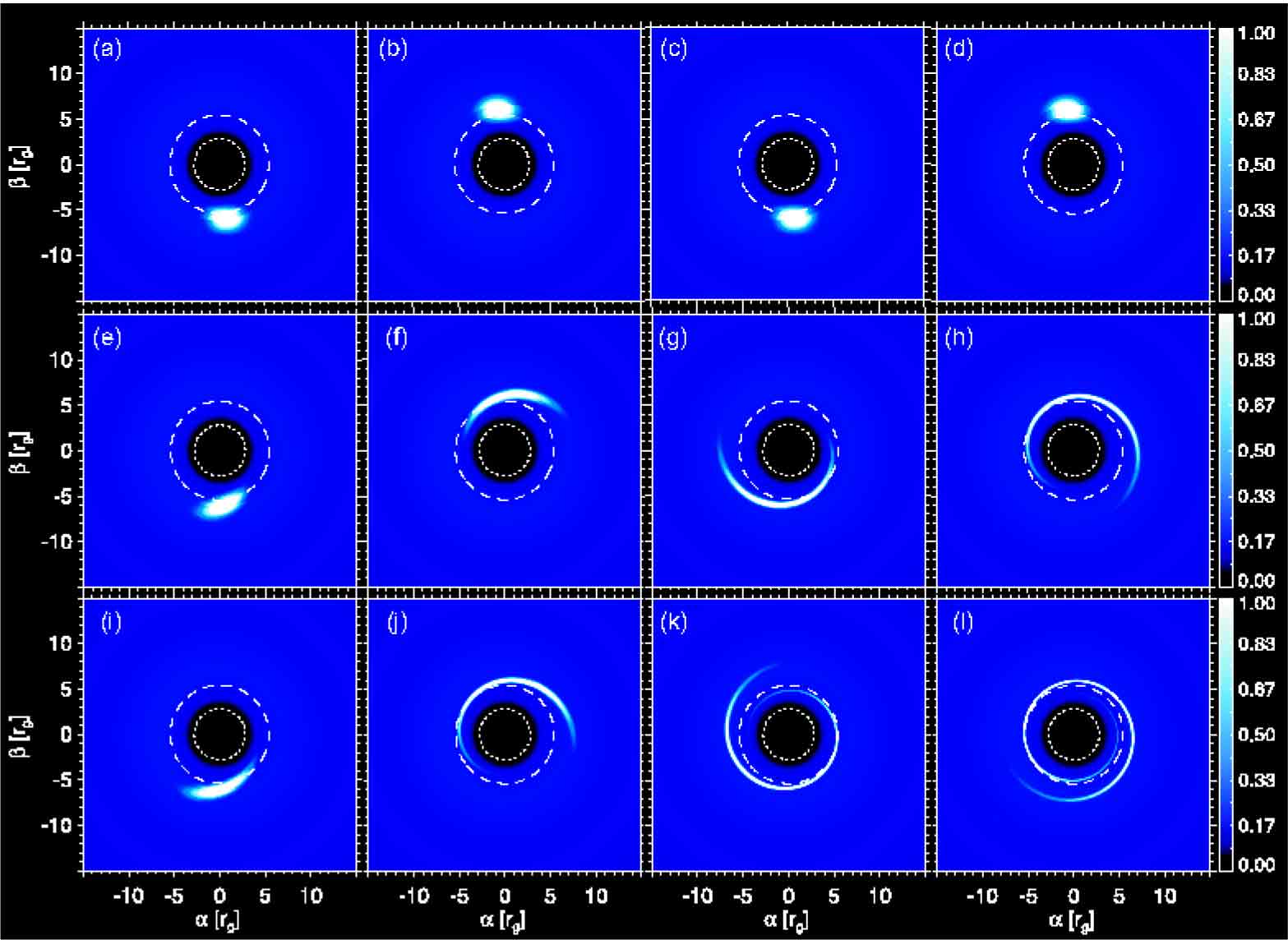}}
\caption{Snapshots of orbiting anomalies inside the accretion disk
as they  appear to a distant observer looking along a line of sight
inclined by $0^{\rm o}$ (relative to the normal to the disk), at
times after $\frac{1}{4}T$, $\frac{3}{4}T$, $\frac{5}{4}T$ and
$\frac{7}{4}T$ (left to right). Each row shows how the event evolves
in time for different values of the characteristic shearing time
scale: $\tau_{sh}=\infty$ (a-d),
 $\tau_{sh}=2.0$ (e-h) and $\tau_{sh}=1.0$ (i-l).
 The spin of the black hole is set to 0.5. The dotted  and dashed
lines indicate the position of the event horizon and marginally
stable orbit, respectively. Each row has been scaled by its maximum
intensity for illustrative purposes. $\alpha$ and $\beta$ are the
projections of the impact parameter of the emitted photons as the
coordinates on the sky of the observer. Both coordinates are labeled
in $r_g(\simeq5\mu$as) units.} \label{spots_shearing}
\end{figure*}

\begin{figure*}[!t]
\begin{minipage}{\textwidth}
\centering{\includegraphics[width=\textwidth]{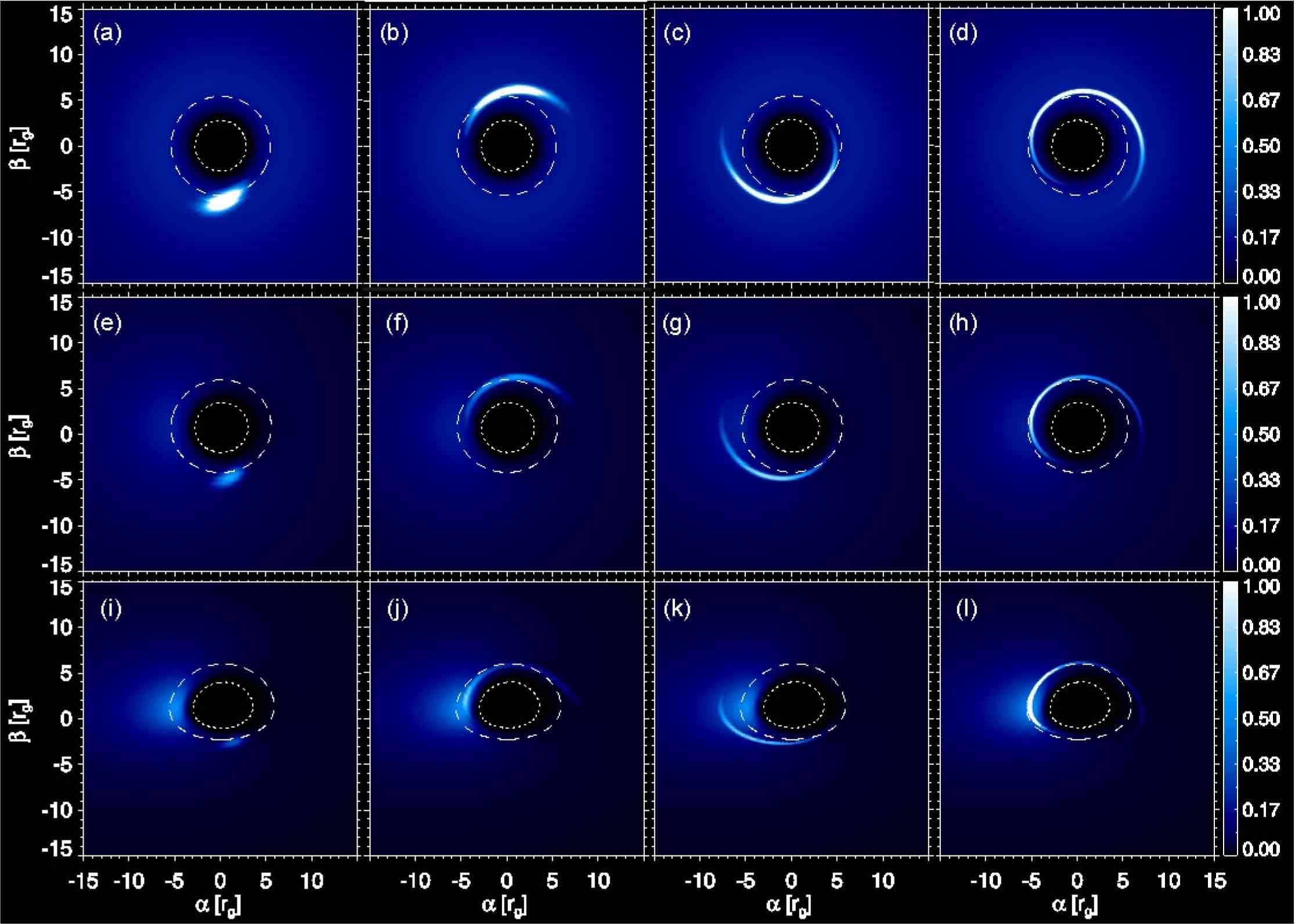}}
\end{minipage}
\caption{Snapshots of an orbiting anomaly inside the accretion disk
 as it appears to a distant
observer looking along a line of sight inclined by $0^{\rm o}$
(a-d), $30^{\rm o}$ (e-h) and $60^{\rm o}$ (i-l) (relative to the
normal to the disk) at times after $\frac{1}{4}T$, $\frac{3}{4}T$,
$\frac{5}{4}T$ and $\frac{7}{4}T$ (left to right). See also the caption
of Fig. \ref{spots_shearing}.} \label{spots_inclination}
\end{figure*}

NIR spectroscopy has shown that a power-law fit,
$F_\nu\propto\nu^{-\alpha}$ (where $F_\nu$, $\nu$  and $\alpha$ are
the flux, frequency and spectral index respectively), can describe
the observed spectrum of NIR flares. Although all the observations
agree with the fact that NIR flares show a soft spectrum
($\alpha>0$), the value of the spectral index is still not well
determined. The first NIR spectroscopy observations in July 2004
(Eckart et al. 2004) proposed the $\alpha$ value to be $\sim0.8-1.3$
during the peak of the flare. In 2006, Gillessen et al. (2006)
observed a correlation between flux and spectral index in their
observations. However, recent observations by Hornstein et al.
(2007) are consistent with a constant spectral index,
$\alpha=0.6\pm0.2$.

The actual value of the spectral index shows its importance in the
modeling of the physical process responsible for the flaring
emission. Some current models (Melia et al. 2001, Liu et al. 2006,
Yuan et al. 2007), predict that during flares a fraction of
electrons near the event horizon of the black hole are accelerated.
This can be described in the simplest form by a power law
distribution in the energy of radiating electrons,
$N(\gamma)=N_0\gamma^{-p}$ where $N(\gamma), N_0, \gamma$ and $p$
are the energy distribution function of electrons, normalizing
constant, Lorentz factor of the electrons and the energy spectral
index respectively. For high values of $\alpha$ one will need  a
sharp cut-off to the energy spectrum of electrons ($\gamma_c$), while
a lower value of $\alpha$ ($\alpha\sim0.6$) allows for a relatively
milder distribution in the energy of electrons. Liu et al. (2006)
have shown that simultaneous NIR and X-ray spectral measurements can
 constrain the parameters of the emitting region well.

Before describing the details of our simulations, here we discuss
how the existing observations limit the possible range of free
parameters. Observationally it is proven that in the Sgr~A* spectrum
  a turn-over frequency in the sub-millimeter to NIR
range exists. By using the turn-over frequency relation,
$\nu_c=2.8\times10^6B\gamma_c^2$GHz, one can put an upper limit on
$\gamma_c^2B$, where $B$ is the magnetic field strength in Gauss.
Here we have used $\gamma_c=100$ and  $B=60$G which give the best
fit to the NIR/X-ray models that already exist (Liu et al. 2006;
Eckart et al. 2008a).

In our simulations we first considered a scenario in which the main
flare is caused by a local perturbation of intensity close to the
marginally stable orbit (via magnetic reconnections, stochastic
acceleration of electrons due to MHD waves, magneto rotational
instabilities (MRI) etc.). These instabilities  spread out and
produce a temporary bright torus around the black hole. In this
scenario, the mentioned variabilities are mainly due to relativistic
flux modulations caused by the presence of an azimuthal  asymmetry
in the torus.

Simulations are dealing with two important velocities: radial
($v_r$) and azimuthal ($v_\phi$). The radial velocity can be
parameterized in the following way:
$v_r\sim\big(\frac{4\beta_P\beta_\nu}{9}\big)\big(\frac{GM}{r}\big)^\frac{1}{2}$
which depends on the ratio of the stress to the magnetic field energy
density, $\beta_\nu$, and the ratio of the magnetic energy density
to the thermal pressure, $\beta_P$ (Melia 2007). The use of the typical
values of $\beta_P$ and $\beta_\nu$ from MHD simulations give us an
estimation ($\beta_P \beta_\nu\sim0.05$). This leads to a radial
velocity of the order of $0.1 \big(\frac{r_g}{min}\big)$. For
the azimuthal velocity, we assumed that above the marginally stable
orbit the plasma is in a Keplerian orbit,
$v_\phi=\big(\frac{r^2-2a\sqrt{r}+a^2}{\sqrt{\Delta}(r^{3/2}+a)}\big)$
where $\Delta=r^2-2r-a^2$, and inside the plunging region the matter
experiences free fall with the same angular momentum as at the
marginally stable orbit.

Furthermore, two important  time scales are at work: heating and
cooling time scales. The heating time scale strongly depends on the
physical processes which act as the engine of the whole event (MHD
instabilities, magnetic reconnections etc.), and one can just put an
observational constraint on that according to the averaged observed
rise time of the events ($\bar{t_{rise}}\sim40$min). Cooling time is
mainly controlled by the Keplerian shearing  and synchrotron loss
time, $t_{syn}=5\times10^5B^{-\frac{3}{2}}\nu^{-\frac{1}{2}}$ min,
where $B$ must be set in Gauss and $\nu$ is in GHz.

\begin{figure}[t]
\centering{\includegraphics[width=0.47\textwidth]{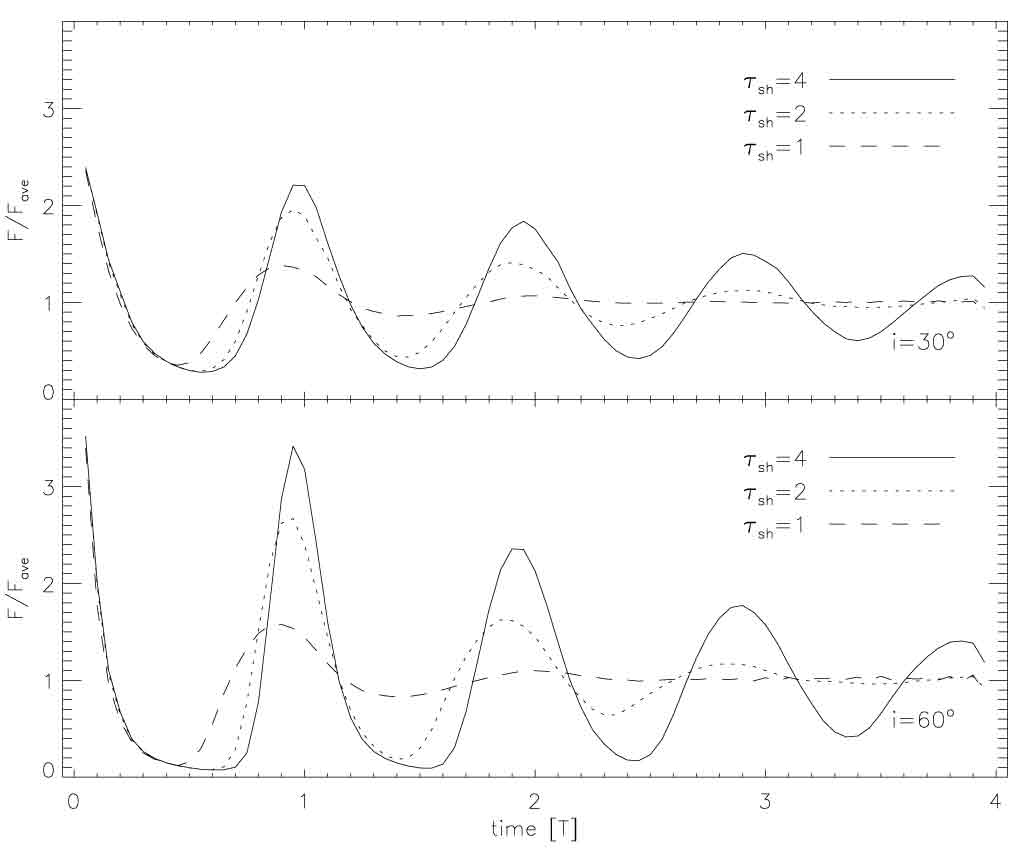}}
\caption{Flux modulations of an evolving perturbation close
to the marginally stable orbit of a Kerr black hole with spin
parameter of 0.5. The light curves show how different values of
shearing parameter and inclination affects the resultant light
curve.} \label{lc_shearing}
\end{figure}

\subsection{Hot spot model and fluctuations of the inner parts of the accretion disk}
Since all these
physical processes happen very close to the black hole and in a very
strong gravitational regime, we  must take into account the effects
of curved space-time. To simulate the changes in paths and
polarization properties of photons from the emitting electrons to
the observer at infinity, we have used the KY ray-tracing code
(Karas et al. 1992; Dov\v{c}iak et al. 2004). KY is able to
calculate all the effects of GR, like light bending and changes in
the emission angle, changes in the polarization angle of photons,
gravitational lensing and redshift, Doppler boosting (since matter
inside the accretion disk is in orbit) and frame dragging (in case
of Kerr black holes) in a thin disk approximation. In the
geometrical optics approximation, photons follow null geodesics, and
their propagation is not affected by spin-spin interaction with a
rotating BH (Mashoon 1973). This means that wave fronts do not
depend on the photon polarization, and so the ray tracing through
the curved space-time is adequate to determine the observed signals.
Since our analysis is focused on high frequency regimes, we have
mainly ignored radiative transfer effects.

\begin{figure}[t]
\centering{\includegraphics[width=0.44\textwidth]{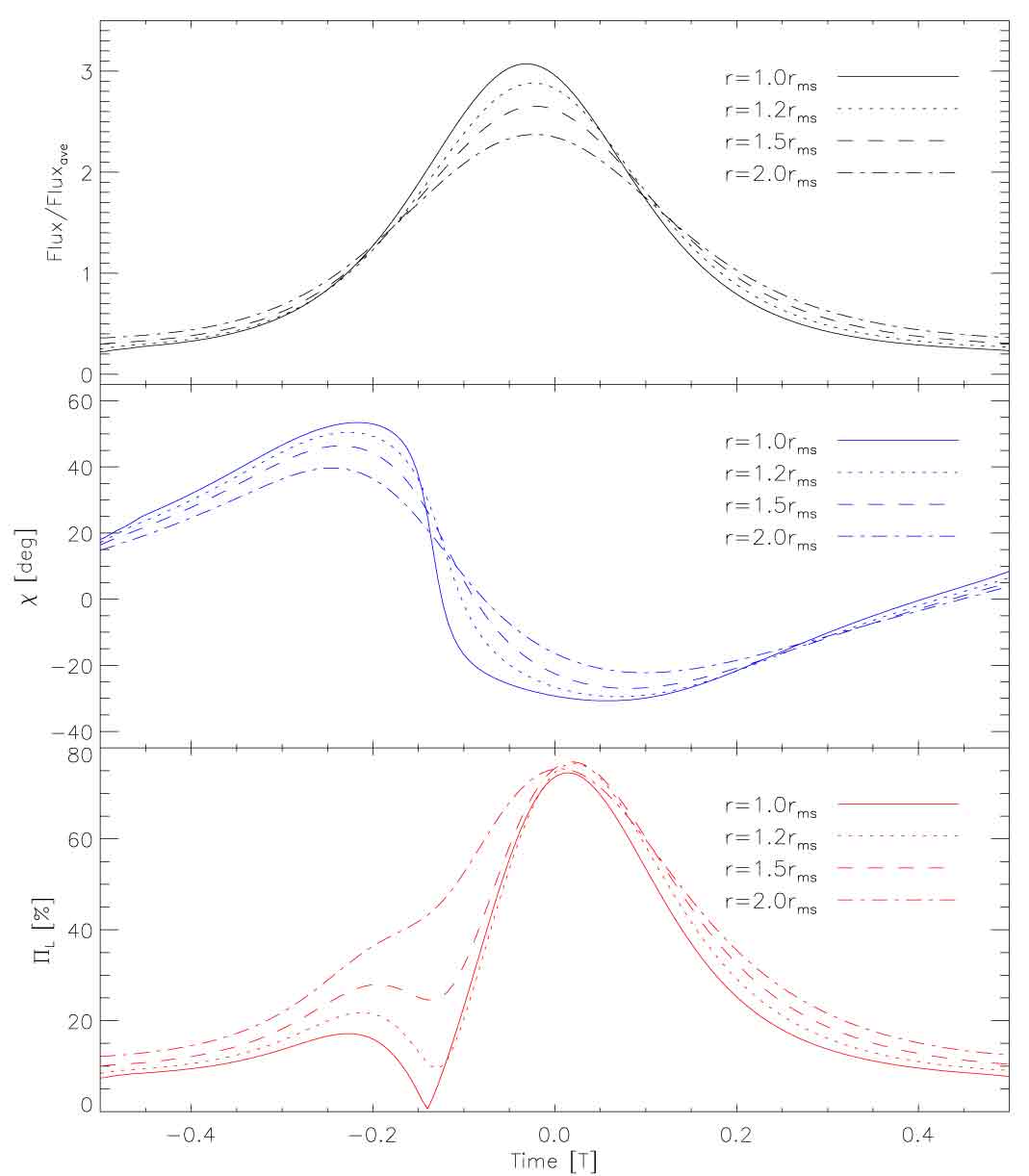}}
\caption{Flux modulation (top), changes in polarization
angle (middle) and degree (bottom) for a  spot on a circular orbit at
$1.0\times r_{ms}$ (solid), $1.2\times r_{ms}$ (dotted), $1.5\times
r_{ms}$ (dashed),$2.0\times r_{ms}$ (dott-dashed) around a Kerr
black hole with a spin parameter of 0.5. The time unit is the orbital time
scale ($T$).} \label{spot_pattern}
\end{figure}

\begin{figure}[!t]
\centering{\includegraphics[width=0.44\textwidth]{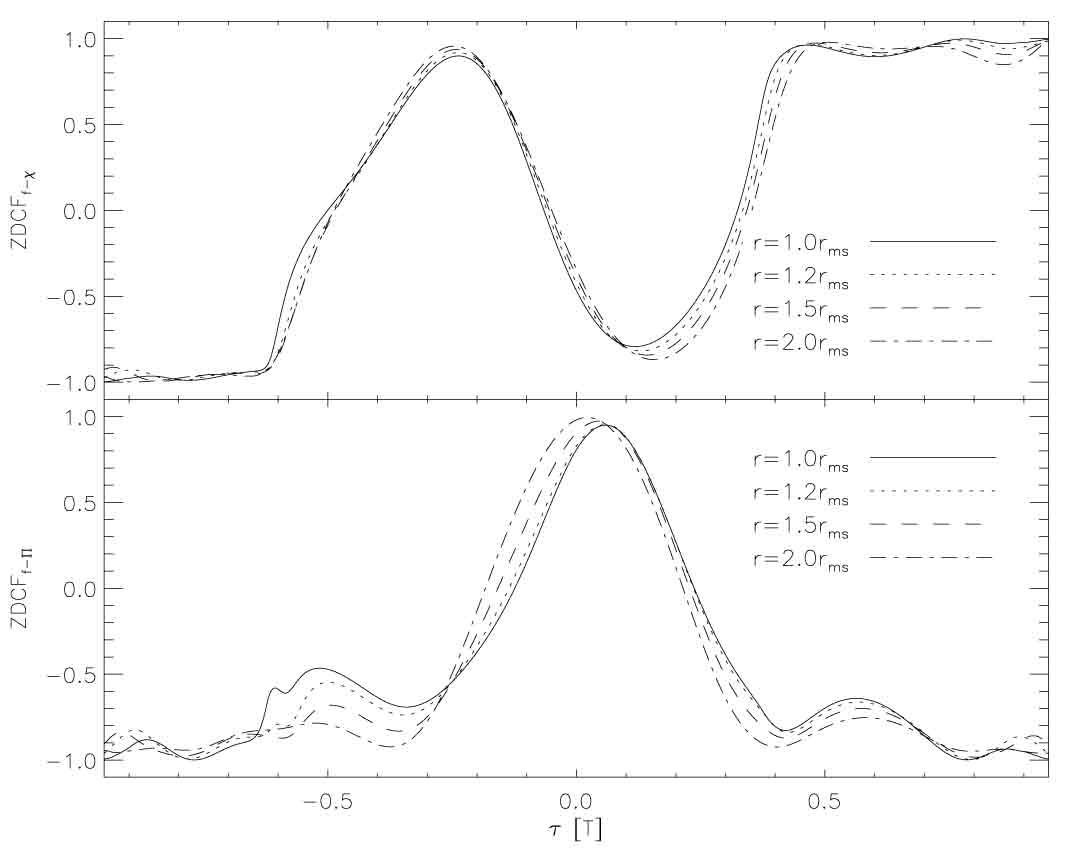}}
\caption{Cross correlation between the changes in flux and
polarization angle (top), and flux and polarization degree (bottom)
for spot circular orbits at $1.0\times r_{ms}$ (solid), $1.2\times
r_{ms}$ (dotted), $1.5\times r_{ms}$ (dashed),$2.0\times r_{ms}$
(dott-dashed) around a Kerr black hole with a spin parameter of 0.5 (Fig.
\ref{spot_pattern}). The time unit is the orbital time scale ($T$).}
\label{spot_pattern_zdcf}
\end{figure}

To make KY work, we must initialize  the properties of radiated
photons at each point of the emitting region. The straightforward
way is to define the intrinsic emission of each point in the context
of the Stokes parameters. For the flux densities (mJy) and
source sizes ($\mu$as) of Sgr~A*, optically thick synchrotron
emission in the NIR can safely be disregarded (see the discussion in
Eckart et al. 2009). Since in this report we focus only on the modeling
of the NIR flares, it is sufficient to pick up a model for the
energy distribution of non-thermal electrons, radiating in an
optically thin regime:
\begin{equation}
N(\gamma)= \left\{\begin{array}{cc}{{N_0\gamma^{-p}}} \  \,
& \gamma \leq \gamma_c \\
    \\
    {{0 }} \ \, &
    \gamma > \gamma_c \end{array}\right.
\label{elecinj}
\end{equation}
which leads to the formulae for polarized emission:
\begin{eqnarray}
I_\nu \ &\propto& \ n \ {(B\sin\theta_e)}^{(\frac{p+1}{2})}\nu^{-(\frac{p-1}{2})} \\
Q_\nu \ &=& \ \Pi_L \cos{(2\chi_e)} \ I_\nu \\
U_\nu \ &=& \ \Pi_L \sin{(2\chi_e)} \ I_\nu\\
V_\nu \ &=& \ \Pi_C  \ I_\nu
\end{eqnarray}
where  $n$, $\chi_e$, $\Pi_L$ and $\Pi_C$ are the number density of the
electrons, the angle between a reference direction and the plane of
an observer co-moving with the disk frame and degree of linear and
circular polarization, respectively. Throughout this paper we assume
that  in the NIR regime the light is not circularly polarized
($\Pi_C=0$). $I_\nu, Q_\nu, U_\nu$, and $V_\nu$ represent the Stokes
parameters. $\theta_e$ is the angle between the direction of the
magnetic field and the direction toward the co-moving observer

\begin{figure*}[!t]
\begin{minipage}{\textwidth}
\centering{\includegraphics[width=\textwidth]{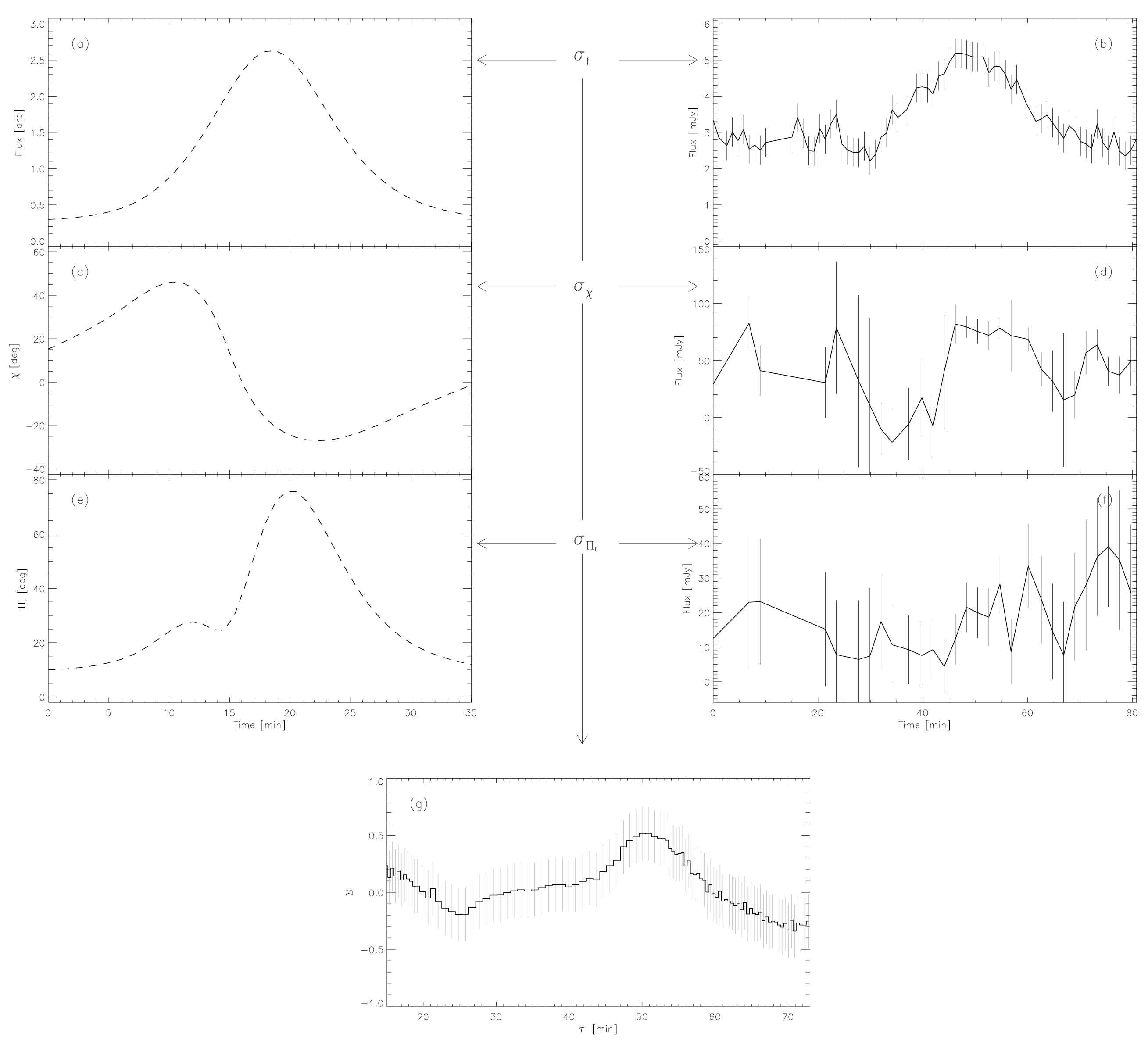}}
\end{minipage}
\caption{A sketch showing how the pattern recognition
coefficient ($\Sigma$) is defined. $\Sigma$ is the multiplication
product of the cross correlation between observed and theoretical flux
($\sigma_f$), observed and theoretical polarization angle
($\sigma_\chi$) and observed and theoretical polarization degree
($\sigma_\Pi$) light curves  following
 Eq. \ref{sigma}. Note that $\Sigma(\tau')$ (g) is defined as a function
  of $\tau'=\tau+\frac{T_{sim}}{2}$  (where $\tau$ is the
time lag in units of minutes) in order to make it easier to match
the position of its peaks with the position of the lensing events in
the observations. } \label{pattern_sketch}
\end{figure*}

\begin{equation}
\theta_e
=\theta_e(\eta,\psi,\phi_e,\delta_e)=\arccos\bigr({\sqrt{\frac{(\mathcal{B}^\alpha\kappa_{e\alpha})^2}{(\kappa_e^\beta\kappa_{e\beta})(\mathcal{B}^\gamma
\mathcal{B}_\gamma)}}} \bigl)
\end{equation}
where we have picked the disk co-moving frame as the reference, so
that the normal to the disk, $\hat{n}$, coincides with the $\hat{z}$
direction. Geometrical orientation of the dominant component of the
global magnetic field vector ($\mathcal{B}$) in this frame can be
defined with a set of angles [$\eta, \psi$] ($0\leq\eta\leq\pi$,
$0\leq\psi\leq2\pi$, see Fig. \ref{magnetic}). $\delta_e$ is the
angle between the direction of the photon momentum ($\kappa$) and
the normal to the disk ($\hat{n}$):

\begin{equation}
\delta_e=\arccos\big(-\frac{\kappa_{e\alpha}n^\alpha}{\kappa_{e\beta}v^{\beta}}\big)
\end{equation}
where $v$ is the four-velocity of matter in the disc. $\phi_e$ is
the azimuthal emission angle which is defined as the angle between
the projection of the three momentum of the emitted photon into the
equatorial plane and the radial tetrad vector:
\begin{equation}
\phi_e = \arctan \big( \frac{\kappa_e^\alpha
\hat{e}_{(\phi)\alpha}}{\kappa_e^\mu \hat{e}_{({r})\mu}}\big)
\end{equation}

\begin{figure*}[!t]
\begin{minipage}{\textwidth}
\centering{\includegraphics[width=\textwidth]{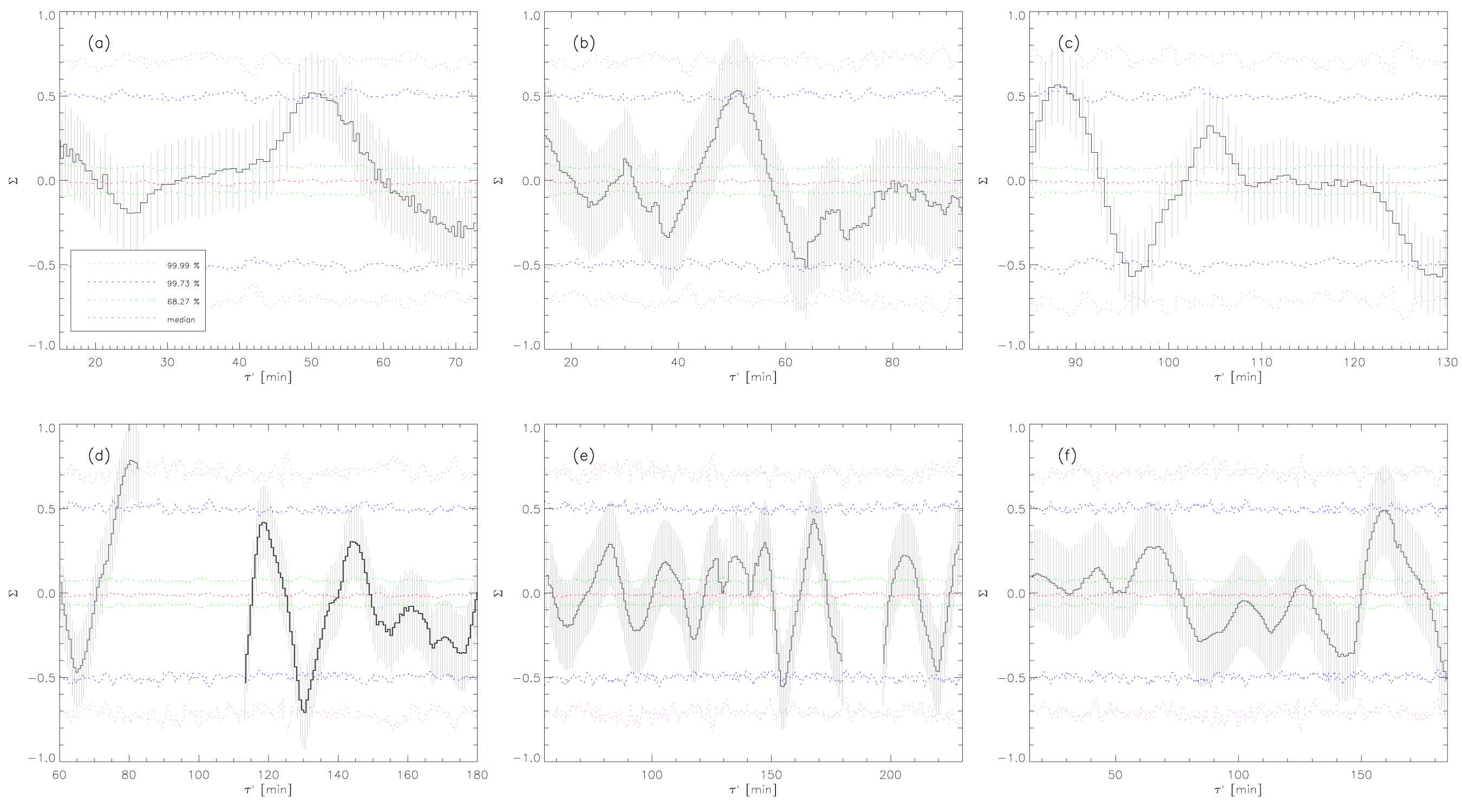}}
\end{minipage}
\caption{Pattern recognition coefficients as a function of
time for our sample of observations [13 June 2004 (a), 30 July 2005
(b), 1 June 2006 (c), 15 May 2007 (d),  17 May 2007 (e) and 28 May
2008 (f)]. Dotted lines show the median (red), $68.3\%$ (green),
$99.8\%$ (blue) and $99.99\%$ (violet) significance levels derived
from $10^4$ random red noise light curves.} \label{pattern_sample}
\end{figure*}

Having all the needed information at hand, by fixing free parameters
like the spatial density  distribution of the emitting electrons,
magnetic field strength, flux spectral index, and the global
configuration of the magnetic field, one can simulate the expected
images  that a distant observer will measure  for different possible
inclinations with respect to the black hole/accretion disk system.
Here we will show how compact azimuthal anomalies inside a uniform
density  distribution of the emitting plasma can reproduce the
observed behavior of Sgr~A* in high frequency regimes.

The synchrotron cooling time at NIR frequencies is only of the order
of a few minutes, which is significantly shorter than the time scale
of the observed flares. One way that a spot can survive long enough
 to be responsible for the observed $\sim4-5$ time flux
modulations is that a SSC mechanism up-scatters the sub-mm seed
photons to the NIR and X-ray frequencies (Eckart et al. 2006a-c,
2008a). The other possibility is that the heating time of the orbiting NIR
component is related to the rise time of the main flare,
$\tau\propto t_{rise}$ where the emissivity profile of the emitting
component follows  $F(t; t_0,\tau)= F_0\exp{(-(t-t_0)^2/2\tau^2)}$.

The above discussion demonstrates that it is essential to consider
the gravitational shearing time scale as a variable in the
simulations. We have implemented this effect in our modeling by
introducing a dimensionless characteristic shearing time scale:
\begin{equation}
\tau_{sh}=\frac{T(r_{spot})}{T(r_{spot}+r_0)-T(r_{spot}-r_0)}
\end{equation}
where for the initial spatial distribution of the relativistic
electrons we have used a spherical Gaussian distribution with its
maximum being located at the radius $r=r_{spot}$ with  FWHM of
$r_0=1r_s$. In our simulations, $\tau_{sh}$ varies between
$\infty$ and $0.8$.  $\tau_{sh}=\infty$ corresponds to the situation
in which the spot preserves its shape for a long time. The mechanism
that stabilizes the spot is not known, although several
possibilities have been proposed and explored in the literature. In
particular, the existence of persistent vortices on accretion disks
has been explored (Abramowicz et al. 1992; Adams \& Watkins 1995).
However, our modeling suggests that shearing effects are indeed
important and can be well represented within the multi-component
scheme or in the spiral pattern scheme generalizing the original
spot scenario.
The $\tau_{sh}=0.8$
corresponds to a pure Keplerian shearing for a spot located at the
marginally stable orbit of an extreme spinning black hole ($a=1$).
One must note that MHD simulations are unable to produce hot spots
with long life-times comparable to the observed flare time scales
(Hawley et al. 2001, Krolik et al. 2002). Schnittman et al. (2006)
used a model for the creation and annihilation of spots with short life
times, distributed by random phase within a belt close to the
marginally stable orbit. They show that their model can resemble the
observed quasi-periodicity in the X-ray light curves. Eckart et al.
(2008a) successfully modeled the simultaneous NIR/X-ray flares of
Sgr~A* by following the same basic idea of a multi-component model.
As mentioned in \S2, following the results of Pech\'{a}\v{c}ek et
al. (2008) these kinds of multi-component models are able to
reproduce the red noise behavior of PSDs while some bright
individuals can show their signatures in polarized light (see Fig.
\ref{multi} and also discussion in Eckart et al. 2008a).

\subsection{Results of the modeling}

Figure \ref{spots_shearing} shows how a hot spot
is created and evolves in time for three different values of
$\tau_{sh}$. A comparison between the rows shows  how pure Keplerian
shearing disrupts the initial shape of the spot and produces an
elongated spiral shape. Figure \ref{spots_inclination} shows the
apparent images of a spot with a mild shearing environment
($\tau_{sh}=2.0$) for three different inclination angles. When we
look face-on at the event ($i\simeq0^{\rm o}$), there are no
modulations by the relativistic effects. For higher inclination
angles ($i=30^{\rm o} \ \mbox{and} \  i=60^{\rm o}$) lensing and
boosting effects play major roles in the observed flux modulations.
Specially for high inclinations one can see how an Einstein arc
develops when the spot passes behind the black hole and how photons
coming from the accretion disk are blue-shifted on the left hand side of the
image according to Doppler boosting. This set-up allows us to
simulate light curves for a wide range of possible free parameters,
mainly by covering the range of all possible inclinations  and
spins of the black hole. Figure \ref{lc_shearing} shows examples of light curves for different
values of inclination and shearing time scale.

\begin{figure}[t]
\centering{\includegraphics[width=0.47\textwidth]{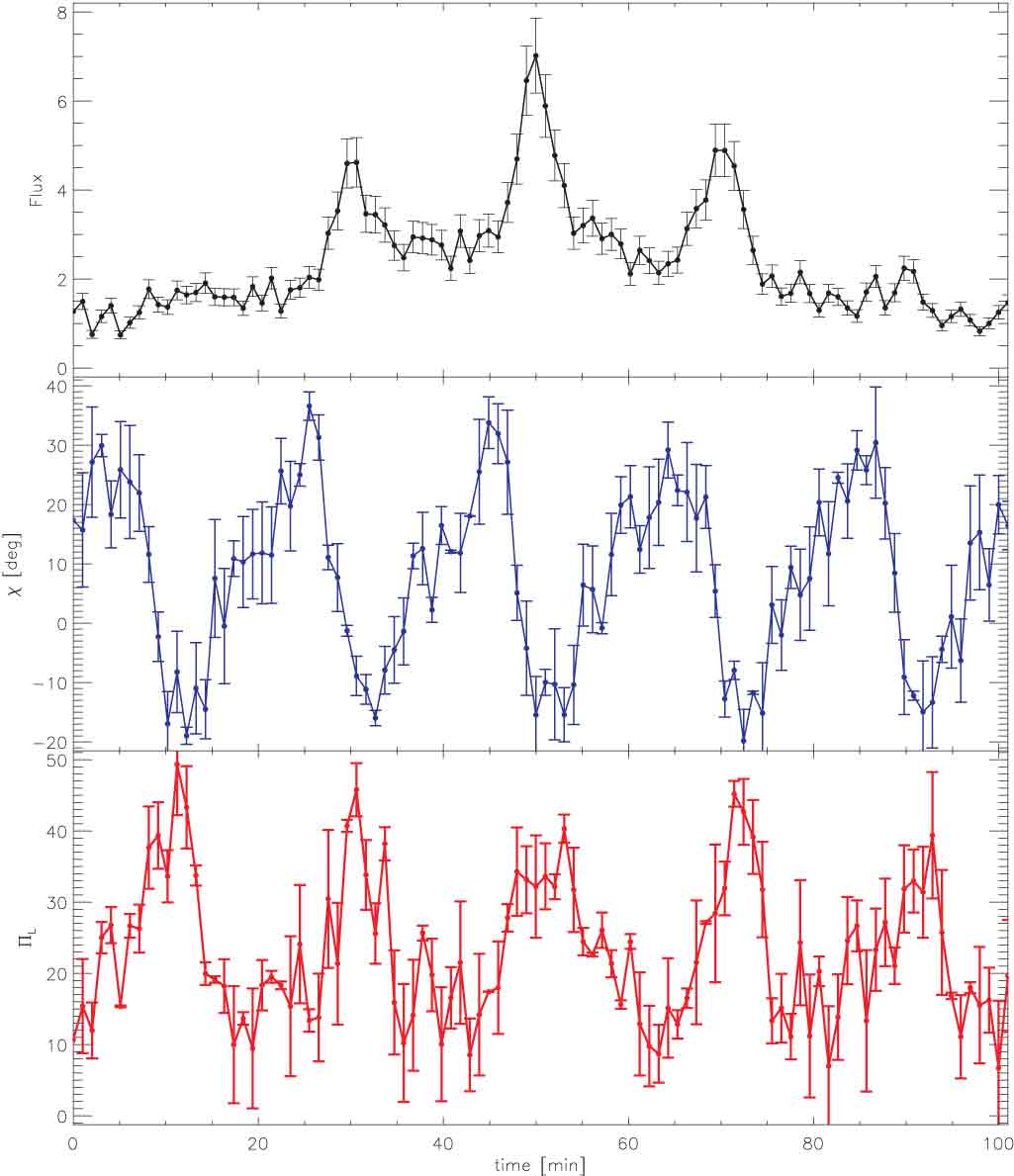}}
\caption{A selected 100 minutes window of the white noise added
simulated light curves of an orbiting spot plus a temporary
variability in the accretion disk. The flux density is presented in
arbitrary units (top). Middle and bottom panels  show the changes in
the angle (degree) of polarization. The spin of the black hole is
set to be $a=0.5$ and the inclination fixed on $i=60^{\rm o}$.}
\label{model}
\end{figure}

\begin{figure}[!h]
\centering{\includegraphics[width=0.45\textwidth]{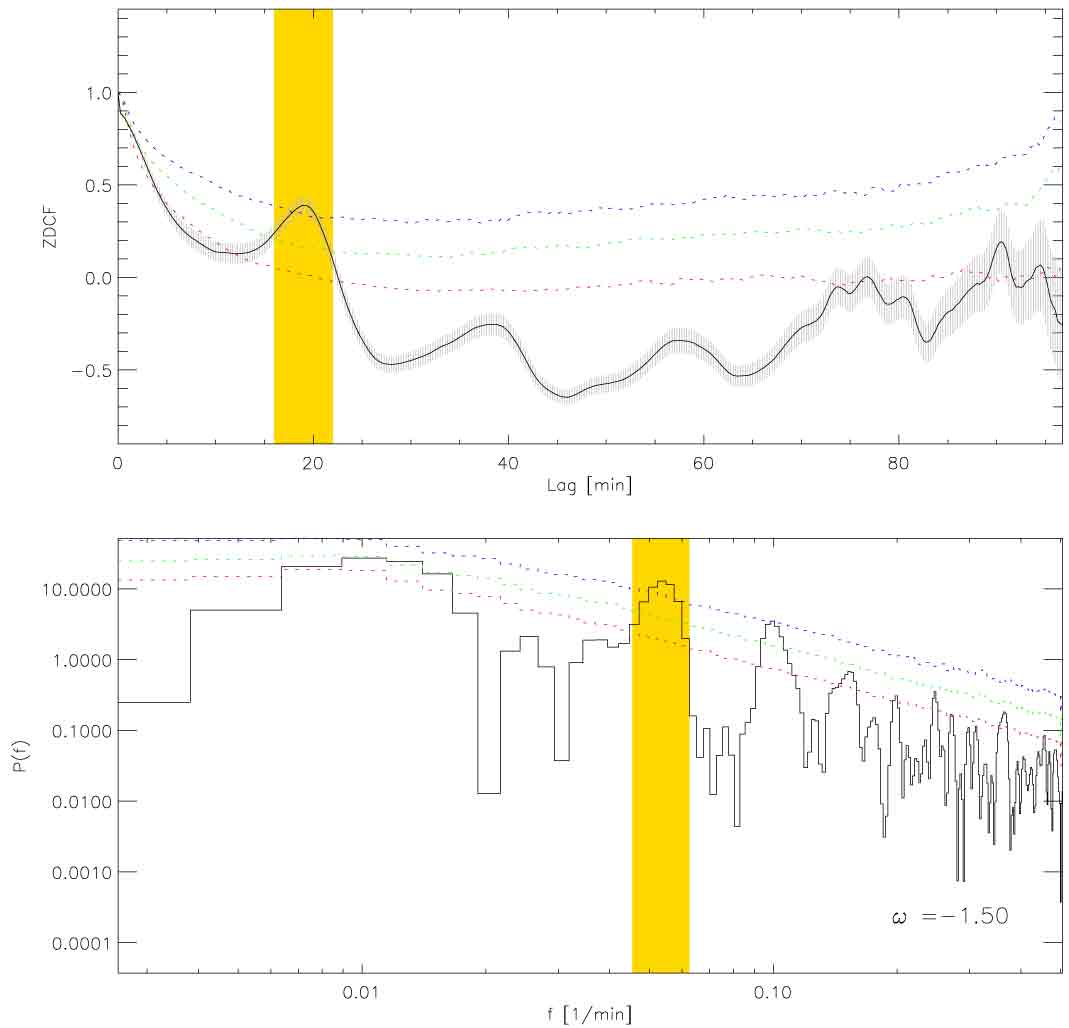}}
\caption{Same as Figs. \ref{zdaf} and \ref{ls} for the
simulated light curves of Fig. \ref{model}. The top shows the
autocorrelation of the flux while the bottom shows the Lomb-Scargle
periodogram. The  colored regions indicate the position of the
 peaks corresponding to the $0.8-1.2 \times r_{mso}$ orbital time scales of a  Kerr black hole with spin 0.5. Dotted lines show
the median (red), $68.3\%$ (green) and $99.8\%$ (blue) confidence
levels of the red noise.} \label{model_zdcf_ls}
\end{figure}

\begin{figure}[!h]
\centering{\includegraphics[width=0.45\textwidth]{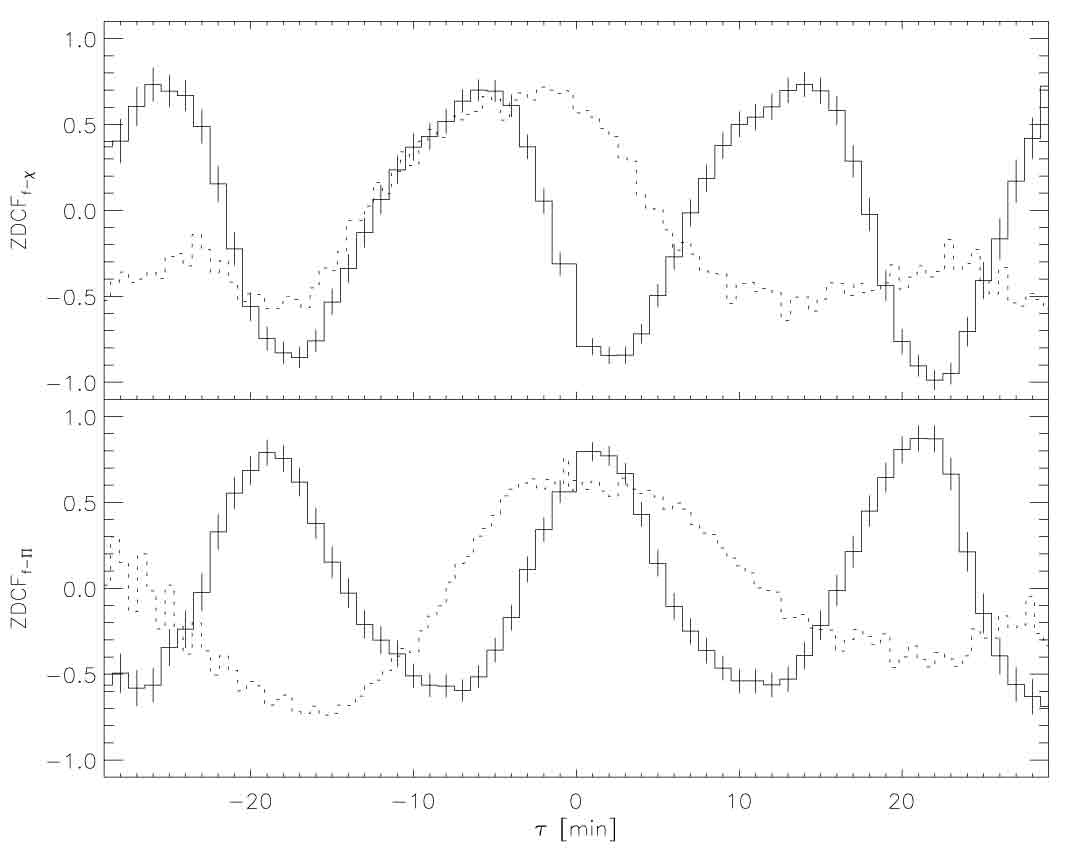}}
\caption{Top (bottom): Cross correlation between the total flux
and polarization angle (degree) of the simulated light curve (Fig.
\ref{model}).
 Dotted lines  indicate the
corresponding cross correlation functions for the 30 July 2005 flare
as the case  most similar to the predictions of  the hot spot model.}
\label{model_lc_zdcf}
\end{figure}

\subsubsection{Pattern recognition analysis: signatures of lensing effects}
Figure \ref{spot_pattern} shows the typical magnification of
flux, polarization angle and degree of an orbiting spot emission as a
function of time. Here we showed the spots located at different
distances from the black hole. These plots indicate the typical
behavior of  light curves when the strong gravitational regime is
prominent and the strong lensing and boosting is happening. As one
can see, the sweep in the polarization angle precedes the peak in
flux magnification, while the peak of the degree of polarization
follows the magnification peak (see Broderick and Loeb 2006 for a
detailed discussion). Figure \ref{spot_pattern_zdcf} shows more clearly this
typical behavior  according to the position of the
corresponding cross-correlation's peaks. It is particularly apparent
that even with changing the position of the emitting source with
respect to the black hole this effect remains more or less the
same.

This constant behavior encouraged us to check whether or not
this type of pattern is manifested in our NIR light curves of
Sgr~A*. For this purpose we used a simple pattern recognition
algorithm mainly via a multiplication of different cross correlation
functions. Similar pattern recognition algorithms are used to
identify gravitational wave signals from noisy data (Pappa et al.
2003, Goggin 2008).

We derived a final pattern recognition coefficient product by multiplying two cross correlation functions
(namely $\sigma_{i}$ and $\sigma_{j}$) as below:
\begin{equation}
 \Sigma_{ij}=\sqrt{(\sigma_i+1)(\sigma_j+1)}-1
\label{sigma}
\end{equation}
By applying this procedure for the cross correlation functions between
the observed flux and the theoretical magnification light curve
($\sigma_f$),  cross correlation functions between observed
polarization angle and the theoretical polarization angle light
curve ($\sigma_\chi$) and the same function for the degree of
polarization ($\sigma_\Pi$), we derived a final pattern recognition
coefficient product ($\Sigma$). Figure \ref{pattern_sketch} shows how
this pattern recognition coefficient is defined.

\begin{figure*}[!t]
\begin{minipage}{\textwidth}
\centering{\includegraphics[width=\textwidth]{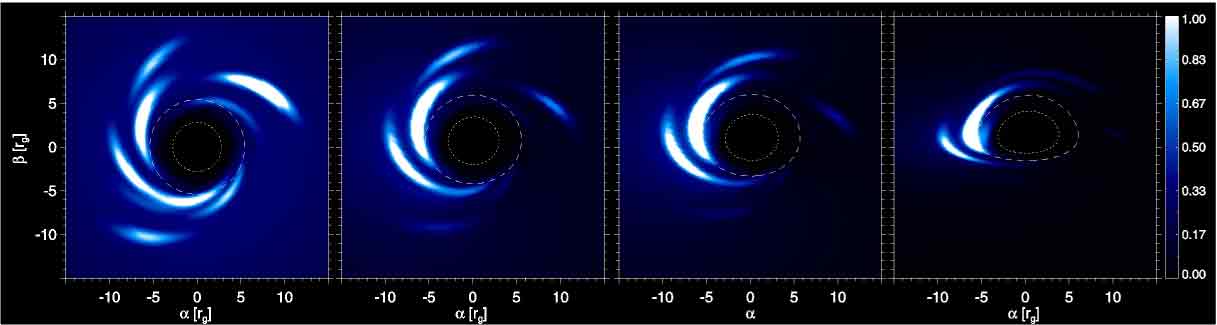}}
\end{minipage}
\caption{Same as Fig. \ref{spots_inclination} for the
multi-component scenario. The observer's inclination is set to $0^{\rm
o}$ (a), $30^{\rm o}$ (b), $45^{\rm o}$ (c) and $70^{\rm o}$ (d)
 (see the discussion in Sect. 3.1 and Table. A.2
in Eckart et al. 2008a).} \label{multi}
\end{figure*}

Figure \ref{pattern_sample} shows the result of our pattern
recognition analysis for the polarized flare events discussed in
this paper. In order to estimate how significant the peaks in the
$\Sigma$ function are, we have repeated the same analysis for $10^4$
random red noise light curves simulated with the same method
mentioned in Sect. 2.2.2. In all but one case the patterns shown in
Fig. \ref{pattern_sample} can be identified at the $>3\sigma$ to
$5\sigma$ level. This shows that strong lensing patterns are
significantly manifested in our sample of NIR light curves.
The strong lensing pattern detected is a further indicator for the existence of
a (clumpy) accretion disk around Sgr~A*.

\subsubsection{A spotted accretion disk?}
Figure \ref{model} shows a selected 100 minutes window of the
resultant light curves of the flux density, polarization degree and
angle for a spot with constant shape ($\tau_{sh}=\infty$), orbiting
around a Kerr black hole ($a=0.5$) close to its marginally stable
orbit ($r=1.1\times r_{mso}$). The line of sight is inclined by 60
degrees ($i=60^{\rm o}$). Gaussian white noise has been added to the
simulated data in order to make the comparison of the periodicity
and cross correlation with corresponding observational results
easier. The level of the noise and the average error bars have been
set from the average rms of the corresponding observed light curves. The
surface brightness of the components follows a profile similar to
$F(t; t_0,\tau)$ with $\tau=25$. The maximum degree of polarization
which can be achieved via synchrotron mechanism is around 70\%.
Since any kind of deviation from the ideal isotropic distribution of
electrons around the magnetic field lines will suppress the degree
of linear polarization, we set the initial value for the  radiation
from the spot to be 50\%. We assumed that the photons originating
from the non-flaring part of the accretion disk
 are weakly polarized ($\sim1\%$), since the main population of its NIR photons have
thermal origin and  relativistic electrons are randomly distributed
around the magnetic field lines.

\begin{figure}[!t]
\centering{\includegraphics[width=0.47\textwidth]{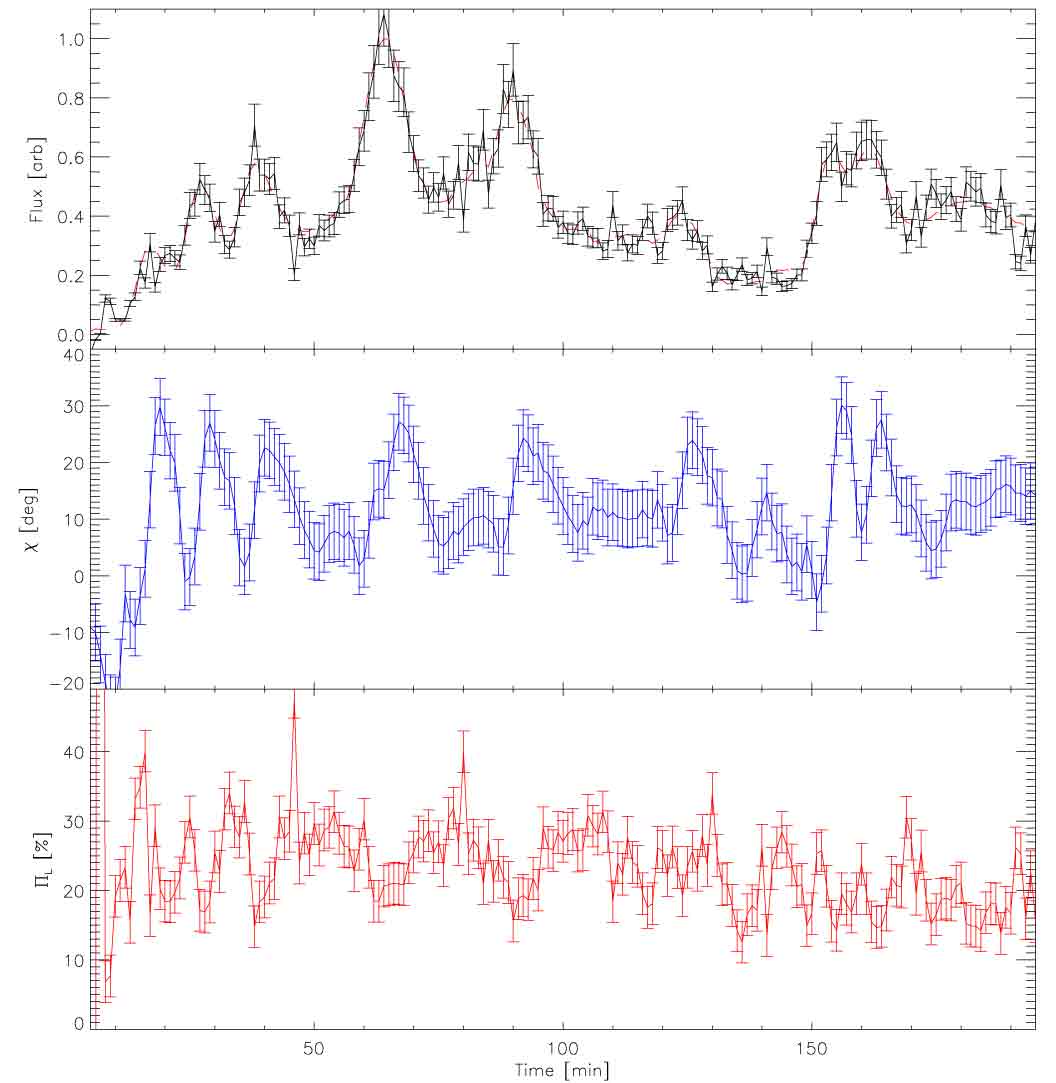}}
\caption{Same as Fig. \ref{model} but a selected 200 minutes-window of the multi component scenario.} \label{multi-lc}
\end{figure}

\begin{figure}[!t]
\centering{\includegraphics[width=0.45\textwidth]{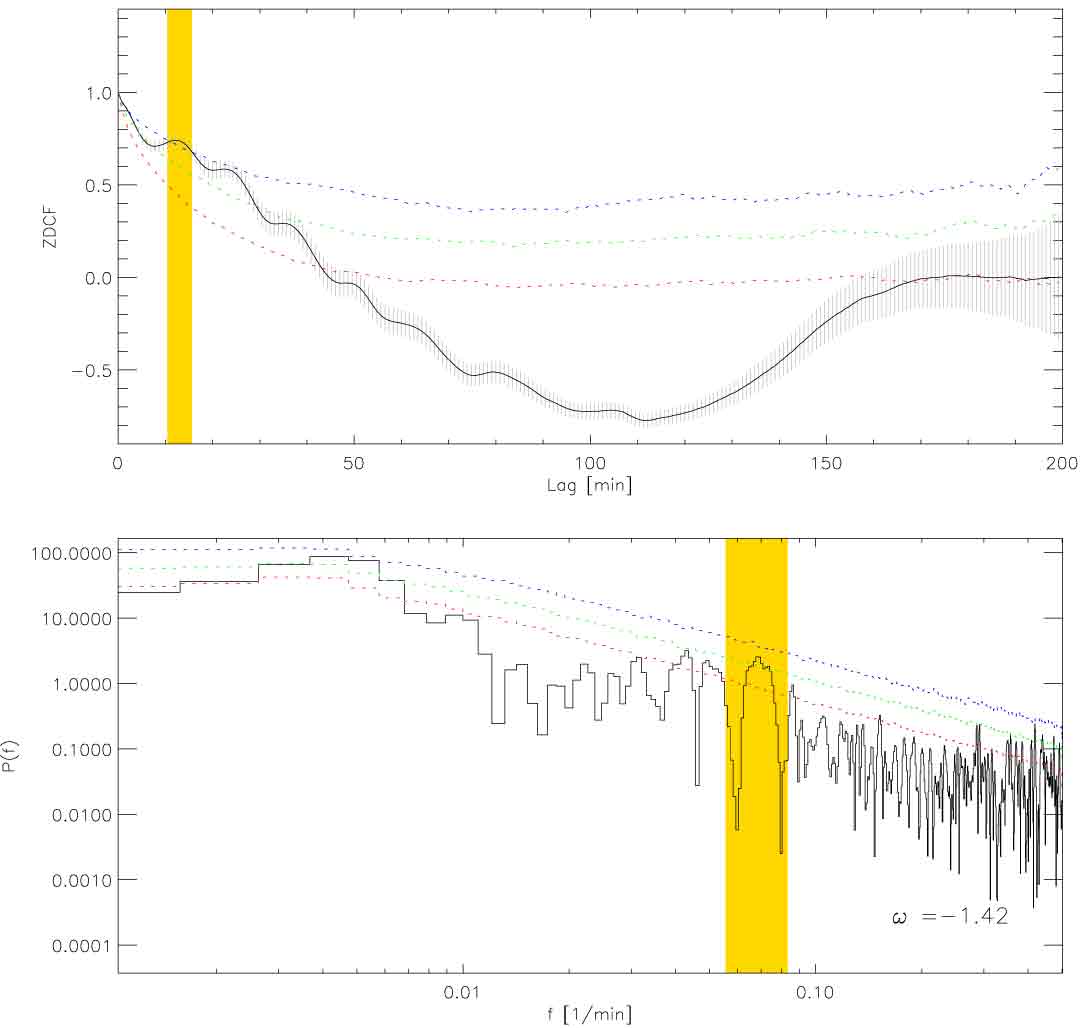}}
\caption{Same as Figs. \ref{model_zdcf_ls} for the multi-component scenario (Fig. \ref{multi-lc}).
 }
\label{multi_zdcf_ls}
\end{figure}

\begin{figure}[!t]
\centering{\includegraphics[width=0.45\textwidth]{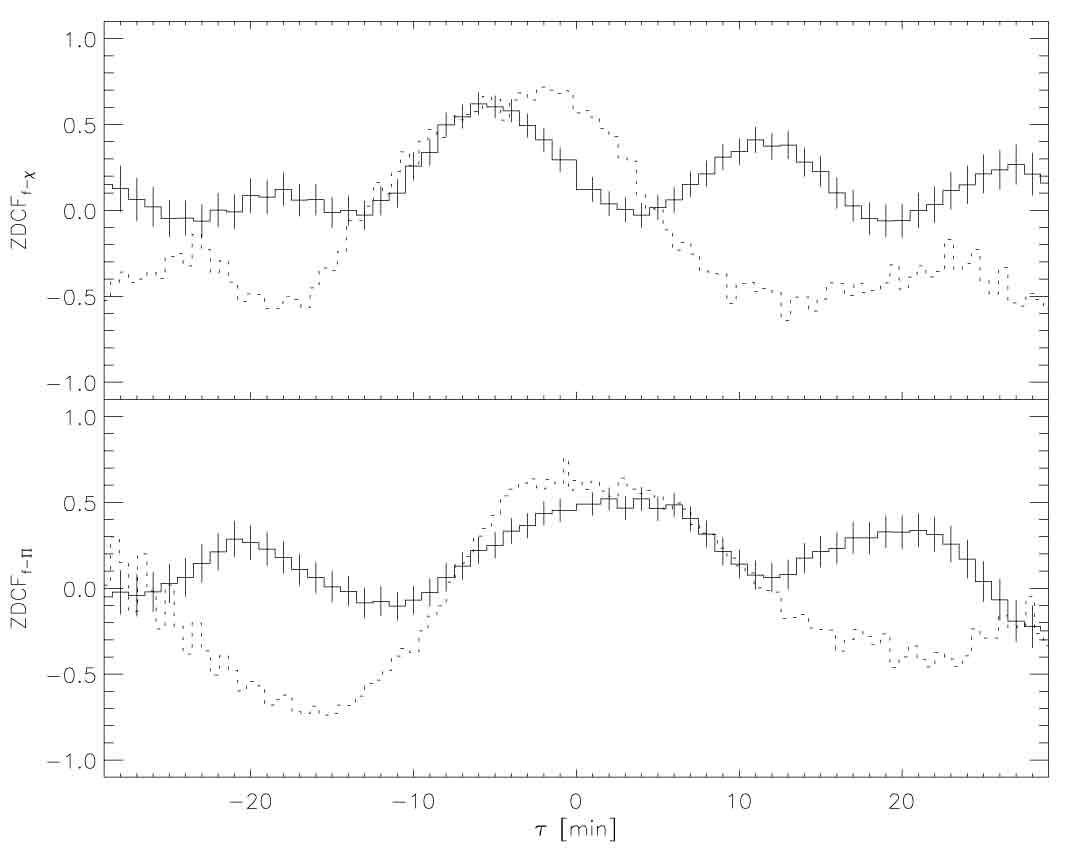}}
\caption{Same as Figs. \ref{model_lc_zdcf} for the multi-component scenario (Fig. \ref{multi-lc}).} 
\label{multi_lc_zdcf}
\end{figure}

\begin{figure*}[t]
\includegraphics[width=\textwidth]{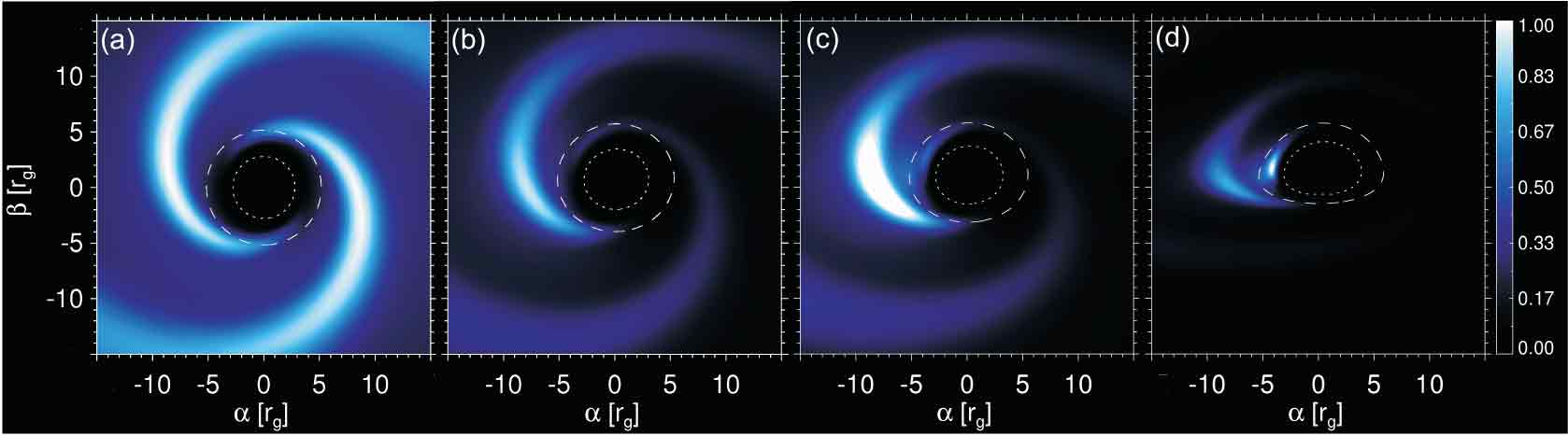}
\caption{Snapshots of the orbiting spiral pattern as they appear to
a distant observer looking along a line of sight inclined by
$0^{\rm o}$ (a), $30^{\rm o}$ (b), $45^{\rm
o}$ (c) and  $70^{\rm o}$ (d)  [relative to the normal to the disk].
See also the caption of Fig. \ref{spots_shearing}}
\label{spiral}
\end{figure*}

Figure \ref{model_zdcf_ls} shows the result from
autocorrelation and Lomb-Scargle analysis (similar to Figs.
\ref{zdaf} \& \ref{ls}), and Fig. \ref{model_lc_zdcf} shows the
results of the cross correlation analysis (similar to Fig.
\ref{zdcf}). The peaks close to the 0 minute time-lag are of special
interest since they have the least dependency on the choice of free
parameters and are mainly related to the basic idea that the flux
modulations  are caused by relativistic effects.  Figures
\ref{model_zdcf_ls} and \ref{model_lc_zdcf} show that an orbiting
spot is able to produce the same cross correlation  pattern observed
in our sample, but the orbital frequency of the spot will be detected 
to be mainly a quasi-periodical signal. The corresponding correlation functions of the 30 July
2005 observation are over-plotted in Fig. \ref{model_lc_zdcf} for a
better comparison. The main peak close to $\tau=0$ coincides very
well for both observation and simulated cross-correlations.

Furthermore, we have simulated a spotted accretion disk. In
this case spots are born, evolve and finally fade away as a function
of time. These anomalies are distributed in the inner part of the
accretion disk in a belt between $1-2\times r_{mso}$. Radial and
azimuthal distribution of the anomalies are completely random and
their distribution in time follows a simple Poisson point process
(see Pechachek et al. 2008).

Figure \ref{multi} shows the snapshots of the spotted disk
scenario as viewed by a distant observer from different inclination
angles. The resulting magnification and polarimetric light curves
are depicted in Fig. \ref{multi-lc}. White noise is added to all
light curves in order to make the comparison between ZDCF and
Lomb-Scargle results with the corresponding observational results
easier. As can be seen in Fig. \ref{multi_zdcf_ls} the random
distribution of the spots strongly suppresses the periodic signal in
the Lomb-Scargle periodogram, while the cross correlation between the
magnification and the changes in the polarized flux still carries
a significant signal from the modulations influenced by strong gravity
(Fig. \ref{multi_lc_zdcf}). As a conclusion one can say that a
general relativistic simulation of turbulences in the inner parts of
an accretion disk  resembles the observed behavior of Sgr~A* very
well.

One must note that for the simulated data, we have chosen typical
values for spin and inclination (consistent with previous results by
Eckart et al. 2006b and Meyer et al. 2006a,b), not trying to fit the
actual data. As we can see, the data analysis and simulations show
that even if there does exist a low-level activity physical process in
Sgr~A* which can be explained as random red-noise, the kind of
observed correlations between the flux and polarization data cannot
be produced via a completely random  process (without taking into
account relativistic effects). Actually, the similarity between
Figs. \ref{model_lc_zdcf} and \ref{multi_lc_zdcf} and what is already
observed in at least three different Sgr~A* flares (Figs. \ref{ls}
and \ref{zdcf}), supports the idea of orbiting matter around the
galactic super-massive black hole.

\begin{figure}[!b]
\centering{\includegraphics[width=0.48\textwidth]{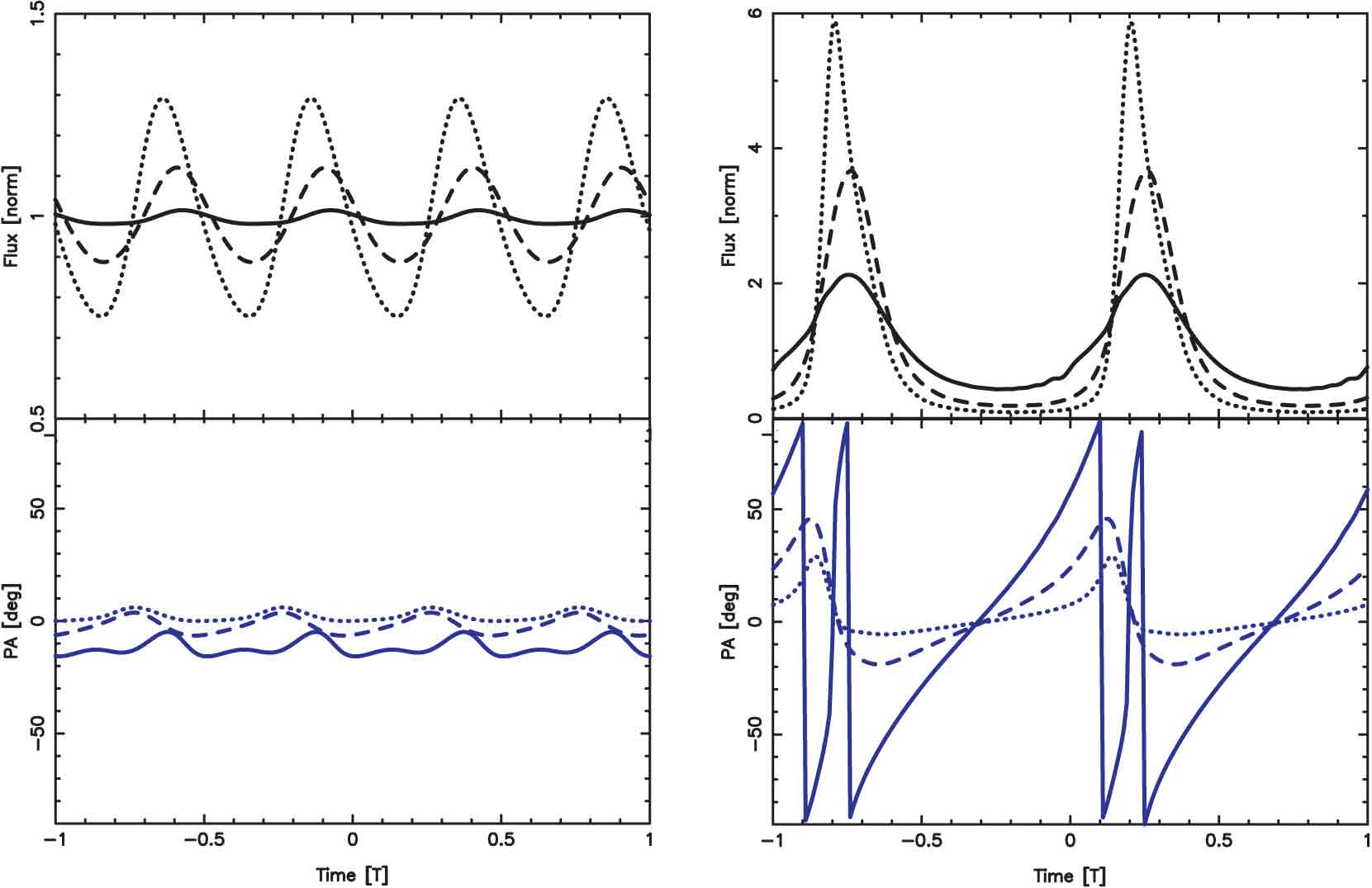}}
\caption{Light curves of the normalized flux (top, black) and
polarization angle (bottom, blue) observed at infinity, obtained
form general relativistic modulations associated with a spiral
pattern (left) or a compact spot (right). Light curves are shown for
three different inclinations: $i=30^{\rm o}$ (solid line),
$i=60^{\rm o}$ (dashed line) and $i=80^{\rm o}$ (dotted line). The
modulations happen twice as often for the spiral pattern than  for the
spot in the same interval, according to the existence of two
symmetrical arms. Flux values are normalized according to their
average values. The spin of the black hole is set to be zero
($a=0$). } \label{spiral_spots}
\end{figure}

\subsection{Alternative models}
As mentioned before, the strength of ZDCF's peaks in Fig.
\ref{zdaf} seems to be correlated with the brightness of flare
events.  This dependency on the brightness of the flares and also
the shape of the resulting ZDCFs are very similar to the expected
autocorrelation if the flux variations originate from  geometric
"light echoes" (Bursa et al. 2007; Fukumura et al. 2009). The
expected periodicities from these models are different from the ones
observed here, but their very close similarity keeps this issue open to
future discussions. If a more detailed modeling of random flares
happening at a distance from the black hole can show a possible
$\sim 20$ minutes QPO, then the light curve observed on 15 and 17
May 2007 could be the first observation of second images created
from separated photon "bunches" (Fukumura 2009).

Furthermore, shocks in relativistic jets can produce a
correlated total flux and polarized intensity. There is a
possibility that under special circumstances (e.g special
inclination,
special magnetic field configuration etc.) a (episodic) relativistic outflow
could produce the same correlation and pattern we have discovered in
our observations.

\section{Geometry of the emitting region}

In this section we discuss  the basic assumption of the existence of
an azimuthal asymmetry in the accretion flow of Sgr~A*. Recently,
Falanga et al. (2007) and Karas et al. (2007) discussed that a
global spiral pattern of disturbance, with an orbiting speed not
directly associated with the underlying Keplerian velocity, can fit
the observed NIR and X-ray modulations of Sgr~A*. Their model has
been used to fit the observed X-ray flare on 31 August  2004
(Falanga et al. 2007). Here we will show  how a combination of
polarimetric observations and the next generation of VLTI
measurements can reveal the geometry of the emitting region of these
high frequency photons.

In order to reproduce the same density profile in the inner part of
the disk, we have used a spiral pattern, characterized by the
emissivity function given below:
\begin{eqnarray}
I_\nu\propto r^{-\gamma}\sin^\beta(\phi+\alpha\log(\frac{r}{r_0}))
\end{eqnarray}
while the power-law index, $\gamma$, describes the overall radial
decrease in the emissivity, $\arctan(\alpha)$ is the pitch angle and
$r_0$ determines the outer radius where the pattern fades away. This
parametrization approximates a spiral pattern evolving in the
background of a Keplerian disk. The main emission mechanism is
assumed to be  synchrotron radiation from  relativistic electrons
with the same energy distribution as Eq. (1). The shape of the
spiral pattern highly depends on $\gamma, \beta, \alpha$ and $r_0$.
We parameterize our model in such a way that it facilitates a
straightforward comparison with Tagger et al. (2006). $\gamma=1,
\alpha=\beta=6$ and $r_0=20$ give the best approximation (see Fig.
\ref{spiral}). Figure \ref{spiral_spots} shows the changes in the flux
and polarimetric quantities, as measured by a distant observer for
different inclinations.
 As mentioned before (and discussed in detail by Falanga
et al. 2007 and Karas et al. 2008), such a spiral pattern can produce
the same typical behavior in flux modulation as that caused by an
azimuthal anomaly in the accretion disk.

\begin{figure}[h]
\centering{\includegraphics[width=0.35\textwidth]{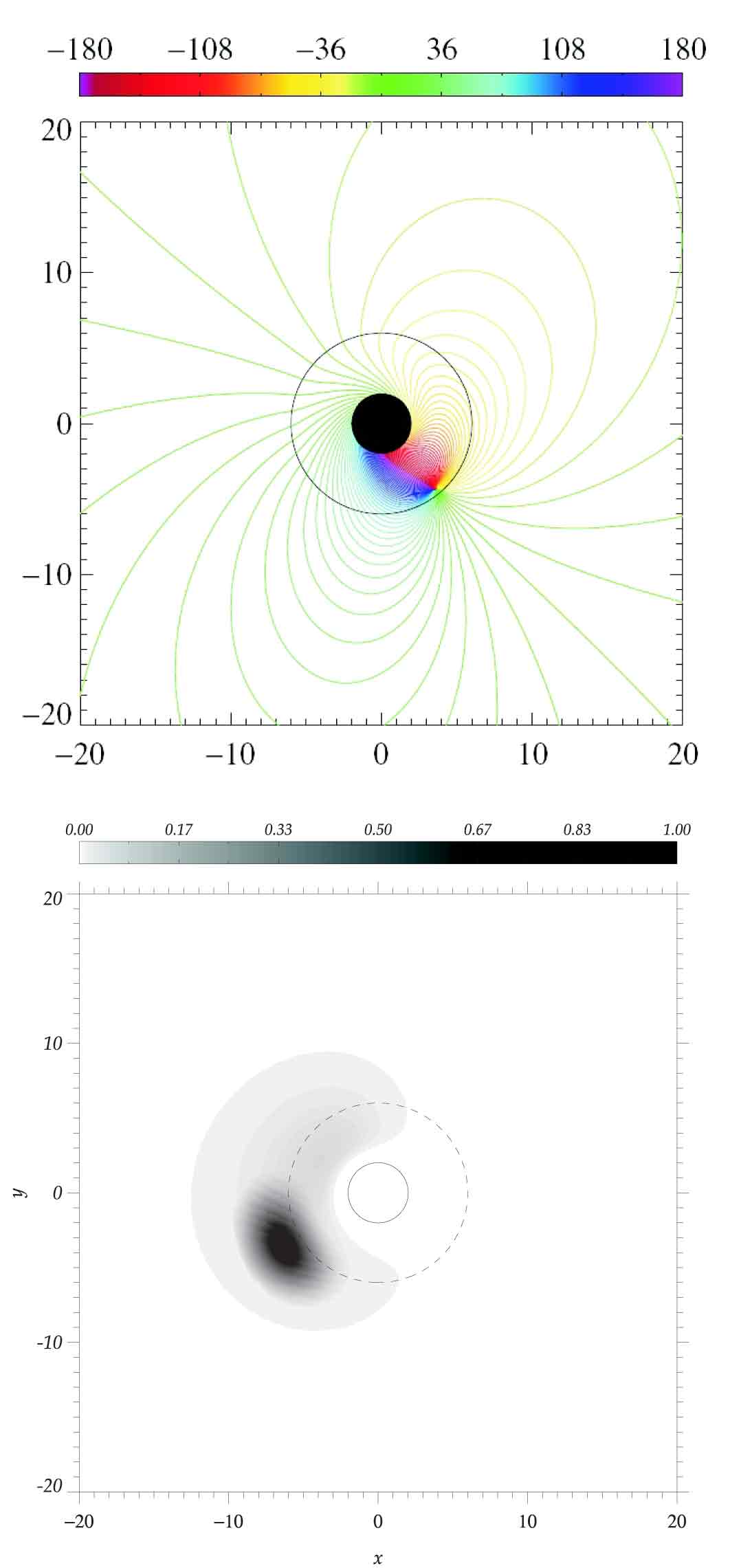}}
\caption{Top: Contour graphs showing the changes in the polarization
angle due to general relativistic effects. Bottom: Spatial
emissivity distribution of a compact hot spot. The observer's
inclination is $i=45^{\rm o}$ and the black hole is assumed to be
spin-less. The observer is located on the top of the pictures. The
innermost stable orbit is shown in both images (solid line in top
image and dashed line in the bottom). The graphs are represented in
the coordinates $x=r\cos\phi, y=r\sin\phi$ in the equatorial plane
where $r$ and $\phi$ are Boyer-Lindquist coordinates. Units are
$r_g$ in the x-y coordinates.} \label{xy}
\end{figure}

The  main idea in our simulations is to include the polarimetric
radiation transfer in curved space-time, which provides the
possibility to compare the behavior of different geometrical
configurations. It seems that even though both of these geometrical
set-ups (spiral shape or compact spot) show the same behavior in
flux, the observed polarized flux will behave significantly
different. In order to reduce the effects of our specific
assumptions about the emission process and magnify the signatures
which result only from different geometrical structures, we have
chosen a toy model for the initial polarized emission: $Q_\nu=I_\nu,
\ U_\nu = V_\nu=0$. Figure \ref{spiral_spots} shows that the spiral
pattern is unable to produce  strong changes in the polarization angle,
while a compact azimuthal source produces a highly variable
polarization angle.  Due to  abberation,
 photons that come from different parts of the
disk are polarized differently,  even for  Schwarzschild black
holes. For the Kerr case, the rotation of the polarization vector
will be added  because of  gravitational frame dragging. The
dependency of these changes on the position of the emitted photon
is depicted in Fig. \ref{xy}. The top panel of Fig.
\ref{xy} shows  how the polarization vector of the emitted photons
will be rotated due to the strong gravity of the central black hole.
As one can see, there is a clear knot visible in
 this contour graph. If the radiating source passes through this
knot, the observed polarization angle will swing dramatically (for a
detailed discussion see Dov\v{c}iak et al. 2008). As the bottom
panel of Fig. \ref{xy} shows, when a compact azimuthal source
orbiting around the black hole is close to its marginally stable orbit,
this nod will be passed. The amount of  change in the polarization
angle depends on the compactness of the source, its position
relative to the black hole, the inclination of the observer and the
spin of the black hole. On the other hand, if the flare-emitting
region is a deformed pattern extending in both radial and azimuthal
directions, the swing in the polarization angle will not be strong.
It is because of the fact that in each  point of the light curve the
polarization angle is the average of photons that come from
different parts of the disk and have a different polarization angle
(Fig. \ref{xy}). As a result, the swings in the angles cancel each
other out, and no significant swing will be observed.

The presence of changes in the degree of the NIR polarization and the swings
in the observed polarization angle simultaneous to the
 flux magnifications supports the idea that the geometrical
shape of the sources is dominated by  compact azimuthal asymmetries
rather than by radially extended spiral patterns (see also discussion
in  Sect. 6 about the differences in the centroid motions of these two
types of geometries). Here we must note that as Tagger et al. (2006)
have mentioned, a second initial configuration was used in their
simulations (simulation no. 2, Tagger et al. 2006). In that set-up,
a clump of matter starts spiraling towards the black hole and
produces the spiral Rossby wave. There, the ratio of the
surface brightness between the hot core moving radially inward and
the tail which is produced plays a critical role in the resulting
light curves. If the hot core is dramatically brighter than the
tail, there is no practical difference between this scenario and the
orbiting spot model. For a complete polarization study of Rossby
wave instabilities, one needs simulated profiles of MHD surface
densities as a function of time, which is beyond the scope of this
paper.

\begin{table*}[t]
\begin{center}
\begin{tabular}{lccccccccccr}\hline \hline
Model  &  $a$  & $\eta$ & $\psi$ & $r_0$ & $r_{spot}$ & $R$  & $\Pi_L$ & $i$ & $\Theta$ & $\chi^2_{red}\pm1$ \\
       &        & [deg]  & [deg]  &  $[r_g]$ & $[r_{mso}]$  &  & [\%] & [deg] &  [deg] &  &  \\
       \hline \hline \\
 I     & 0.5   &   $90$  &  $90$ &  4 & 1.1 & 5 & $50\pm10$ & $60\pm15$ & $15\pm20$ & $1.63\pm1$ \\
      \\
       &        &          &       &       \\
 II    & 0.5   &   $0$  &  $0$ &  4 & 1.1 & 5  & $50\pm10$ & $65\pm10$ & $110\pm15$ & $2.95\pm1$\\
             &        &          &       &      \\
       &        &          &       &       \\
III    & 0.5   &   $78.56\pm12$  &  $72\pm18$ &  4 & 1.1 & 5 & $60\pm10$ & $50\pm20$ & $30$ & $1.54\pm1$\\
             &        &          &       &      \\
\hline \hline
\end{tabular}
\end{center}
\caption{Final parameters resulting from  the best $\chi_{red}^2$ fit
to the NIR flare observed on 30 July 2005.}
\label{table2}
\end{table*}

\section{Magnetic field structure and geometrical orientation of
the system}
\begin{figure}[t]
\includegraphics[width=0.45\textwidth]{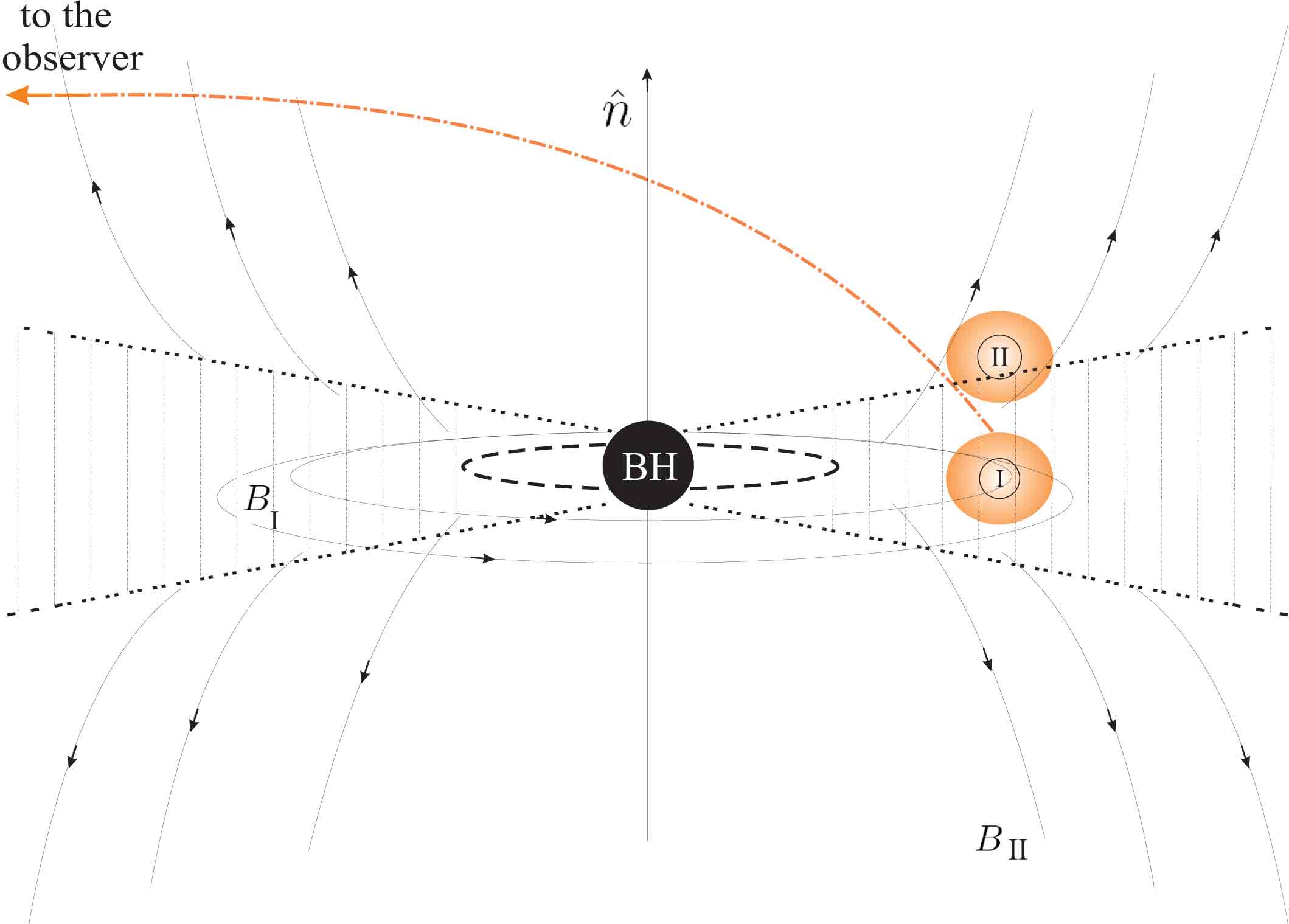}
\caption{A sketch of the model (not to scale) of two extreme cases
of global magnetic field configuration. Model I: the localized flare
happens inside the accretion disk, where the dominant component of
the magnetic field is toroidal ($B_{I}$). Model II: over-density of
the accelerated electrons happens at the tip of a possible short jet
(wind). Magnetic field lines  ($B_{II}$) are assumed to be elongated
toward the axial symmetry axis of the system, parallel to the normal
to the disk ($\hat{n}$.)} \label{magnetic_field}
\end{figure}

In our simulations the structure of the magnetic field lines
according to the black hole/accretion disk system can be controlled
by a set of parameters [$\eta$, $\psi$]. Of special interest are two
extreme configurations in which the envelope of the accelerated
electrons is  located inside the accretion disk with a global
toroidal magnetic field (model I: [$\eta=\frac{\pi}{2},
\psi=\frac{\pi}{2}$]) or inside a region  with  magnetic field lines
perpendicular to the  disk (model II: [$\eta=0, \psi=0$]). The
latter can be interpreted as the spot being located at the tip of a
jet with magnetic field lines aligned parallel to its symmetry axis
(see Fig. \ref{magnetic_field}).

\begin{figure}[t]
\includegraphics[width=0.45\textwidth]{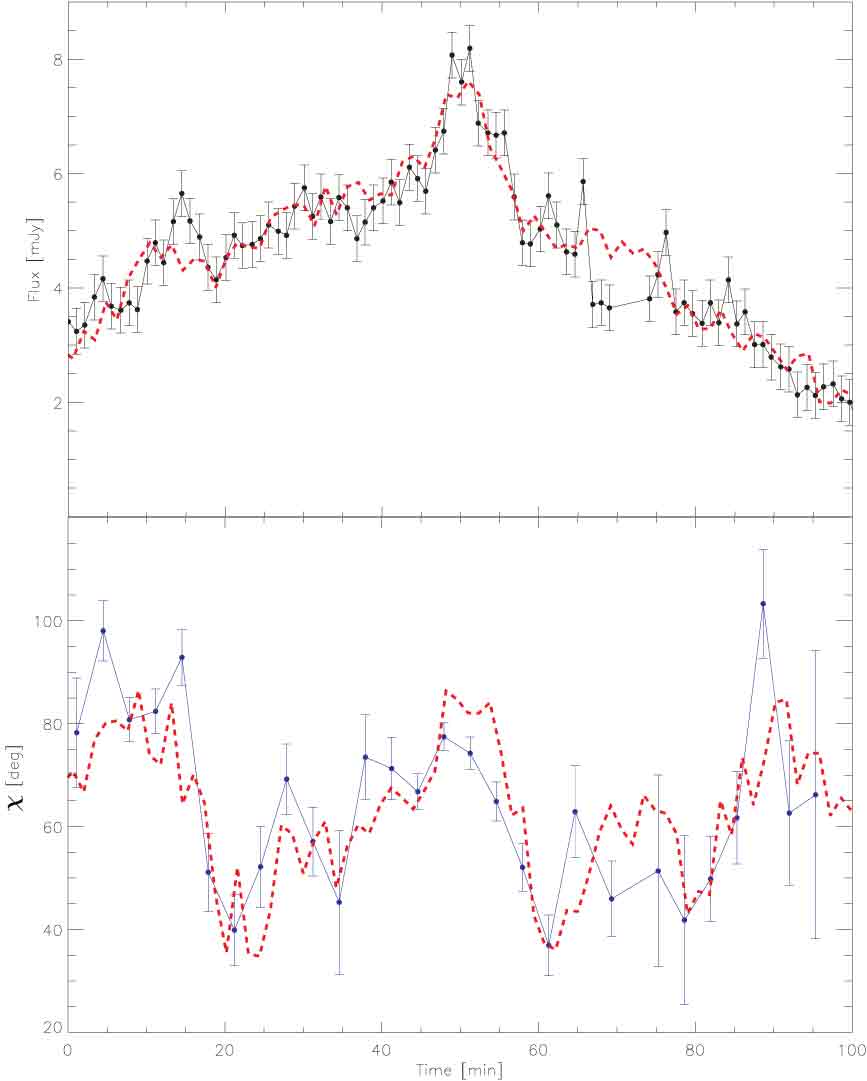}
\caption{Best fit  achieved for the flux and polarization angle of the
30 July 2005 flare. The points and error bars represent the
observation, red dashed lines show the model with Gaussian noise
being added. The parameters are presented in Table 2.} \label{fit}
\end{figure}

We assumed that the magnetic field inside the accretion disk is
dominated by the toroidal component. This structure is compatible
with the results of several MHD  simulations (Hawley \& Balbus 1991;
Hirose et al. 2004; De Villiers et al. 2003). However, in our case
it is not clear if the cloud of accelerated electrons is located
inside the accretion disk, somewhere above it in the corona, or even
inside the tip of a possible short jet/wind. In fact, there is a
possibility that a source structure in which an accretion disk is
associated with a short jet/collimated outflow  can explain the
multi-wavelength behavior of Sgr~A* (Markoff et al. 2001;  Yuan et
al. 2002; Eckart et al. 2005, 2006a, 2008a-c). Here we discuss how
different orientations of the magnetic field lines inside the
emitting region affect the resultant physical parameters that one
can extract via fitting the model to the observed data.

\begin{figure}[h]
\includegraphics[width=0.5\textwidth]{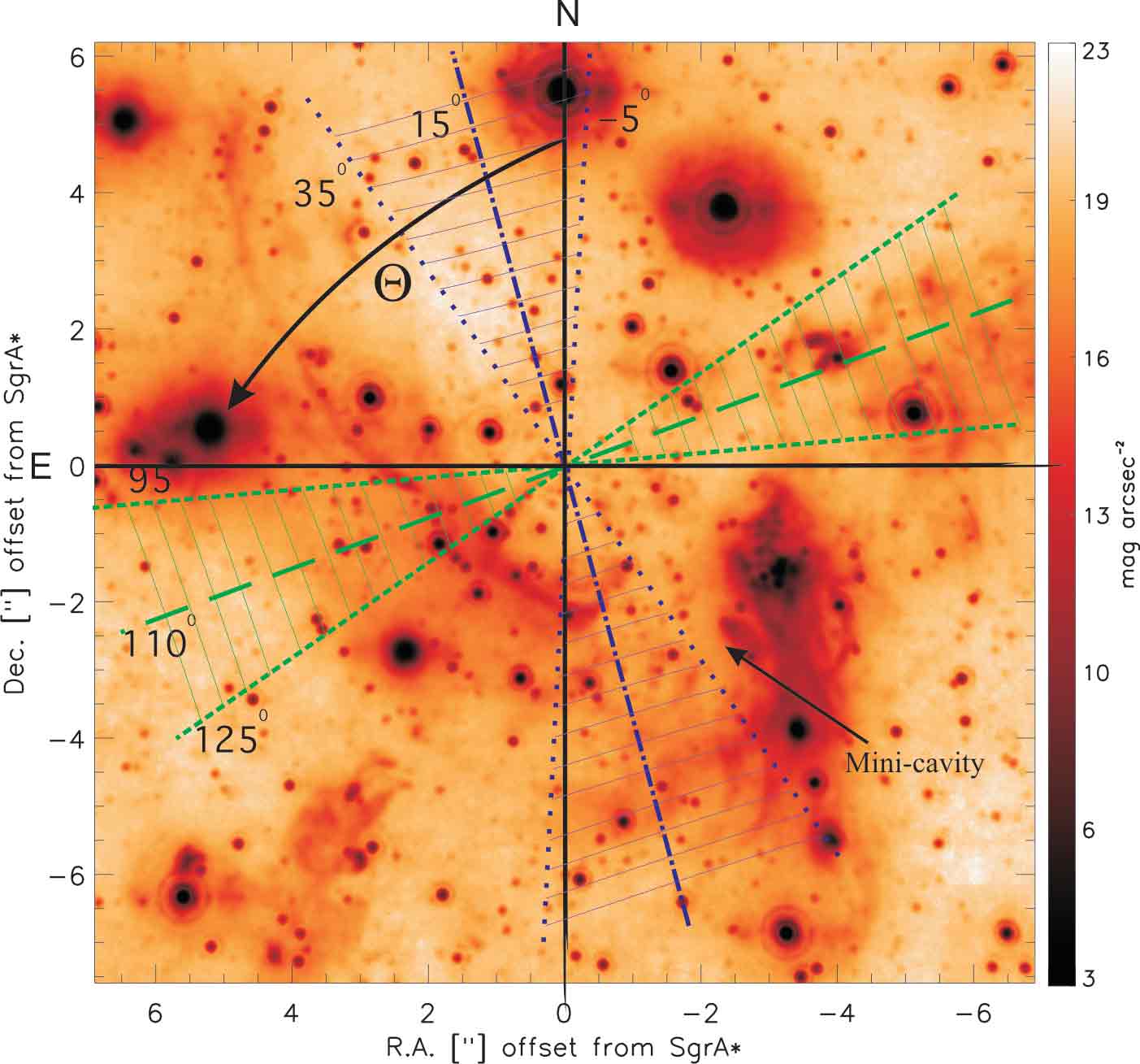}
\caption{The resulting orientations for the direction of  possible
collimated outflow from  Sgr~A* ($\Theta$) for model I (blue dashed-dotted line)
 and model II (green long dashed line). The shaded
regions indicate the range of  possible values of $\Theta$ for
$\chi^2_{red}~\pm~1$ interval. The background image shows the
Galactic Center environment in NIR $L'$ band. Sgr~A* is located at
the center of the image. The position of the mini-cavity (indicated
by an arrow) coincides well with the predictions of model I. }
\label{jet_disk}
\end{figure}

The KY code allows us to simulate light curves for a wide range of
free parameters, so  one  can fit the model parameters to the actual
observed data (Meyer et al. 2006a,b, 2007; Eckart et al. 2006a,b,
2008). For this purpose, simulations have been carried out, as they
cover a wide enough range of possible inclinations ($0.1^{\rm o}\leq
i\leq85^{\rm o}$), the initial degree of linear polarization
($0\%\leq \Pi_L\leq70\%$), possible combinations of $\eta$ and
$\psi$, and different orientations of the whole system on the sky
($0^{\rm o}\leq\Theta\leq 90^{\rm o}$). Here $\Theta$ defines the
direction of the normal to the accretion disk projected on the sky.
The spot has an initial radius of $1R_s$ and orbits very closely to
the marginally stable orbit of a Kerr black hole ($a=0.5$). The
dimensionless shearing time scale is fixed to be 2.0
($\tau_{sh}=2.0$), and the ratio of the surface brightness of the
spot to the torus is set to five. We have chosen these values according
to  existing results of  several fits to the NIR flares (Meyer et
al. 2006a,b; Eckart et al. 2006b, 2008).

As a first step, we focus on  model I ($[\eta=\frac{\pi}{2},
\psi=\frac{\pi}{2}]$). In order to find the fit with best
$\chi^2_{red}$ we have carried out a grid search in a
($i-\Theta-\Pi_L$) parameter space. The steps for $i$, $\Theta$ and
$\Pi_L$ have been chosen to be $5^{\rm o}$, $5^{\rm o}$ and $5\%$,
respectively. The least $\chi^2_{red}$ value was achieved for a high
inclination angle ($i=60^{\rm o}$), $\Theta=15^{\rm o}$ and highly
polarized source ($\Pi_L=50\%$) (see Fig. \ref{fit} and Table 2).
The grid search has been performed
on $\Pi_L$, $i$ and $\Theta$ for the models I and II. The free
parameters in model III are $\Pi_L$, $i$, $\eta$ and $\psi$.
Figure \ref{jet_disk} shows the resultant possible range for the
orientation of the correlated outflow/wind in this scenario.
Interestingly, the observed position of the mini-cavity lies well
within this interval. Mu\v{z}i\'{c} et al. (2007) have shown how a
collimated outflow from the position of Sgr~A* can describe the
observed motion of filamentary structures and the mini-cavity,
according to their observations in the NIR $L'$-band. This
observational evidence supports the idea that the NIR flares
actually originate in accelerated electrons within an accretion
disk.

For  model II ([$\eta=0, \psi=0$]), the envelope of relativistic
electrons has been located somewhere in the bottom of an outflow.
The least $\chi^2_{red}$ achieved for this model and related
parameters are presented in Table 2. Possible directions of this jet
structure are depicted in Fig. \ref{jet_disk}. As one can see,
$\chi^2_{red}$ values and observational facts support the idea
that the NIR photons originate in the inner parts of  an accretion
disk rather than in an outflow (Thorne 1994, Blandford 2001).

One can also invert the question and try to constrain the magnetic
field structure of the emitting region by fixing the direction of
the outflow in the plane of the sky (model III). For this purpose, a
$\chi^2_{red}$ search has been  carried out, assuming that the
normal to the  disk is fixed at $\Theta=30^{\rm o}$. This is the
best value that can fit the observed position of the mini-cavity
(Mu\v{z}i\'{c} et al. 2007). By a good approximation, the best fit
is achieved for a toroidal magnetic field structure (see Table 2).
This shows that if any future observations  reveal  traces of an
outflow from Sgr~A*, NIR polarimetry could be used as a tool to
constrain its magnetic field structure.

\begin{figure*}[!htb]
\centering{\includegraphics[width=1.\textwidth]{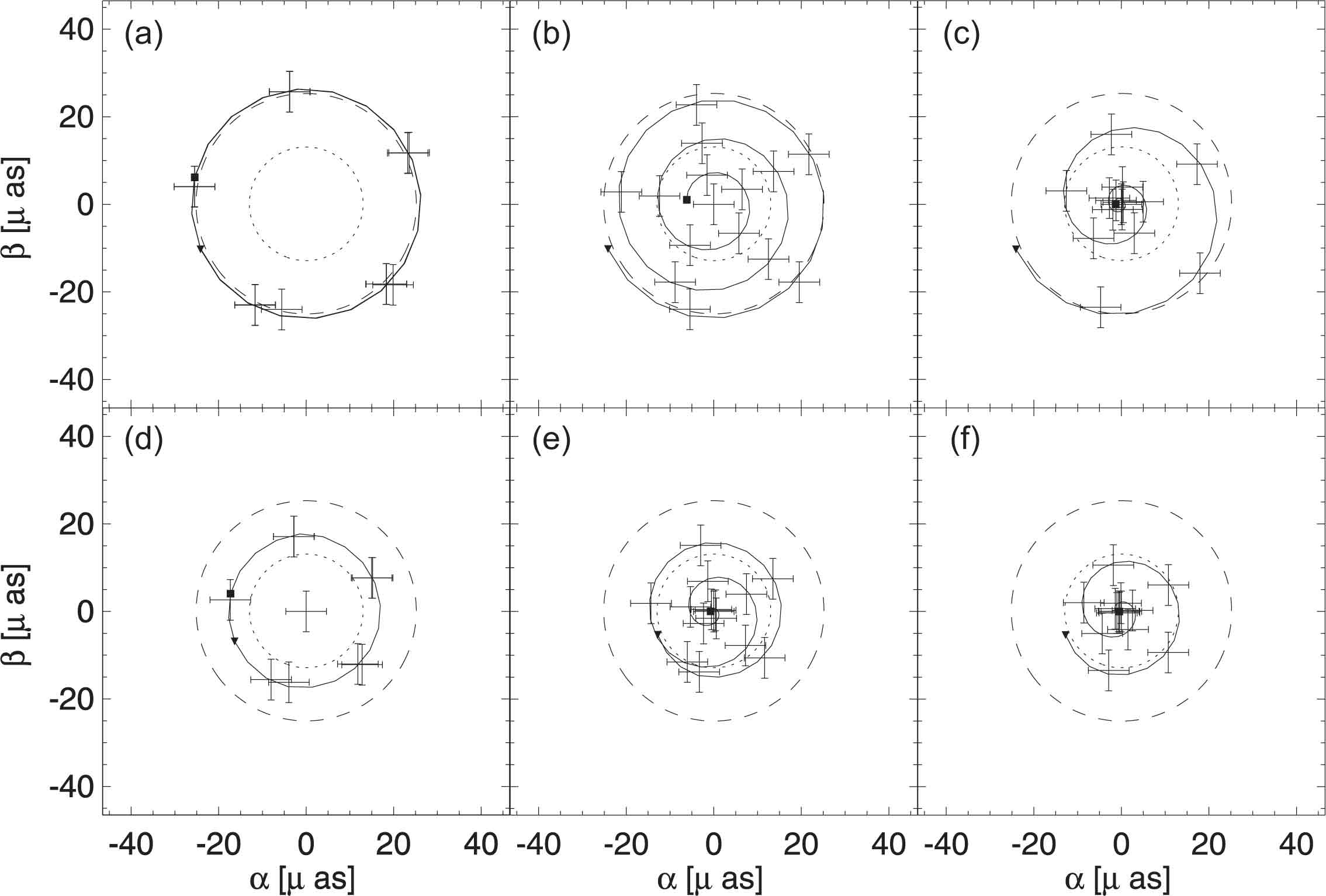}}
\caption{The  centroid motion of NIR images viewed from
$0^{\rm o}$ above the orbital plane for a Kerr black hole with a spin of
0.5. The first row shows the paths for the background-subtracted
images for different values of the gravitational shearing time scale.
Each column shows $\tau_{sh}$ values $\infty$, 2.0 and 1.0 from
right to left. The second row shows the same paths for the images
without the background subtraction.  The error bars show the
simulated positions as they are expected to be observed by GRAVITY 
(FWHM $\sim 8.5 \mu$as) with a time resolution of five
minutes. The whole paths belong to three orbital times. A triangle
indicates the beginning and a square the end of the track. Both
coordinates are labeled in $\mu$as units. The dotted and dashed
lines indicate the position of the event horizon and the marginally
stable orbit respectively.} \label{center_ratio}
\end{figure*}

\begin{figure*}[!htb]
\centering{\includegraphics[width=1.\textwidth]{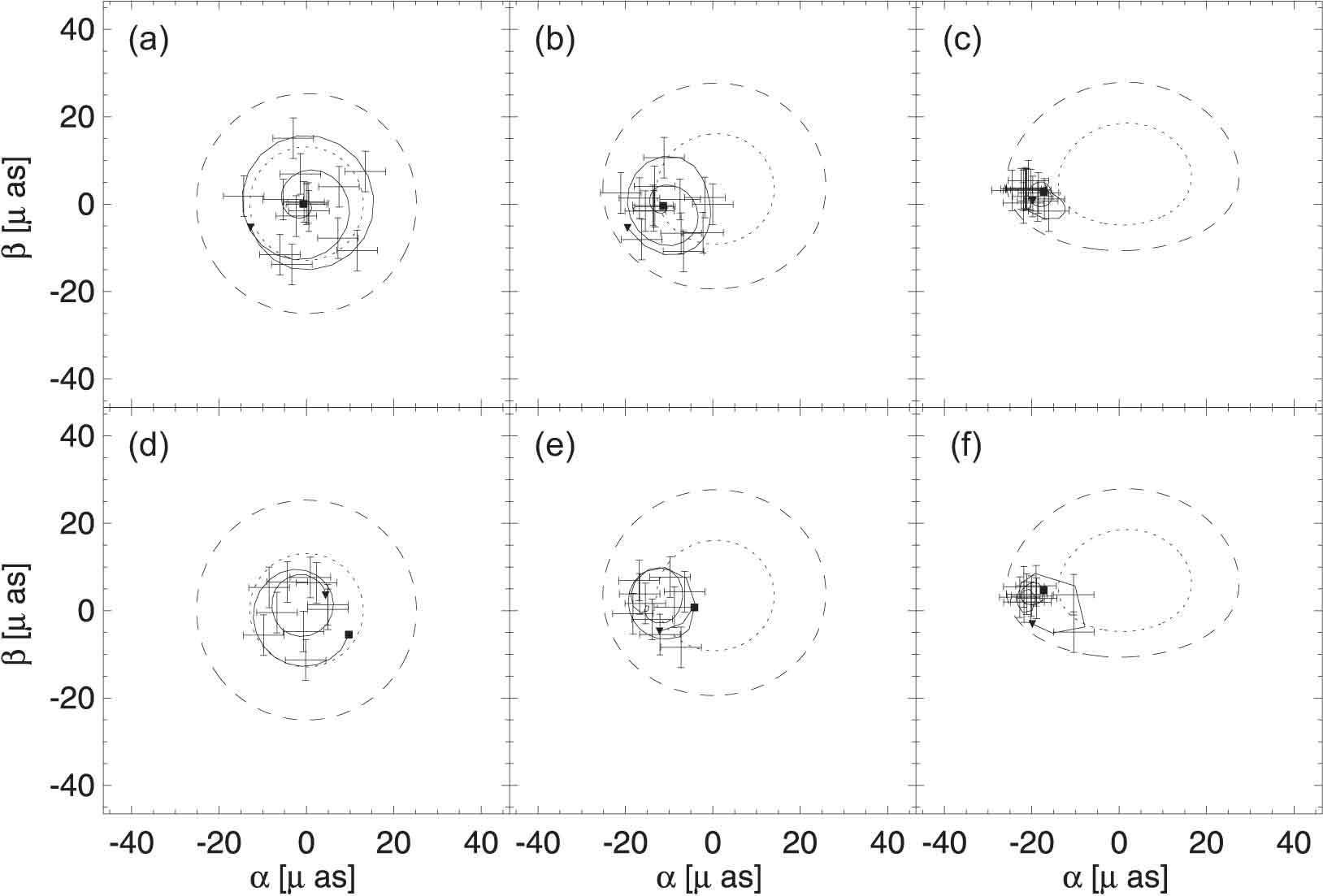}}
\caption{Similar to Fig. \ref{center_ratio} where different
columns represent different values of inclination (from left to
right: $i=0.1^{\rm o}$, $i=30^{\rm o}$, $i=60^{\rm o}$). (a-c) shows
the centroid track of an orbiting spot with  $\tau_{sh}=2.0$. (d-f)
shows the centroid for a multi-component model (see Fig.
\ref{multi}). } \label{center_incl}
\end{figure*}

\section{Next generation of NIR interferometry}
According to its proximity, Sgr~A* is known as  the best candidate
for studying the details of physical processes on the event horizon
scales (Falcke et al. 2000; Bromley et al. 2001). VLT and Keck
telescopes have already  achieved angular resolutions as high as
40mas in their imaging mode (Genzel et al. 2003; Eckart et al. 2004,
2006a,b; Ghez et al. 2005). In interferometric mode these telescopes
can observe with resolutions of the order of only a few mas. The
next generation of VLT Interferometer instrument, namely GRAVITY,
would be able to achieve resolutions of the order of $10\mu$as in
its phase-reference tracking mode.  GRAVITY will try to measure 
the position of the lumicenter 
of the flares with respect to a reference source. For such a measurement 
the 1$\sigma$ uncertainty is $\pm10 \mu$as 
with an integration time of five minutes on an assumed $K=15$ pointsource. 
The real necessary integration time for that precision is of course directly dependent on the magnitude of the flare.
This resolution is high enough to track the
centroid motion of NIR images of Sgr~A* with a resolution of the
order of one Schwarzschild radius (Eisenhauer et al. 2005, 2008;
Gillessen et al. 2006).  These unique
features make the next generation of VLTI a perfect instrument for
high accuracy astrometry of the matter around Sgr~A*. Recently,
Doeleman et al. (2008) proposed a VLBI configuration with a high
enough time resolution to possibly track the centroid motion
of sub-mm images of Sgr~A* with even higher angular resolution. In
this section we present the results of our study on centroid motion
of the NIR image of Sgr~A*.

Some predictions of the centroid motions related to the hot spot
model have already been studied by several authors (Broderick \&
Loeb 2006a,b; Paumard et al. 2006; Zamaninasab et al. 2008b; Hamaus
et al. 2009). In previous works it has been mainly assumed that the
spots preserve their shape in time. This assumption has been 
based mostly on the presence of strong magnetic field lines or "unknown
mechanisms" (Hamaus et al. 2009). We will show  here that  it
is actually very crucial that the effects of gravitational shearing inside
the accretion disk are  taken into account, since the predicted
results can  change dramatically. The other very important fact to
be considered in the simulations is due to the NIR photons which
come from the body of the accretion disk and confuse with the spot's
emission. Since GRAVITY can only achieve its 10$\mu$as resolution in
the phase-reference tracking mode and not in the imaging mode, even
a constant confusion from the (thermal) electrons inside the
accretion disk can not be reduced.

Figure \ref{center_ratio} represents  how these two 
parameters mentioned can affect the expected centroid tracks and may present a
complication in detecting the plasma structure close to the event
horizon of Sgr~A*. The simulated paths in Fig. \ref{center_ratio}
belong to a configuration where the observer is aligned
approximately face-on ($i\simeq0^{\rm o}$). This is the inclination
in which the clearest wobbling of the centroid is expected to be
seen. Fig. \ref{center_ratio}a shows that the position wanders of
the center of the images are large enough to be detected easily if
GRAVITY could achieve its five minutes time-resolution (simulated
points). Panels b \& c in Fig. \ref{center_ratio} show how
gravitational shearing changes the centroid track.  One
can see in particular that in the case of pure Keplerian shearing the change in the
center is detectable for approximately only six points. The second
row (Fig. \ref{center_ratio}, panels {\it d-f}) shows the result of
the simulated tracks including the effect of confusion due to the
NIR photons coming from the inner parts of the accretion disk.

\begin{figure*}[t]
\centering{\includegraphics[width=\textwidth]{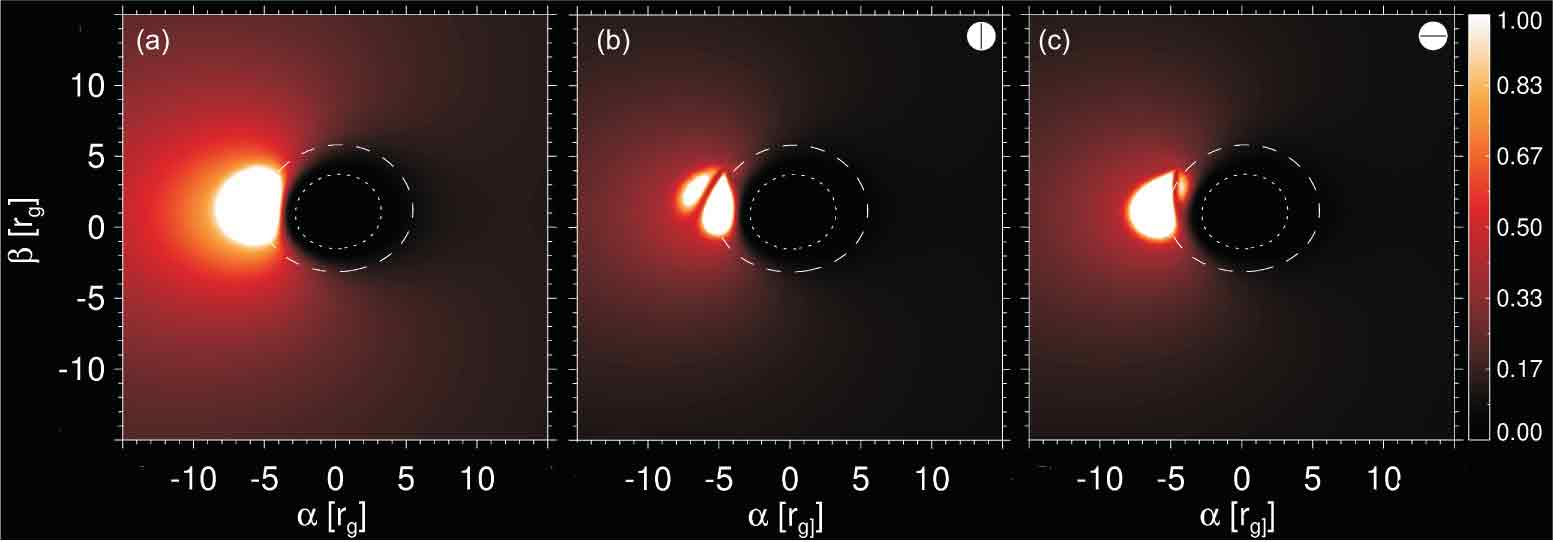}}
\caption{Apparent images of a flare event for a distant
observer looking along a line of sight inclined by $45^{\rm o}$ in
total flux (a), ordinary (b) and extra-ordinary polarized channels
(c).} \label{polari}
\end{figure*}

\begin{figure*}[t]
\centering{\includegraphics[width=\textwidth]{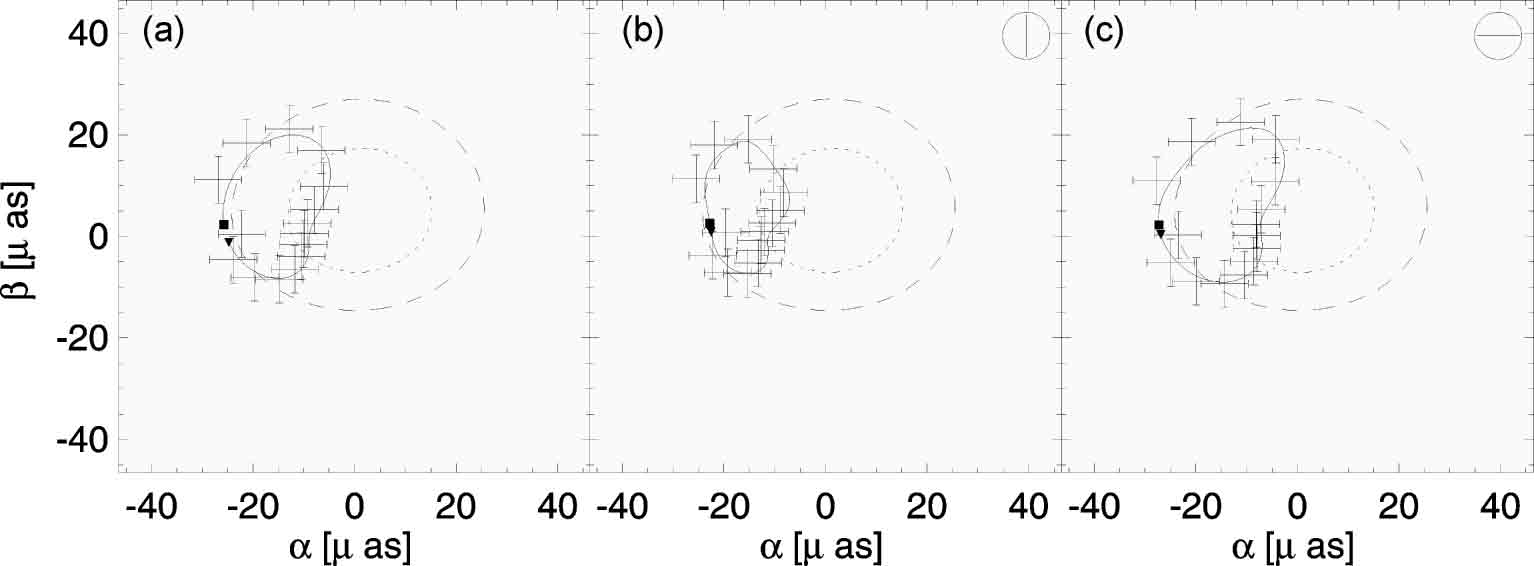}}
\caption{Centroid paths for the same inclinations as Fig.
\ref{polari} in the total flux (a), ordinary (b)
 and extra-ordinary (c) polarized channels. See also caption of Fig.
\ref{center_ratio}.} \label{polari_centr}
\end{figure*}

Figure \ref{center_incl} shows the centroid paths of spots
orbiting in a mild shearing environment ($\tau_{sh}=2.0$) for three
different inclinations ($i=0.1^{\rm o}, 30^{\rm o}, 60^{\rm o}$) in
presence of confusion from the torus. Panel {\it c} shows that the
detection of any position wander becomes even more difficult in the
case of a shearing spot observed at high inclinations.
The bottom row in Fig. \ref{center_incl} shows the centroid motion
resultant from simulations including several emitting components
(Fig. \ref{multi}).
The photons originated from the spot are highly polarized, while the
radiation from the torus is weakly polarized. This fact may lead us
to a solution for compensating for this kind of confusion.
Figure \ref{polari} shows how different the apparent images of the
flares are when they are  observed in total flux or two
orthogonal polarized channels. One sees that the different centroid
paths can be revealed if GRAVITY could achieve its $10\mu$as
resolution in the polarimetry mode Fig.\ref{polari_centr}.

To summarize the results of this section, one can say that a
non-detection of a position wander in NIR images of Sgr~A* will not
rule out a spotted disk scenario. On the other hand, a clear detection
of such a position wander by GRAVITY would lead us to strongly favor the  spot
model and would open a new era in studying the physics of high
gravitational regimes.

\section{Summary and conclusion}
We used a sample of NIR flares of Sgr~A* observed in
polarimetry mode to study the nature of the observed variability.
Using the z-transformed discrete correlation function algorithm, we
found a significant correlation between changes in the measured
polarimetric data and  total flux densities. This provides  evidence
that the variations probably  originate from inner parts of an
accretion disk while a strong gravity's effects are manifested inside
them.

In order to obtain this information polarization data is
indispensable since the corresponding signals are very difficult or
even impossible to extract from simple total intensity light curves
only (Do et al. 2008). The presence of significant signals from
orbiting matter then calls for detailed modeling of the NIR
polarized light curves in order to analyze the distribution of
the emitting material and the magnetic field structure within the
accretion disk (e.g. Eckart et al. 2006a, 2008a, Meyer et al.
2006ab,2007).

In addition, the pattern recognition algorithm we employed in
this paper is an efficient tool to search for flare events that
carry the signature of strong gravity in light curves which are
significantly longer than the orbiting time scale over which such an
event can typically occur. Other methods that make simultaneous use
of the entire light curve (like e.g. the Lomb-Scargle algorithm)
tend to dilute these signals and lower the possibility for
significant detections dramatically (Do et al. 2008, Meyer et al.
2009).

In order to constrain the physical properties of the emitting region we employed
a relativistic disk model with azimuthal over-densities of
relativistic electrons. A combination of a synchrotron mechanism and
relativistic amplifications allows us to fit the real observed data
and make predictions about astrometric parameters of the accretion
disk around Sgr~A*. The modeled light curves show the same correlation
between the flux and polarimetric data as
the one deduced from observations.

Simulations have been carried out in a way that they cover a wide range of
parameters, including the effects of gravitational shearing inside
the accretion disk, the heating and cooling time scales, the inclination and
the spin of the black hole. It is discussed how the observed swings in the
polarization angle support the idea of a compact source for the
emission, instead of  radially extended spiral shapes. Furthermore,
we present a model in which the observed NIR polarization angle can
lead to confining the expected region for the expected outflow/wind
from Sgr~A*. The model also predicts that when observations will be able
to resolve the position of such an outflow, the magnetic field
structure inside the accretion disk could be confined.

Finally, the centroid paths of the NIR images are discussed. In
comparison with the results by Broderick \& Loeb (2006a,b) and
Hamaus et al. (2009), we have shown that the geometrical structure
of the emitting region (elongation of the spot according to the
Keplerian shearing, multi-component structures, spiral arms,
confusion caused by the radiation from the hot torus) can affect the
expected centroid tracks. While all the mentioned geometries are
able to fit the observed fluxes, we show how the future NIR
interferometer GRAVITY on the VLT can break these degeneracies. The
results of simulations propose that focusing GRAVITY observations on
the polarimetry mode could reveal a clear centriod track of the
spot(s). We conclude that even though a non-detection of centroid
shifts can not rule out the multi-components model or spiral arms
scenarios, a clear position wander in the center of NIR images
during the flares will  support the idea of  bright long-lived spots
orbiting occasionally around the central black hole. This possible
detection opens a new window for testing the physics very close to
the edge of infinity.

\begin{acknowledgements}
The authors would like to thank anonymous referee for helpful comments.
Part of this work was supported by the German \emph{Deut\-sche
For\-schungs\-ge\-mein\-schaft, DFG\/} via grant SFB 494. M.
Zamaninasab, and D. Kunneriath are members of the International Max
Planck Research School (IMPRS) for Astronomy and Astrophysics at the
MPIfR and the Universities of Bonn and Cologne. N. Sabha
acknowledges support from the Bonn-Cologne Graduate School of Physics
and Astronomy (BCGS). V.K. and M.D. acknowledge the Czech Science
Foundation (ref. 205/07/0052).
\end{acknowledgements}


\begin{thebibliography}{}

\bibitem[Abramowich 1991]{abb91} Abramowicz, M. A., Bao, G., Lanza, A., Zhang,
X.-H., 1991, A\&A 245, 454

\bibitem[Abramowich 1992]{abb92}
    Abramowicz, M. A., Lanza, A., Spiegel, E. A., Szuszkiewicz, E.,
    1992, Nature 356, 41-43

\bibitem[]{} Adams, F. C., Watkins, R., 1995, ApJ v.451,
314

\bibitem[Alexander 1997]{Alexander} Alexander, T., 1997, ASSL 218, 163

\bibitem[Armitage 2003]{Armitage} Armitage, P. J., Reynolds, C. S., 2003, MNRAS 341, 1041

\bibitem[Baganoff et al.(2001)]{Baganoff01} Baganoff, F. K., Bautz, M. W., Brandt, W. N., et al. 2001, Nature, 413, 45

\bibitem[Bardeen]{bardeen} Bardeen, J. M., Press, W. H., Teukolsky, S. A., 1972, ApJ 178, 347

\bibitem[2003]{blanger03} Belanger, G., Terrier, R., de Jager, O. C., Goldwurm, A., Melia, F., 2006, JPhCS 54, 420

\bibitem[2003]{} Blandford, R. D., 2001, Progress of Theoretical Physics Supplement 143, 182

\bibitem[]{} Broderick, A. E., Loeb, A., 2006a, ApJ 636L, 109

\bibitem[]{} Broderick, A. E., Loeb, A., 2006b, MNRAS 367, 905

\bibitem[]{} Bromley, B. C., Melia, F., Liu, S., 2001, ApJ 555L, 83

\bibitem[]{} Bursa, M., 2007, IAUS 238, 333

\bibitem[]{} \v{C}ade\v{z}, A., Calvani, M., Gomboc, A., Kosti\'{c}, U., 2006, AIPC 861, 566

\bibitem[]{} Chan, C., Liu, S., Fryer, C. L., et al. 2006,
ApJ 701, 521-534

\bibitem[2009]{} Chan, C., 2009, ApJ 704, 68-79

\bibitem[1973]{cunn73}  Cunningham, J. M., Bardeen, C. T., 1973, ApJ 183, 237

\bibitem[]{} Devillard, N., 1999, ASPC 172, 333

\bibitem[]{} De Villiers, J, Hawley, J. F., 2003, ApJ 589, 458

\bibitem[]{} Diolaiti, E., Bendinelli, O., Bonaccini, D., et al. 2000, ASPC 216, 623

\bibitem[]{} Do, T., Ghez, A. M., Morris, M. R., et al. 2009, ApJ 691, 1021

\bibitem[]{} Doeleman, S. S., Fish, V. L., Broderick, A. E., Loeb, A., Rogers, A.
E. E., 2008, arXiv0809.3424

\bibitem[dovciak04]{04} Dov\v{c}iak, M., Karas, V., Yaqoob, T., 2004, ApJS 153, 205

\bibitem[dovciak07]{07}  Dov\v{c}iak, M., Karas, V., Matt, G., Goosmann, R.
W., 2008a, MNRAS 384, 361

\bibitem[]{} Dov\v{c}iak, M., Muleri, F., Goosmann, R. W., Karas, V., Matt,
G., 2008b, MNRAS 391, 32

\bibitem[Eckart \& Genzel 1996]{eckart96} Eckart, A., \& Genzel, R. 1996, Nature 383, 415

\bibitem[1997]{eckart97} Eckart, A., \& Genzel, R. 1997, MNRAS
284, 576

\bibitem[2002]{eckart02} Eckart, A., Genzel, R., Ott, T. and Schoedel, R. 2002, MNRAS, 331, 917

\bibitem[2004]{eckart04} Eckart, A., Baganoff, F. K., Morris, M., et al. 2004, A\&A 427, 1

\bibitem[2006a]{eckart06a} Eckart, A., Baganoff, F. K., Sch\"odel, R., et al.  2006a, A\&A 450, 535

\bibitem[2006b]{eckart06b} Eckart, A., Sch\"odel, R., Meyer, L., et al.  2006b, A\&A 455, 1

\bibitem[2006c]{eckart06c} Eckart, A., Sch\"odel, R., Meyer, L., et al., 2006c, ESO Messenger 125, 2

\bibitem[2008a]{eckart08a} Eckart, A., Baganoff, F. K., Zamaninasab, M., et al. 2008a, A\&A 479, 625

\bibitem[2008b]{eckart08b} Eckart, A., Sch\"odel, R., Baganoff, F. K., et al. 2008b,
JPhCS 131a, 2002

\bibitem[2008c]{eckart08c} Eckart, A., Sch\"{o}del, R., Garc\'{\i}a-Mar\'{\i}n,
M., et al. 2008c, A\&A 492, 337

\bibitem[]{}
    Eckart, A., Baganoff, F. K., Morris, M. R
M., et al. 2009, A\&A 500, 935-946


\bibitem[2003]{eisenhauer03} Eisenhauer, F., Sch\"odel, R., Genzel, R., et al. 2003, ApJ 597, L121

\bibitem[2005a]{eisen05a} Eisenhauer, F., Genzel, R., Alexander, T.,
2005a, ApJ 628, 246

\bibitem[]{} Eisenhauer, F., Perrin, G., Rabien, S., et al. 2005b, AN 326, 561

\bibitem[]{} Eisenhauer, F., Perrin, G., Brandner, W., et al. 2008, SPIE 7013E, 69

\bibitem[]{} Falanga, M., Melia, F., Tagger, M., Goldwurm, A., B\'{e}langer,
G., 2007, ApJ 662L, 15

\bibitem[]{} Falanga, M., Melia, F., Prescher, M., B\'{e}langer, G., Goldwurm,
A., 2008, ApJ 679L, 93

\bibitem[]{} Falcke, H., Melia, F., Agol, E., 2000, ApJ 528L, 13

\bibitem[]{} Fukumura, K., Kazanas, D., Stephenson, G., 2009, arXiv0901.2858

\bibitem[Genzel et al.(2003)]{Genzel03} Genzel, R., Sch\"odel, R., Ott, T., et al. 2003, Nature, 425, 934


\bibitem[2000]{ghez00} Ghez, A., Morris, M., Becklin, E.E., Tanner, A. \& Kremenek, T.,
2000, Nature 407, 349

\bibitem[2005]{ghez05a} Ghez, A. M., Salim, S., Hornstein, S. D., et al. 2005a, ApJ 620, 744

\bibitem[]{} Ghez, A. M., Hornstein, S. D., Lu, J. R., et al. 2005b, ApJ 635, 1087

\bibitem[2008]{Gierlinski08}  Gierli\'{n}ski, M., Middleton, M., Ward, M., Done,
C., 2008, Nature 455, 369

\bibitem[]{} Gillessen, S., Perrin, G., Brandner, W.,  et al. 2006, JPhCS 54, 411

\bibitem[2009]{gillessen09} Gillessen, S., Eisenhauer, F., Trippe,
S., et al. 2009, ApJ 692, 1075

\bibitem[]{} Goggin, L. M., 2008,  PhD Thesis, California
Institute of Technology


\bibitem[]{} Hamaus, N., Paumard, T., M\"{u}ller, T., et al. 2009, ApJ 692, 902

\bibitem[Hawley 1991]{Hawley}  Hawley, J. F., Balbus, S. A., 1991, ApJ 376, 223

\bibitem[2004]{hirose} Hirose, S., Krolik, J. H., De Villiers, J.-P., Hawley, J. F., 2004, ApJ, 606, 1083

\bibitem[holly]{holly} Hollywood, J. M., Melia, F., Close, L. M., McCarthy, D. W. Jr., Dekeyser, T.
A.,  1995, ApJ 448L, 21

\bibitem[Hornstein 2007]{Hornstein} Hornstein, S. D., Matthews, K., Ghez, A. M., et al. 2007 ApJ  667, 900

\bibitem[Karas]{karasa}  Karas, V. \& Bao, G., 1992, A\&A 257, 531

\bibitem[Karas]{karasb}  Karas, V., Vokrouhlicky, D., Polnarev, A.
G., 1992, MNRAS 259, 569

\bibitem[]{} Karas V., Dov\v{c}iak M., Eckart A., Meyer L.
2007, in Proceedings of the Workshop on the Black Holes and Neutron
Stars, eds. S. Hled\'{\i}k and Z. Stuchl\'{\i}k (Silesian
University, Opava), pp. 99-108 (arXiv:0709.3836)


\bibitem[]{} Krolik, J. H., Hawley, J. F., 2002, ApJ 573, 754

\bibitem[]{} Kosti\'{c}, U., \v{C}ade\v{z}, A., Calvani, M., Gomboc, A., 2009,  arXiv:0901.3447

\bibitem[Liu 2002]{Liu02} Liu, Siming, Melia, F., 2002, ApJ 566L, 77

\bibitem[Liu 2006]{Liu06} Liu, S., Melia, F., Petrosian, V., 2006, ApJ 636, 798

\bibitem[]{} Lyubarskii, Y. E., 1997, MNRAS 292, 679

\bibitem[markoff]{markoff} Markoff, S., Falcke, H., Yuan, F., Biermann, P.
L., 2001, A\&A 379L, 13

\bibitem[marrone08]{marrone08} Marrone, D. P., Baganoff, F. K., Morris, M., et al. 2008, ApJ 682, 373

\bibitem[]{} Mashoon, B., 1973, Phys. Rev. D 7, 2807

\bibitem[]{} Melia, F., Bromley, B. C., Liu, S., Walker, C. K., 2001, ApJ 554, 37

\bibitem[]{} Melia, F., The Galactic Supermassive Black Hole, Princeton University Press, 2007

\bibitem[2006a]{meyer06a} Meyer, L., Eckart, A., Sch\"odel, R., et al. 2006a, A\&A 460, 15

\bibitem[2006b]{meyer06b} Meyer, L., Sch\"odel, R., Eckart, A., et al. 2006b, A\&A 458, L25

\bibitem[2007]{meyer07} Meyer, L., Sch\"odel, R., Eckart, A., et al. 2007, A\&A 473, 707

\bibitem[]{}  Mu\v{z}i\'{c}, K., Eckart, A., Sch\"{o}del, R., Meyer, L., Zensus,
A., 2007, A\&A 469, 993

\bibitem[nayakshin]{nayakshin} Nayakshin, S., Cuadra, J., Sunyaev,
R., 2004, A\&A 413, 173

\bibitem[]{} Nishiyama, S., Tamura M., Hatano, H., 2009,
ApJL 702, L56-L60

\bibitem[1998]{Nowak} Nowak, M. A., Lehr, D. E., 1998 npad.conf, 233

\bibitem[]{} Papa, M. A., Schutz, B. F., Sintes, A. M.,
Proceedings of the conference "Gravitational waves: a challenge to
theoretical astrophysics", (June 5-9 2000, Trieste), ICTP Lecture
Notes Series

\bibitem[]{} Paumard, T., Eisenhauer, F., Genzel, R., et al. 2006, via conf., 257

\bibitem[]{} Pech\'{a}\v{c}ek, T., Karas, V., Czerny, B., 2008, A\&A 487, 815

\bibitem[Porquet et al.(2003)]{Porquet03} Porquet, D., Predehl, P., Aschenbach, et al.  2003, A\&A 407, L17

\bibitem[Porquet et al.(2008)]{Porquet08} Porquet, D., Grosso, N., Predehl,
P., et al. 2009, A\&A 488, 49

\bibitem[Press]{Press} Press, W. H., Rybicki, G. B.,  1989,
ApJ 338, 277

\bibitem[]{} Reid, M. J., Broderick, A. E., Loeb, A., Honma, M., Brunthaler,
A., 2008, ApJ 682, 1041

\bibitem[]{} Schnittman, J. D., Krolik, J. H., Hawley, J. F., 2006, ApJ 651, 1031

\bibitem[2002]{shoedel02} Sch\"odel, R., Ott, T., Genzel, R., et al. 2002, Nature 419, 694

\bibitem[2002]{shoedel02} Sch\"{o}del, R., Eckart, A., Alexander, T.,
et al. 2007, A\&A 469, 125

\bibitem[]{} Schulz, M., Mudelsee, M.,  2002,  Computers \&
Geosciences 28, 3

\bibitem[2002]{tagger} Tagger, M., Melia, F., 2006, ApJ 636L, 33

\bibitem[2002]{tagger} Timmer, J., K\"{o}nig, M., 1995, A\&A 300,
707

\bibitem[2002]{} Thorne, K. S., Black holes and time warps : Einstein's outrageous legacy, 1994,
Norton \& Company

\bibitem[]{} Vaughan, S., Edelson, R., Warwick, R. S., Uttley, P., 2003, MNRAS 345, 1271

\bibitem[2002]{yuan02} Yuan, F., Markoff, S., Falcke, H., 2002, A\&A 383, 854

\bibitem[2004]{yuan04} Yuan, F., Quataert, E. \& Narayan, R., 2004, ApJ  606, 894

\bibitem[2006a]{yusef-zadeh06a} Yusef-Zadeh, F., Bushouse, H., Dowell, C. D.,  et al., 2006a, ApJ 644, 198

\bibitem[2006b]{yusef-zadeh06b} Yusef-Zadeh, F., Roberts, D., Wardle, M., Heinke, C. O., Bower, G. C., 2006b,  ApJ 650, 189

\bibitem[2007]{yusef-zadeh07}   Yusef-Zadeh, F., Wardle, M., Cotton, W. D., Heinke, C. O., Roberts, D.
A., 2007, ApJ 668, 47

\bibitem[2008]{yusef-zadeh08}  Yusef-Zadeh, F., Wardle, M., Heinke,
C., et al. 2008, ApJ 682, 361

\bibitem[]{} Zamaninasab, M., Eckart, A., Kunneriath, D., et al. 2008a, MmSAI 79, 1054

\bibitem[]{} Zamaninasab, M., Eckart, A., Meyer, L., et al. 2008b, JPhCS 131a

\end{thebibliography}
\end{document}